# 飞秒激光光学频率梳的精密相位控制与相干脉冲合成的研究

**Precise control of optical phase and coherent synthesis in femtosecond laser based optical frequency combs**

一级学科：光学工程

学科专业：光学工程

作者姓名：田昊晨

天津大学精密仪器与光电子工程学院

二〇二〇年五月

# 摘要

光学频率梳是指能够产生离散、等间距、高相干度且拥有梳状光谱结构的激光光源。光学频率梳技术的诞生为光频稳定度与射频稳定度之间的传递建立了桥梁。该技术在极大地促进精密光谱学发展的同时，也对光学时间/频率基准之间的传递、光学分频技术、时间/频率基准的远距离传输、高精度距离测量等应用的发展做出了非常重要的贡献。飞秒激光器得益于其宽光谱输出的特点，成为了实现光学频率梳运转的最佳选择。对飞秒激光器输出的脉冲序列的重复频率和载波-包络偏移频率进行高精度控制后，便可以得到时域和频域均稳定的飞秒激光光学频率梳。本文围绕飞秒激光光学频率梳的重复频率与载波-包络相位的精密控制及相干脉冲合成进行了研究，本文的工作概括如下：

1、系统研究了飞秒激光光学频率梳输出脉冲的定时抖动、载波-包络相位噪声和梳齿噪声的理论模型，分析了噪声来源。研究了飞秒激光光学频率梳稳定性评价方法。功率谱分析法和方差分析法分别从频域和时域的角度对相位锁定后的光学频率梳进行稳定性评价。研究了在光电探测过程中，光电探测器产生的热噪声和散粒噪声对噪声探测的信噪比产生的影响。

2、首次实现了两台独立的掺镱光纤飞秒激光器连续 5 天的高精度时间同步。同步后，环内光学互相关系统输出的剩余时间误差信号的均方根值为 103 as，环外光学互相关系统输出的剩余时间误差信号的均方根值为 733 as。环内交叠的艾伦方差的 $1.31\times10^5$ s 稳定度为 $8.76\times10^{-22}$。环外交叠的艾伦方差的 $1.31\times10^5$ s 稳定度为 $1.36\times10^{-20}$。环内噪声功率谱在 10 kHz 至 10 MHz 频率范围的积分值为 430 as。研究了光纤引入的额外定时抖动，5 米长的单模光纤引入的定时抖动的积分值为 64.7 as。提出了基于电学降频法和互相关算法的光学外差探测法。

3、首次使用光学外差探测法实现了超高范围、超高精度的载波-包络相位噪声的功率谱测量。在 5 mHz 至 8 MHz 傅立叶频率范围内，实现了动态范围高于 270 dB 的载波-包络相位噪声功率谱测量。相位噪声测量基底低于 1 μrad/√Hz。使用功率谱分析、Hadamard 方差分析和 Kendall 互相关分析法研究了载波-包络相位噪声的来源。首次测量了孤子分子对中的两个光学孤子的相对相位噪声。两个光学孤子的相对相位噪声的的积分值仅为 3.5 mrad，相对线宽仅为 μHz 量级。

4、首次使用非对称光纤延迟线干涉仪对掺铒光纤飞秒激光器的光学梳齿的频率噪声进行了高精度测量。系统研究了掺铒飞秒激光光学频率梳中 $f_{rep}$、$f_{ceo}$ 和 $\nu_n$ 的频率噪声功率谱。分析了光学频率梳中定时抖动和所有梳齿的噪声来源，研究了 $n\times f_{rep}$ 噪声与 $f_{ceo}$ 噪声之间的反相关特性。系统验证了 $f_{rep}$、$f_{ceo}$ 和 $\nu_n$ 的锁相

环环路之间的串扰。

5、首次实现了两台独立的掺镱光纤飞秒激光器的长期稳定相干脉冲合成。使用平衡光学互相关系统锁定重复频率，使用腔外声光调制器的反馈控制实现相对载波-包络偏移频率的锁定，实现相干脉冲合成。合成后，两脉冲的相对剩余定时抖动的积分值为380 as，剩余的相对载波-包络噪声的积分值为375 mrad。使用 Mach-Zehnder 干涉仪对相干合成后的脉冲的相干性进行评估，合成后的光谱干涉的对比度为58%。60分钟内，两脉冲的相对载波-包络相位漂移的均方根值为165 mrad。并且两激光器的光斑有明显的干涉条纹。

# ABSTRACT

Optical frequency combs are laser sources which are capable of generating discrete, equal-spaced and highly coherent comb modes. Optical frequency comb technique provides a significant bridge to transfer the stability between optical frequency and radio frequency. The advances of this technology greatly promote the development of precision spectroscopy, optical time/frequency transfer, optical frequency division, long-distance transfer of time/frequency references and high-precision distance measurement. Benefiting from the wide spectral outputs, femtosecond lasers have become the best choice for the fulfilment of optical frequency combs. Within the precise control of the repetition frequency and carrier-envelope offset frequency of the pulse train from femtosecond lasers, a stable optical frequency comb both in the time domain and frequency domain can be obtained. Based on the precise control of repetition rate, carrier-envelope offset frequency and conherent pulse synthesis in optical frequency combs, the major contents in this synthesis are summarized as follows:

1. The theoretical models of timing jitter, carrier-envelope phase noise and comb line noise are systematically studied. The origins of noise are analyzed. The methods of frequency stability evaluation of femtosecond laser have been studied. Power spectrum analysis and the variance analysis could reveal the stability of optical frequency combs in frequency domain and time domain, respectively. The impacts of the thermal noise and shot noise generated by the photodetector during the photo-electric detection on the signal-to-noise ratio of signal have been studied.

2. The precise optical synchronization between two independent yetterbium-doped fiber optical frequency combs for 5 consecutive days has been achieved for the first time. After synchronization, the rms value of residual time error signal from in-loop optical cross-correlation system is 103 as. The rms value of residual time error signal from out-of-loop optical cross-correlation system is 733 as. Overlapping allen variance stability of in-loop error signal is $8.11\times10^{-22}$ in $1.31\times10^{5}$ s. Overlapping allen variance stability of out-of-loop error signal is $1.36\times10^{-20}$ in $1.31\times10^{5}$ s. The integration of the in-loop noise power spectrum in the frequency range from 10 MHz to 10 kHz is 430 as. The excess timing jitter induced from optical fiber has been studied. Integrated timing jitter of 5-meter long fiber is 64.7 as. The optical heterodyne method based on three-mixer design and cross-correlation algorithm has been studied.

3. The optical heterodyne detection method is applied to achieve ultra-broad range and ultra-high-precision power spectrum measurement of carrier-envelope phase noise for the first time. The carrier-envelope phase noise power spectrum measurement with a dynamic range exceeding 270 dB in the Fourier frequency range from 5 mHz to 8 MHz is achieved. The phase noise measurement resolution is below 1 μrad/√Hz. Using power spectrum analysis, Hadamard variance analysis and Kendall cross-correlation analysis, the origin of noise has been revealed. The relative phase noise between two optical solitons has been characterized for the first time. The integrated relative phase noise of the two optical solitons is only 3.5 mrad and the relative line width is only at μHz level.

4. The power spectra have been characterized using an asymmetric fiber delay line with high precision for the first time. The frequency noise power spectra of $f_{rep}$, $f_{ceo}$ and $\nu_n$ in the optical frequency comb are measured. The origins of timing jitter and comb-line noise have been studied. The anti-correlation characteristics between $n\times f_{rep}$ noise and $f_{ceo}$ noise are discovered. The system verified the crosstalk between $f_{rep}$, $f_{ceo}$ and $\nu_n$ stabilization phase-locked loops.

5. The long-term stable coherent pulse synthesis of two independent yetterbium-doped fiber has been realized for the first time. A balanced optical cross-correlation system and an extra-cavity acousto-optic frequency shifter are applied to lock the repetition frequencies between two lasers and the relative carrier-envelope offset frequency respectively, to achieve coherent pulse synthesis. After synthesis, the integration value of the residual time jitter of the two pulses is 380 as and the integration value of the residual relative relative-envelope noise is 375 mrad. The Mach-Zehnder interferometer was used to evaluate the coherence of the two pulses after coherent synthesis. The contrast of the spectral interferemetric fringes after synthesis is 58%. Within 60 minutes, the rms value of relative carrier-envelope phase drift between the two pulses is 165 mrad. And the spots of the two lasers have obvious fringes.

# 目录

# 第1章 绪论

“年，熟谷也。”

古代时期人们通过太阳的出没、月亮的盈亏和庄稼成熟的物候形成了对时间的认知。从机械钟表到原子钟，人类对时间/频率的计量的准确度都在不断地提高。时间，作为七个基本物理量之一，其计量单位是秒。在第十三届国际计量大会上，国际计量委员会将秒定义为铯 133 原子基态的两个超精细能级之间跃迁时所辐射的电磁波周期的9,192,631,770倍的时间。2018年，国际计量委员会通过“修订国际单位制”决议，对四项基本单位的定义进行了更新。质量单位、电流单位、温度和物质的量均转为由自然常数定义。特别的是，这些定义都间接关联于时间单位秒。由此可见，时间/频率基准定义的重要性日益提升。

2005 年，美国国家标准与技术研究院（NIST）的 John L. Hall 与马克思普朗克量子光学研究所的 Theodor W. Hänsch 获得了诺贝尔物理学奖，以表彰他们对光学频率梳技术和基于激光的精密光谱学做出的贡献[1-3]。光学频率梳技术的出现在极大地促进精密光谱学发展的同时，作为高精度的时间基准与频率基准，光学频率梳也对光学时间频率基准之间的传递、光学时间频率基准向射频的传递、时间频率基准的远距离传输、距离测量等其他应用的发展起到了至关重要的作用。

本文将围绕飞秒激光光学频率梳的重复频率与载波-包络相位的精密控制与相干脉冲合成进行研究：系统研究了飞秒激光光学频率梳的定时抖动特性，首次实现了两台飞秒激光器的时间同步；系统研究了飞秒激光光学频率梳的载波-包络相位噪声特性，首次实现了飞秒激光器的载波-包络相位噪声的高精度测量；首次实现了两台独立的掺镱光纤激光器的长期稳定相干脉冲合成；最后，系统研究了掺铒光学频率梳中 $f_{rep}$、$f_{ceo}$ 和 $\nu_n$ 的频率噪声功率谱。这些成果从理论和实验方面为低噪声的光学频率梳运转提供了基础。

## 1.1 飞秒激光光学频率梳中噪声的测量与稳定

### 1.1.1 飞秒激光光学频率梳输出脉冲的定时抖动

飞秒激光器输出脉冲序列的重复频率的噪声在时域上表现为脉冲时域包络

的定时抖动。脉冲的定时抖动由多方面噪声源引入，如激光器腔内的放大自发辐射、激光腔的非零色散、脉冲的强度波动以及可饱和吸收体的恢复时间。定时抖动的理论模型将在第二章中详细介绍。

相比于传统的微波振荡器，飞秒激光器输出的脉冲序列有着极低的定时抖动特性。因此，需要一种高精度、高分辨率的测量技术对飞秒激光器的定时抖动性能进行分析。在 1986 年，一种通过直接对飞秒激光器输出脉冲序列的重复频率的射频谱进行分析，从而得到定时抖动的方法被提出[4]。使用高速光电探测器（> 1 GHz 带宽）对脉冲序列重复频率的高次谐波（$n \times f_{rep}$）进行测量，通过分析高次谐波的线宽，即可得到脉冲序列的定时抖动功率谱密度。需要注意的是，由于自由运转的飞秒激光器的定时抖动不是一个静态过程，因此该此方法无法准确测量低频范围的定时抖动功率谱[5]。为了提高定时抖动的测量精度，通常会将另一台飞秒激光器作为参考激光器。参考激光器的定时抖动比待测飞秒激光器的定时抖动更低或者相近。由于两台激光器是完全独立的，其定时抖动完全不相关。两台激光器相对的定时抖动功率谱除以 2 即可得到单台激光器的定时抖动功率谱。在基于光电探测的相位鉴别法中，需要对待测激光器和参考激光器的重复频率高次谐波进行提取，分别将两台激光器的高次谐波滤出后进行相位鉴别，最后得到激光器重复频率的相位噪声谱。2001 年 Scott 等人使用光电探测相位鉴别法对钛宝石飞秒激光器的定时抖动进行了测量[6]。随后，使用同样的方法，其他多种类型的光纤飞秒激光器，如孤子锁模激光器[7]、自相似孤子锁模激光器[8]、碳纳米管锁模激光器[9]、高重复频率飞秒激光器[10]、$MoS_2$ 锁模激光器[11]的定时抖动也依次被测量。近年来，由于商用化相位噪声分析仪的快速发展，Keysight 公司、Rohde&Schwarz 公司均研发出了高精度基于射频波段的相位噪声分析仪。相位噪声分析仪的原理是将仪器内部的低噪声振荡器相位锁定至待测信号，通过多路互相关分析得到待测信号的相位噪声功率谱。相位噪声分析仪可以直接测量高速光电探测器输出的高次谐波信号的相位噪声，实现定时抖动相位谱的测量。相比于光电探测相位鉴别法，由于省去了参考激光器，实验装置被很大程度地简化。然而，使用光电探测器对脉冲序列直接探测的过程中，光电探测器的热噪声和散粒噪声会将功率谱的测量范围限制在 1 kHz 至 10 kHz 偏频[12]。同时，如果入射的光功率过高，光电探测器会产生高阶非线性效应，使得功率谱产生畸变[13]。

为了克服光电探测器的限制，进一步提高定时抖动的测量分辨率，光学平衡互相关技术（Balanced optical correlation，BOC）在 2003 年被 Schibli 等人提出[14]。在光学平衡互相关技术中，同样需要将一台激光器作为参考激光器，一台激光器作为待测激光器。两台激光器输出的脉冲序列经过一块非线性晶体产生和频信号，该和频信号的强度与两脉冲的时间位置信息相关，由此可以实现阿秒（attosecond，

$10^{-18}$ s）量级分辨率的定时抖动测量。通过使用平衡探测器对和频信号进行探测，激光器自身强度噪声的干扰会被消除。在 2007 年，光学平衡互相关技术被首次应用于光纤激光器的定时抖动测量。结果表明，在高于 10 kHz 偏频部分，掺铒光纤激光器的定时抖动仅为 5 fs[12]。在 2012 年，商用钛宝石激光器的定时抖动由平衡光学互相关技术测得，在 100 Hz 至 41 MHz 频率范围内，定时抖动的均方根值仅为 13 as[15]。之后，平衡光学互相关技术已经被广泛应用到各种类型的、不同重复频率的固体激光器[16-17]、光纤激光器[18-23]以及光纤放大器[24-25]的定时抖动测量中。特别地，在 2015 年，Jung 等人从锁相环的纠正信号中提取出了锁定带宽之内的功率谱，从而实现了动态范围高达 340 dB 的定时抖动功率谱测量[26]。随后，Xin 等人在光学平衡互相关技术上进行了改进，提出了偏振噪声抑制的光学平衡互相关（Polarization-noise-suppressed balanced optical correlation，PNS-BOC）技术[27]。通过在光学平衡互相关之前加入一块双折射晶体，可以有效地降低不同偏振态脉冲的串扰，提高误差信号的信噪比。2015 年，Hou 等人提出了基于光学外差探测的激光器定时抖动测量技术[28]。该技术将两台激光器输出的两个不同波长分别做拍，通过将两个拍频信号混频，可以消除两台激光器相对载波-包络偏移频率的影响，得到仅与两台激光器重复频率差相关的误差信号。相比光学平衡互相关技术，光学外差法中避免了脉冲的二阶非线性过程。因此，依赖于一阶线性过程的光学外差探测法可以允许更高的测量精度。光学外差探测技术可以实现幺秒（yoctosecond，$10^{-24}$ s）量级的定时抖动测量分辨率。为了避免参考激光器的引入，韩国高等技术研究院（KAIST）的课题组使用包含非对称光纤延迟线的干涉仪对一台激光器的定时抖动进行了直接测量[29]。与光学外差法的原理基本相同。该技术使用公里量级长的光纤延迟线对激光器输出的脉冲进行延迟，由于脉冲的定时抖动在激光腔内是积累的，光纤延迟线起到了对定时抖动放大的作用，将延迟后的脉冲与未延迟的脉冲进行比较，即可得到激光器的定时抖动信息。与光学平衡互相关技术、光学外差探测技术相比，非对称光学延迟线干涉仪技术是一种更实用、更加低成本的定时抖动测量技术。该技术甚至可以被应用到超连续光谱的定时抖动测量中[30]。

在实现了高精度的定时抖动功率谱测量后，如何降低飞秒激光器输出脉冲的定时抖动同样是一个重要的研究课题。降低激光器输出脉冲的定时抖动主要有两类方法。第一类是从飞秒激光器本身入手，通过优化激光器参数来降低腔内的量子噪声。2011 年，Song 等人报道了在孤子锁模激光器、自相似锁模激光器和色散管理孤子锁模激光器中，激光器的输出脉冲对应不同特性的定时抖动功率谱[20]。进一步地，该课题组报道了在不同的腔内净色散条件下，脉冲的定时抖动功率谱存在差异，并且得到了当腔色散接近于零色散时，Gorden-Haus 抖动最低

的结论[18]。这与文献[31-32]中的理论模型相吻合。另一方面，通过在激光器腔内加入一窄带滤波器也可以达到降低 Gorden-Haus 抖动的效果。该技术的主要原理是，Gorden-Haus 抖动是由脉冲的光谱的频率波动通过腔内非零色散引入的，腔内窄带滤波器降低了脉冲的光谱的频率波动，继而可以降低 Gorden-Haus 定时抖动。文献[33]首次报道了在碳纳米管锁模的飞秒激光器中加入带宽为 0.7 nm 的窄带滤波器降低定时抖动。2015 年，带有腔内窄带滤波器的飞秒激光器的定时抖动特性被更加系统地研究[34]。在腔中加入滤波器后，由于 Gorden-Haus 抖动被很大程度地去除了，因此，腔内净色散对定时抖动的影响变得微乎其微。但是，当腔内的滤波器带宽过窄，小于 1 nm 时，由于腔内脉冲脉宽的变宽，定时抖动性能会被恶化[35]。在高重复频率的非线性偏振旋转（Nonlinear polarization rotation，NPR）锁模激光器中[36]，100 kHz 以下的频率对应的定时抖动主要由脉冲的强度噪声引入。在该激光器中，可以通过激光器的强度噪声锁定来降低脉冲的定时抖动。在文献[37]中，通过泵浦电流调制对输出脉冲的强度进行稳定后，定时抖动功率谱在 3 kHz 至 30 kHz 频率范围内有 10 dB 的降低。

第二类定时抖动的降低方法是使用锁相技术，将飞秒激光器输出脉冲的重复频率锁定至低噪声的射频参考源上，实现脉冲重复频率的稳定。射频参考源可以是低噪声的信号发生器或者蓝宝石负载谐振腔振荡器[38]。最常用的技术手段是使用高速光电二极管与微波混频器实现激光器和射频参考源实现相位鉴别。除此之外，平衡光学-微波相位探测（Balanced optical-microwave phase detector，BOM-PD）技术[39]、光纤环路光学-微波相位探测（Fiber-loop optical-microwave phase detector，FLOM-PD）技术[40]和 Mach-Zehnder 调制器技术[41]均可以实现更高精度的、长时间的光学脉冲与射频信号的相位鉴别。为了实现脉冲重复频率的稳定，需要对激光振荡器的腔长进行快速地调节。将激光腔的一个端镜置于压电陶瓷上，改变加载在压电陶瓷的电压，即可实现激光器腔长的调节。并且，通过对压电陶瓷底座的特殊设计，可以将重复频率的调节带宽提高到 100 kHz 以上[42]。对于光纤激光器，可以将光纤缠绕在压电陶瓷上。文献[43]中报道了几种特殊的光纤缠绕方法，可以实现高速调节激光器的重复频率。另一方面，在激光器腔内插入一个电光调制器，通过改变施加在电光晶体上的电压，晶体的折射率被改变，也可以对重复频率进行调节[44]。相比于压电陶瓷，电光调制器的调制带宽更高，很容易达到 100 kHz 以上甚至接近于 1 MHz 的调制带宽。电光调制器的缺点是，其调节激光器的重复频率范围较小。因此，可以将压电陶瓷作为大调节范围的低速锁定器件，电光调制器小调节范围的高速锁定器件，两者相结合，实现高带宽、长期稳定的重复频率锁定，降低输出脉冲的定时抖动[45]。

低定时抖动的飞秒脉冲作为高精度的时间基准，可以被应用到诸多应用中。

其中，低定时抖动的激光源是实现高精度时间基准的远距离传输的重要前提。当飞秒脉冲作为时间基准在公里量级的光纤链路或者自由空间传输过程中，使用光学平衡互相关、光纤环路光学-微波相位探测或者光学外差探测等技术均可以对传输链路引入的额外定时抖动进行实时的探测，通过加入可调节的光纤延迟或者空间延迟，可以实现对传输过程中引入的定时抖动的高精度的、实时的补偿。在文献[46-51]中，均实现了在公里长度范围内，精度在阿秒量级、1000 s 稳定度在 $10^{-20}$ 量级、时间达到天量级的稳定的时间基准传输。高精度、高稳定度的时间基准的远距离传输技术在大尺度科学仪器，如 X 射线自由电子激光器中的时间基准分布[52-53]、相干孔径合成[54]等技术中有着至关重要的作用。另外的，具有低定时抖动特性的飞秒激光器也是高精度光学采样和光子模拟数字转换器[55]、激光雷达[56]、光学互联技术[57-59]的重要光源。

### 1.1.2 飞秒激光光学频率梳的载波-包络相位

飞秒激光器输出脉冲序列的载波-包络相位表现为脉冲电场的峰值与包络的峰值在脉冲传输过程中周期性的不重合。载波-包络相位的存在主要是由于飞秒激光器内非零的净色散导致光学脉冲的群速度与相速度产生差异，继而产生脉冲的包络与包络下电场在时域相互扫描的现象。飞秒脉冲的载波-包络相位由多方面噪声源引入，如激光器腔内的放大自发辐射、激光腔的损耗、激光器泵浦源的抖动、激光器腔长的波动等因素。飞秒脉冲的载波-包络相位噪声的理论模型将在第二章中详细介绍。

对于飞秒激光器输出脉冲的载波-包络相位探测，最常用的方法是基于 $f$-$2f$ 干涉仪的自参考法[60]。$f$-$2f$ 干涉仪的基本原理见图 1-1（a），将飞秒激光器输出的脉冲使用高非线性光纤进行超连续光谱展宽后，得到覆盖范围超过一个倍频程的光谱。该光谱中包含 $\nu_n = n\times f_{rep} + f_{ceo}$ 和 $\nu_{2n} = 2n\times f_{rep} + f_{ceo}$ 的部分。将第 $n$ 个频率梳齿通过非线性晶体倍频，得到 $2\nu_n = 2n\times f_{rep} + 2f_{ceo}$。将 $\nu_{2n}$ 与 $2\nu_n$ 做拍，即可得到飞秒激光器载波-包络频率信号 $f_{ceo}$。图 1-1（b-d）中展示了几种常见的 $f$-$2f$ 干涉仪的结构示意图。在早期的钛宝石飞秒激光器的 $f_{ceo}$ 信号探测中，通常使用非共路（Non-common path，NCP）的 $f$-$2f$ 干涉仪结构[61]，如图 1-1（b）所示。非共路的结构的主要缺点是环境扰动会很容易对干涉仪的两臂产生非共模的影响。$f_{ceo}$ 信号信噪比降低的同时，非共模噪声也会耦合到 $f_{ceo}$ 噪声的探测当中。因此，使用共路（Common path，CP）的干涉仪可以消除非共模噪声的影响，见图 1-1（d）。将高非线性光纤输出的光谱的高频与低频部分使用色散补偿光纤进行色散补偿，保证高频与低频脉冲在时域上重合的同时，消除非共模噪声的影响。该设

计常见于全保偏光纤的光学频率梳系统中[43]。共路干涉仪有结构简单，不包含空间光路等特点。但是，由于高频和低频部分的脉冲的空间相干长度十分有限，$f_{ceo}$ 信号的信噪比对两脉冲的时域重合程度十分敏感。然而由于切割、熔接色散补偿光纤的精度是十分有限的，无法十分精确地调节光谱的高频与低频部分在时域上的延迟，因此该方法并不能得到高信噪比的 $f_{ceo}$ 信号。出于此原因，准共路（Quasi-common path，QCP）的 $f$-$2f$ 干涉仪结构被设计出来[62]，结构见图 1-1（c）。在共路设计的基础上，采用一个双色镜将光谱的高频与低频部分分开，随后调节双色镜后的反射镜位置，使得两脉冲在时域上精确重合。由于准共路的设计兼备了延迟可调节和结构简单的特点，已经被广泛应用于各种激光器的 $f_{ceo}$ 信号探测中。

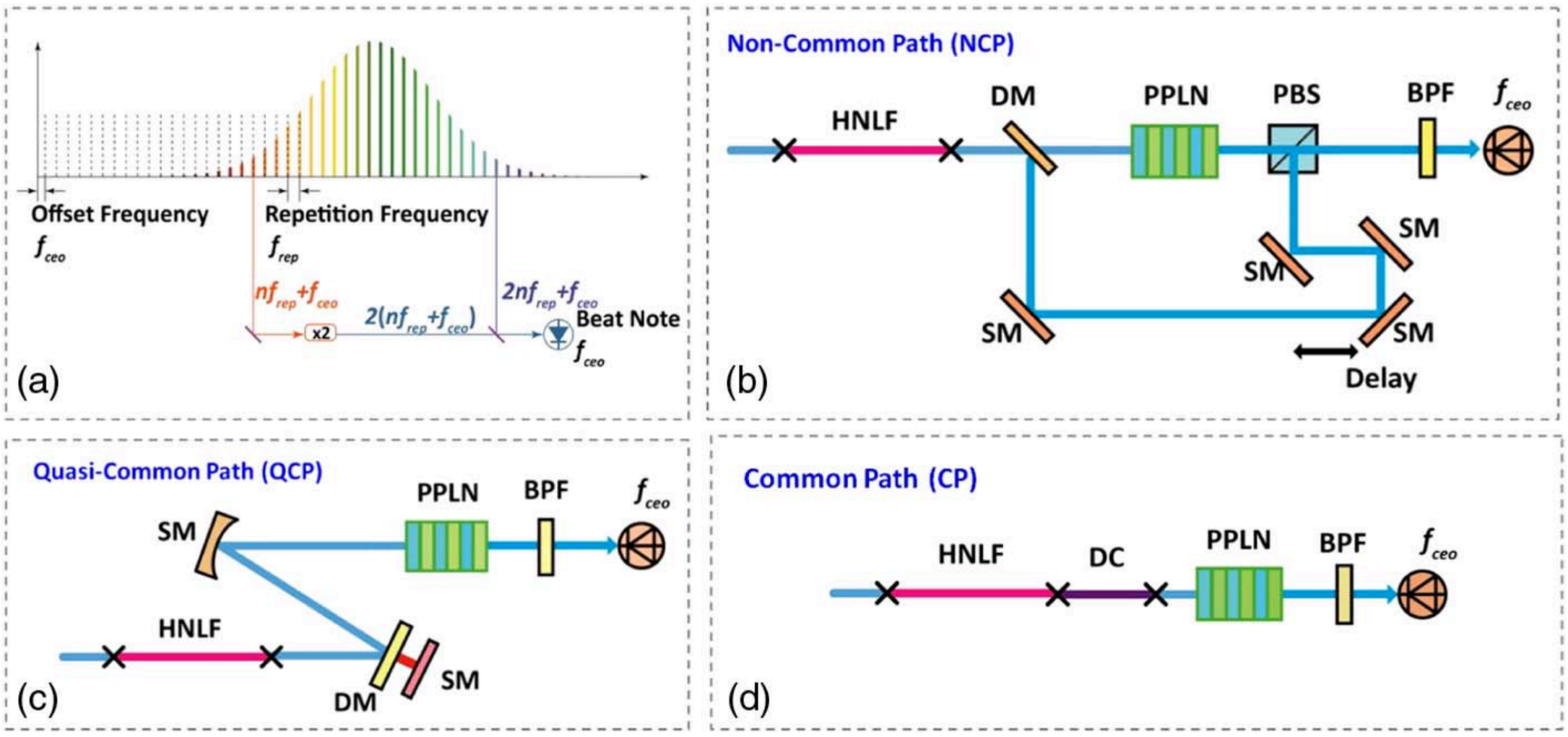

图 1-1 （a）$f$-$2f$ 干涉仪原理示意图。（b）非共路 $f$-$2f$ 干涉仪。（c）准共路 $f$-$2f$ 干涉仪。（d）共路 $f$-$2f$ 干涉仪。

高信噪比的 $f_{ceo}$ 信号对于该信号频率噪声的测量和稳定都是至关重要的。在超连续的产生过程中，由于振幅-相位调制等多方面因素的影响，超连续光谱的相干性会下降。这使得探测 $f_{ceo}$ 信号时，探测基底变高，信噪比降低。因此通过优化高非线性光纤的设计，使用全正色散光纤的自相位调制效应进行光谱展宽，避免超连续产生过程中孤子波和色散波的产生，优化超连续的过程，可以使得光谱的相干性得到保持。另一方面，$f_{ceo}$ 信号的信噪比一般受限于光电探测器产生的散粒噪声[63]。在光电探测器之前使用具有几纳米带宽的光学滤波器进行窄带通滤波，可以有效地降低探测器的散粒噪声，提高 $f_{ceo}$ 信号的信噪比。此外，使用平衡的光电检测可以消除超连续光谱振幅的波动，提高 $f_{ceo}$ 信号的信噪比[64-65]。通常，使用 $f$-$2f$ 干涉仪对 $f_{ceo}$ 信号进行探测，1 MHz 偏频以上的功率谱部分开始受限于探测器的散粒噪声或者热噪声基底。在 100 kHz 分辨率带宽下，光纤激光

器的 $f_{ceo}$ 信号的信噪比可以达到 40 dB 左右[43, 66-70]。对于固体飞秒激光器，由于其本身的低噪声特性，$f_{ceo}$ 信号的信噪比最高可以达到 60 dB[71-72]。另外需要指出的是，在文献[73]中，将 $f$-$2f$ 干涉仪的 1040 nm 部分使用掺镱光纤放大器放大，由于 1040 nm 部分在非线性晶体上的倍频效率与入射脉冲的光功率成指数相关，相比于未被放大的脉冲，有着更高的倍频效率。该手段同样可以达到提高 $f_{ceo}$ 信号的信噪比的目的。但是，1040 nm 部分脉冲需要耦合进入光纤放大器中，在此过程中，信号的信噪比已经有了 3 dB 左右的损失。并且放大器在放大光学脉冲的过程中，由于放大器品质因数的问题，也会使得信噪比下降至少 3 dB。该损失在后续放大过程中是无法得到弥补的。因此，在使用该技术时，需要考虑放大器损失的信噪比和非线性过程提升的信噪比孰重孰轻的问题。在更多情况下，该技术更适用于不易产生高于 40 dB 信噪比的 $f_{ceo}$ 信号的激光器，例如半导体飞秒激光器[74]、微环谐振腔[75]等光源。而对于已经可以产生很高 $f_{ceo}$ 信号信噪比的激光器，该技术并不适用。

除了 $f$-$2f$ 干涉仪技术，还可以使用电学混频技术对 $f_{ceo}$ 信号进行探测。在 2015 年，Brochard 等人将连续波激光器与掺铒光纤飞秒激光器做拍，通过将分频后的拍频信号与飞秒激光器的重复频率信号的第 60 次谐波混频，可以得到 $f_{ceo}$ 信号的噪声功率谱[76]。在使用直接电学混频技术得到的噪声功率谱中，高于 100 kHz 偏频部分，测量开始受限于仪器本底噪声的基底。2018 年，Tian 等人提出了使用光学外差法探测两台光纤激光器相对载波-包络相位噪声的方法[77]。在该方法中，首先使用光学平衡互相关技术将两台光纤飞秒激光器的重复频率进行锁定。然后将两台光学频率梳齿直接做拍，就可以得到两台光纤激光器相对载波-包络相位噪声。使用腔外声光调制器将该 $f_{ceo}$ 信号锁定至 3 MHz 频率基准，通过测量分析锁相环输出的纠正信号和误差信号，可以得到 2 mHz 至 2 MHz 傅里叶频率范围内 $f_{ceo}$ 信号的噪声功率谱图。除了在频域探测以外，Kim 等人还探索了一种时域探测 $f_{ceo}$ 信号的方法[78]，该方法在使用光学平衡互相关锁定激光器重复频率的基础上，使用干涉互相关（Interferometric cross-correlation，IXCOR）从时域上探测掺铒光纤飞秒激光器的 $f_{ceo}$ 信号。将来自干涉互相关的误差信号进行傅立叶变换后，可以实现偏移频率达到 2 MHz 的 $f_{ceo}$ 信号功率谱测量。

同样的，在实现了 $f_{ceo}$ 信号的高精度探测后，对飞秒激光器输出脉冲的载波-包络相位的精密锁定是得到低噪声光学载波的重要前提。由于激光器的载波-包络偏移频率对泵浦激光器的功率波动十分敏感，在早期的钛宝石飞秒激光光学频率梳中，使用声光调制器对激光器的泵浦光功率进行调制，进而飞秒脉冲的载波-包络偏移频率得到调节，从而达到载波-包络相位锁定的目的[60, 79-81]。随后，在光纤飞秒激光光学频率梳中，也采用了相同的技术手段，直接对激光器泵浦二极

管的电流进行调制，锁定飞秒脉冲的载波-包络相位[43, 82]。使用泵浦功率调制锁定载波-包络相位的主要瓶颈是，激光器中增益介质的上能级粒子寿命会将载波-包络相位的稳定带宽限制在 kHz 量级，若想提高锁定带宽，需要对比例积分微分伺服控制器的参数进行优化，加大锁相环路的微分增益。在文献[45]中，通过特殊设计泵浦电流调制电路，优化锁相环参数，实现了~ 1 MHz 带宽的载波-包络相位锁定。在 2010 年，Koke 等人提出在激光器腔外使用声光频移器将光学频率梳齿的整体频率频移，从而实现整个光学频率梳的载波-包络偏移频率锁定[83]。由于该技术是在激光器腔外进行调制，且声光调制器的调制带宽很容易达到 MHz 量级，使用前馈的锁相技术，该课题组实现了剩余噪声为 45 mrad、对应剩余定时抖动为 12 as 的载波-包络相位稳定。进一步的，该课题组通过优化 *f-2f* 中的非线性晶体的长度，并与泵浦功率反馈相结合，实现了剩余噪声为 20 mrad、对应剩余定时抖动仅为 10 as 的载波-包络相位稳定结果[84]。并且在 2012 年，Lucking 等人实现了对载波-包络相位的长时间稳定[85]。另一方面，调制飞秒激光器腔内的损耗同样可以达到锁定载波-包络相位的效果。并且，损耗调节不会受到增益介质上能级粒子寿命的限制。2015 年，Kuse 等人通过调制掺铒光纤飞秒激光器中石墨烯上的电压，达到高带宽调节激光器腔损耗的目的，进而实现带宽在 MHz 量级的载波-包络偏移频率的锁定[68]。并且，该手段被广泛应用于不同的光纤激光器，如全保偏非线性放大环形镜激光器[67]、掺铥光纤激光器[70]的载波-包络相位稳定中。以上技术手段都需要载波-包络偏移频率锁定至一个射频参考源。因此，得到的脉冲序列的脉冲包络下电场是呈周期性的。2018 年，Okubo 等人提出了一种将载波-包络偏移频率锁定至重复频率的方法[86]。通过平衡探测技术，将 $f_{rep}+f_{ceo}$ 信号和 $f_{rep}-f_{ceo}$ 信号均锁定至脉冲的重复频率，实现了无参考的载波-包络偏移频率锁定。用此技术得到的脉冲序列中每一个脉冲包络下电场都是完全相同的。2011 年，Krauss 等人提出了一种被动锁定掺铒光纤激光器的载波-包络相位的方法[87]。通过将掺铒光纤激光系统输出的脉冲使用高非线性光纤进行光谱展宽后，将得到的孤子波和色散波进行差频。由于孤子波和色散波源于同一个激光源，两脉冲之间几乎没有相对的载波-包络相位噪声。差频后得到的 1550 nm 波段的飞秒脉冲的载波-包络相位是自稳定的。该自稳定载波-包络相位系统的相位噪声很低，在 10 μHz 至 50 MHz 频率范围内，噪声功率谱密度的积分值只有 250 mrad[88]。

当飞秒脉冲的时域宽度小于或者接近于单个光学周期时，脉冲的电场相位的稳定就变得十分重要。即要保证飞秒脉冲序列中，脉冲与脉冲之间有着良好的复现性。因此飞秒脉冲的载波-包络相位的稳定在多路相干脉冲合成产生少周期脉冲[89-90]、产生任意波形电场[91]等应用中起着非常重要的作用。另一方面，在飞

秒脉冲高次谐波产生[92]、X 射线波段的阿秒脉冲产生[93-95]、阿秒脉冲与物质相互作用的瞬态过程测量[96]、少周期脉冲的极限非线性过程[97]等研究中，飞秒脉冲的相位稳定也扮演着至关重要的角色。与此同时，用有稳定载波-包络相位的激光脉冲还可以作为一种高精度的测量工具。2016 年。Feng 等人在钛宝石飞秒激光器腔内插入一个电光晶体，在对电光晶体施加周期性调制电压后，通过测量激光器载波包络偏移频率的变化情况，可以精确测量电光晶体克尔常数[98]。后续的，Rustige 等人实现了激光器腔镜的周期性变化导致的腔内脉冲的多普勒效应的测量[99]。

### 1.1.3 飞秒激光光学频率梳的梳齿线宽

光学频率梳中，单根频率梳齿的噪声的测量通常是将窄线宽连续波激光器与光学频率梳中的某根梳齿进行光学外差探测[43, 45, 100]。通常，使用待测飞秒激光器与窄线宽单频激光器做拍，使用高速光电探测器可以探测到在射频波段的拍频信号。该拍频信号的噪声同时与参考激光器和待测光学频率梳的线宽相关。当参考激光器的噪声很低时，该信号的主要噪声来源为待测光学频率梳。由该技术可以得到 MHz 偏移频率的梳齿噪声功率谱。但是，窄线宽连续波激光器仅存在于某些波长，并且通常造价昂贵。其中，基于 Pounder-Driver-Hall 技术稳频至光学超稳腔的高频率稳定度连续波激光器的成本更加高昂。为了研制出一种无需参考激光、低成本的梳齿线宽噪声的测量方法，光学延迟线自外差探测法在 1990 年代左右被研发出来，并且首先被应用于测量连续波激光器的线宽[101-102]。该方法利用非对称的光纤干涉仪作为相位鉴别器，将被测激光器的频率波动转换为射频波段的电学信号的相位波动，该方法无需任何外部参考激光器。使用该方法已经成功实现了光学频率梳中梳齿噪声功率谱的测量，并且对激光线宽的分辨率在 kHz 量级[103-105]。2015 年，Coluccelli 等人将光学频率梳参考至一个低精细度的 Fabry-Pérot 腔，将 Fabry-Pérot 腔的透射曲线作为高精度的相位鉴别器，可以测得梳齿噪声在 1 MHz 偏移频率的功率谱密度[106]。

通过同时锁定光学频率梳的两个自由度，即脉冲的重复频率和载波-包络偏移频率可以实现梳齿噪声的稳定。在单根梳齿被稳定后，梳齿的相位相干性可以传递至光学频率梳中的所有梳齿。载波-包络偏移频率通过泵浦功率的反馈控制来稳定，而重复频率则通常将激光器腔长调节锁定至超稳腔稳定的连续波激光器。使用高速的腔长调节装置，例如高带宽压电陶瓷[42]或者电光调制器[107-108]，可以将窄线宽参考激光器的低噪声特性传递到光学频率梳中的所有梳齿，实现 Hz 量级的线宽稳定。需要指出的是，由于从射频域到光频域，信号的相位噪声会被放

大 $n^2$ 倍（$n$ 为梳齿模式数）。因此，如果将激光器重复频率锁定到射频参考源，射频源的噪声特性传递到光频过程中，噪声会被放大 $n^2$ 倍，无法实现梳齿线宽的压缩。在飞秒激光器中每个光学梳齿的噪声都被有效地抑制后，便在光频域得到了窄线宽的光学频率梳。光学频率梳的众多应用将在 1.2.2 小节中详细介绍。

### 1.1.4 飞秒激光光学频率梳的相对强度噪声

飞秒激光器输出脉冲的功率波动为脉冲的相对强度噪声。使用光电探测器与低频滤波器即可实现飞秒激光器输出脉冲相对强度噪声的测量。2009 年，Budunoğlu 等人首次对工作在不同色散域的飞秒激光器输出脉冲的相对强度噪声进行了系统的分析[109]。该工作指出相对强度噪声与激光腔品质因数大小相关，当品质因数越高时，脉冲的相对强度噪声越低。在 NPR 锁模的飞秒激光器中，当腔内净色散接近于零时，可获得最小的相对强度噪声[110]。另外，在 NPR 锁模的飞秒激光器中，通过在激光腔内加入一带宽为 7 nm 的窄带滤波器，也可以有效地降低激光器的相对强度噪声[34]。并且，滤波器的引入使得相对强度噪声与激光腔的净色散无关。类似地，在碳纳米管锁模的飞秒激光器和全正色散光纤飞秒激光器中，同样可以使用腔内滤波器来降低相对强度噪声[9, 111]。另外，文献[112]中提出了通过腔外滤波来实现降低相对强度噪声的方法。

同样地，可以通过反馈控制来稳定激光器输出脉冲的强度。一种是基于声光调制器的反馈技术，该技术控制声光调制器的透射率以补偿强度波动[107, 113]。另一种方法是基于泵浦激光器电流反馈以调节脉冲的强度波动[114-115]，可以将脉冲的强度抖动降低至-150 dBc/Hz 量级。

## 1.2 光学频率梳的发展及其应用的国内外现状

### 1.2.1 光学频率梳的发展概述

**固体飞秒激光光学频率梳：**将飞秒激光器输出脉冲的两个维度：脉冲的重复频率和载波-包络偏移频率同时得到稳定后，激光器输出的光谱中的每个光学梳齿便得到了稳定，可以得到稳定运转的光学频率梳。钛宝石飞秒激光器作为发展历史较为悠久，技术最为成熟的飞秒激光器，其窄脉冲、宽光谱和低噪声等特性是其他种类激光器至今都无法比拟的。基于钛宝石飞秒激光器的光学频率梳于 2000 年左右被研制出来。Ye 等人使用微结构光纤将钛宝石飞秒激光器输出的脉冲进行扩谱后，得到覆盖范围超过一个倍频程的光谱，将该光谱中的 1064 nm 部

分和 532 nm 部分分别与窄线宽的 Nd:YAG 激光器做拍，分别通过调节钛宝石激光器腔内压电陶瓷的长度和棱镜后反射镜的角度，可以将得到的两个拍频信号锁定至同一个射频参考源上，得到频率锁定运转的光学频率梳[116]。同年，Jones 等人首次使用 *f-2f* 干涉仪对钛宝石激光器的载波-包络偏移频率进行探测，通过激光腔内棱镜后端镜的角度调节和压电陶瓷的长度调节，实现了稳定运转的光学频率梳[117]。从此，基于光学频率梳的光学、射频频率比对及其他众多应用拉开了序幕。随后的几年里，NIST 的课题组实现了光学频率梳与 $I_2$ 稳频激光器[118]、$Hg^+$光钟、Ca 光钟[119]、$Al^+$光钟、Yb 光钟的比对[120]，以及光学频率梳之间的比对[121]。在射频频率基准向光学频率基准的传递，单个光学频率基准向光学频率梳的传递，射频基准与光学频率基准同时向光学频率梳的传递等应用中，光学频率梳作为一个有效的传递桥梁，将多种不同频率的基准连接起来。2006 年，Fortier 等人将钛宝石飞秒激光频率梳的重复频率提高到 1 GHz，在该激光器中，由于钛宝石激光器输出的光谱覆盖一个倍频程，因此，腔外无需额外的光谱展宽，即可实现载波-包络偏移频率的探测与锁定[120]。2009 年，Bartels 等人进一步将钛宝石飞秒激光光学频率梳的重复频率提高到了 10 GHz。该激光器的腔长只有若干厘米，并且，直接用光谱仪测量激光器输出的光谱，即可观察到梳状的结构[122]。

**光纤飞秒激光光学频率梳：**尽管钛宝石飞秒激光器有着低噪声的优良特性，伴随而来的是复杂的水冷系统以及昂贵的激光泵浦源。然而，光纤激光器，特别是输出波长在通信波段的掺铒光纤激光器，由于其较为低廉的激光泵浦源和紧凑集成的系统，受到了广泛的关注。2004 年，Washburn 等人将一台 8 字腔锁模的掺铒光纤飞秒激光器的重复频率和载波-包络偏移频率分别锁定至两个射频参考源上，实现了首台掺铒光纤激光器的光学频率梳的运转[123]。随后，由于光纤激光器可以实现高度集成化，掺铒光纤激光器如雨后春笋般迅速发展起来。在文献[124]中，首次实现了掺铒光纤激光器与光钟和 $I_2$ 稳频激光器的比对。文献[43]中，通过使用高度集成化的半导体可饱和吸收镜-波分复用器，和特殊的压电陶瓷设计结构，可以实现重复频率为 200 MHz 的全保偏光纤的光学频率梳[125]。全保偏光纤结构的光学频率梳有着很好的环境稳定性，可以很好地阻隔环境中声学噪声以及机械振动的影响[126]。甚至，经过特殊的设计，可以将所有光学器件和电学锁定器件高度集成起来，实现光纤光学频率梳的仪器化。经过特殊的环境抗干扰设计，该光纤频率梳甚至可以在颠簸的货车内仍然保持良好的性能[127]。随后，掺铒光纤频率梳的发展趋势向大范围重复频率调谐[128]、高重复频率[129]及低功耗[130]的方向发展。需要特别指出的是，在文献[130]中，使用特殊设计的光学波导代替高非线性光纤获得倍频程光谱，由于光波导有着更高的非线性效应，因此，对入射脉冲的能量要求更低。这大大降低了光纤放大器的负担，实现了低功

耗的光学频率梳。使用光波导作为 *f-2f* 干涉仪中的核心器件，进行载波-包络偏移频率的探测，可以极大的促进光梳的低功耗和集成化，必然是今后光学频率梳发展的大方向。于此同时，基于其他增益介质的飞秒激光光学频率梳也在蓬勃发展。在这其中，有掺铬橄榄石固体激光光学频率梳[131]、掺镱光纤激光光学频率梳[42, 132-135]、掺铥光纤激光光学频率梳[70]等等。

**其他新型光学频率梳：**除了基于飞秒激光器的光学频率梳，采用电光调制器[136-137]对 CW 激光器进行调制产生的光学频率梳、基于量子级联激光器（Quantum cascade laser，QCL）的光学频率梳[138]、基于半导体飞秒激光器[139]、基于微环谐振腔（Microresonator）的光学频率梳[140-141]均被研制出来。特别地，在微环谐振腔中产生稳定时域脉冲的研究是非常前沿的课题[142-145]。这些新型的光学频率梳覆盖的光谱范围不仅仅限于通信波段，已经拓展到 2 μm 以上的中红外波段。这些工作为中红外波段的分子指纹谱探测提供了可靠的光源[146]。

### 1.2.2 光学频率梳的应用概述

**光学频率基准之间的传递：**光学频率梳的出现为实现不同频率的光钟之间进行比对、射频基准向光学频率传递以及光学频率基准向射频传递提供了不可或缺的桥梁。2011 年，Nakajima 等人使用腔内电光调制器作为高带宽的激光器腔长的调节器，将两台光学频率梳同时锁定至同一个高稳定度的连续光激光器上，实现了相对线宽仅为 7.6 mHz 的两台光纤光学频率梳的运转[147]。并且，通过将两台光学频率梳输出的频率梳齿分别进行光纤放大，将不同路放大之后的梳齿进行稳定度的比对，得到了 1 s 稳定度为 $10^{-16}$ 量级，1000 s 稳定度为 $10^{-19}$ 量级的结果，该工作成功地实现了光学频率基准在不同光学频率梳之间的高稳定度传递。2017 年，Leopardi 等人通过对掺铒光纤光学频率梳中光纤部分的特殊设计，实现了单台光纤光学频率梳与原子钟的比对，1000 s 稳定度达到 $10^{-19}$ 量级[148]。

**光学频率基准向射频的传递—光生微波：**由于从射频域到光频域，信号的相位噪声会被放大 $n^2$ 倍。因此，如果将射频波段的频率基准通过光学频率梳传递至光学频率，其稳定度会严重下降。但是，相反地，如果将光频波段的频率基准通过光学频率梳传递至射频波段，光学频率基准的稳定度便会被大大保留下来，基于光学分频（Optical frequency division，OFD）的光生微波技术便使用了此原理。首先，将一个可调谐的连续光激光器锁定至一个光学超稳腔上，得到一个高稳定度的光学基准。将飞秒激光器的重复频率锁定至该光学基准，将载波-包络偏移频率锁定至一射频基准后，实现高稳定度的光学频率梳运转，使用光电探测器对光学频率梳输出的脉冲序列进行探测，便可以将光学频率分频，产生的高次

谐波信号可作为高稳定度的射频频率基准。2011 年，Fortier 等人将钛宝石飞秒激光光学频率梳作为光源，经过光学分频后产生了在 100 kHz 偏频处，噪声为-168 dBc/Hz 的 10 GHz 微波信号[149-150]。同年，Quinlan 等人使用掺铒光纤光学频率梳作为光源，经过光学分频后产生了在 100 kHz 偏频处，噪声为-145 dBc/Hz 的 10 GHz 微波信号[151]。2013 年，Meyer 等人使用 Yb:KYW 晶体为增益介质的光学频率梳作为光源，将光学分频后产生的微波信号与钛宝石光学分频的结果进行比较，得到了在 100 kHz 偏频处，噪声为-140 dBc/Hz 的 10 GHz 微波信号[152]。在上述光学分频的介绍中，产生的低噪声微波的噪声主要受限于光电探测器的热噪声和散粒噪声。降低二极管热噪声和散粒噪声影响的手段是提高入射激光的光功率，但是，对于大多数光电探测器来说，过高的入射光功率会导致光电探测器产生非线性效应，输出的电学信号产生畸变，恶化所产生的微波的噪声特性。因此，光学分频技术性能的进一步提升需要一种可以承受更高光功率的、拥有更高线性度的、高带宽的光电探测器。相对于普通的 P-I-N 光电探测器，单向载流子传输（Uni-travelling carrier，UTC）光电探测器中由于减轻了空间电荷效应，可以实现更高功率、高速和高线性度的光电探测。改进的单向载流子传输（Modified uni-travelling carrier，MUTC）光电探测器通过引入“悬崖”层来控制吸收层和集电极区域的相对电场强度，进一步提高了空间电荷的耐受性。实现了较高响应度（0.92 A/W @ 1550 nm）、更高功率（35 dBm）、高速（> 20 GHz）和高线性度（180 mA 饱和电流）的光电探测[153-155]。使用 MUTC 作为光学分频技术中的光电探测器，可以有效地提升微波信号的噪声特性。2013 年，Fortier 等人通过使用 MUTC 对钛宝石激光光学频率梳输出的脉冲信号进行探测，并且使用互相关算法对两套光学分频系统输出的微波信号进行处理，得到了在 10 MHz 偏频处，噪声基底在-177 dBc/Hz 的低噪声微波信号[156]。2014 年，Quinlan 等人首先使用光纤交错器对光纤激光光学频率梳输出脉冲的重复频率在腔外多次倍频，随后使用 MUTC 对光学频率梳输出的脉冲信号进行探测，并且使用互相关算法将光纤激光光学频率梳与钛宝石激光光学频率梳光学分频系统输出的微波信号进行对比，得到了在 10 MHz 偏频处，噪声基底在-175 dBc/Hz 的低噪声微波信号[157]。此外，还可以将光纤激光光学频率梳的两个梳齿分别锁定至两个不同波段的高稳定度连续光激光器上，对光学频率梳的脉冲进行光学分频[158]。2015 年，Jung 等人将光学频率梳锁定至光纤延迟线上，实现了低噪声微波信号的产生，使用光纤延迟线替代昂贵的超稳腔，可以大大降低产生低噪声微波的成本[29]。2018 年，Endo 等人使用自由运转的单片固体激光器作为光源，进行光学分频产生低噪声微波。利用单片固体激光器特有的低噪声特性，该文章中彻底摆脱了复杂、昂贵的重复频率和载波-包络偏移频率的探测和锁相系统，产生了 100 kHz 偏频处-180 dBc/Hz 的低噪声微波[41]。基于光学分频系统的光生微波技术可以产生低噪声性能微波，可以被应用于激光雷达、GPS 基准等应用中。并且可以作为直接数字频率合成器的时钟源，产生可调谐的、波长范围覆盖 L 波段、S 波段 C 波段甚至 X 波段的低噪声微波[159-160]。

**光学频率基准的远距离传输：**在获得了高稳定度的光学频率基准之后，如何将该光学频率传输至远端，实现光学频率基准的比对，同样也是一个非常重要的课题。2007 年，Newbury 等人使用声光频移器补偿光纤链路引入的噪声，首次实现了长度为 251 公里的光学频率基准传输[161]。2009 年，Grosche 等人实现了光学频率基准在 146 公里的光学链路中的传输，传输精度为 $10^{-19}$ 量级[162]。2012 年，德国马克思普朗克研究所与德国联邦物理技术研究院合作，实现了长度为 920 公里的光学频率基准传输[163]。该频率基准的传输链路遍布德国境内十余个城市，系统包含了十分复杂的光学放大基站、色散补偿、噪声补偿系统，最终实现了 1000 s 门时间内稳定度为 $10^{-19}$ 量级的光学频率传递。2013 年，NIST 研究所研制出了光学双向时间频率基准传递技术（Optical two-way time and frequency transfer，O-TWTFT）[164]。该技术使用具有 kHz 量级重频差的两台光学频率梳作为光源，使用光学降采样技术实现了光学频率基准在 2 公里自由空间中的传输。光学频率传输的 1000 s 稳定度达到 $10^{-19}$ 量级。随后，基于此技术，该研究所实现了更远距离的、更高精度的自由空间中的光学基准传输[165-167]。2018 年，意大利国家计量研究院实现了光学频率基准在海底的高精度传输[168]。

**双飞秒激光光学频率梳：**将两台具有重复频率差的光学频率梳在空间合束，两台激光器输出的脉冲便在时域相互降采样，该装置被称为双飞秒激光光学频率梳。双飞秒激光光学频率梳系统本质上是一种改进的飞秒激光泵浦探测装置。在传统的基于飞秒激光的泵浦探测技术当中，通常需要使用机械扫描装置来实现两飞秒脉冲在时域上的相互扫描，进而从采样信息中提取信号脉冲中包含的待测样品的相关特性信息。样品的测量速度会受限于机械扫描速度。双飞秒激光光学频率梳的出发点是利用两脉冲的重复频率差来替代机械扫描，实现高速的、实时的飞秒激光泵浦探测。双飞秒激光频率梳最为广泛的应用是气体的光谱学的测量[169-170]。得益于光学频率梳低噪声特性，使用该装置进行气体的光谱学测量的光谱分辨率可以达到 GHz 量级。2007 年，Diddams 等人首次使用单台钛宝石激光光学频率梳实现了 $I_2$ 蒸汽的吸收谱测量[171]。2008 年，Coddington 等人首次提出使用双飞秒激光光学频率梳实现光谱学测量[172]。信号脉冲经过待测气体后，与参考脉冲在空间上合束，对合束后的时域信号做傅里叶变换处理后，即可得到待测气体的吸收谱，测量的分辨率可达到 kHz 量级。在随后，基于双飞秒激光光学频率梳的光谱学测量技术蓬勃发展。腔增强的光谱学测量技术[173]、消除多普

勒效应的光谱学测量技术[174]、自适应的实时光谱学测量技术[175]等被相继研究出来。与此同时，关于该技术的基础理论，例如探测灵敏度的提高[176]、测量信噪比的提高[177]、光学频率梳的噪声对测量的影响[169]等也一直不断地在被深入地挖掘。使用三飞秒激光光学频率梳，可以进一步提升光谱学测量的纬度[178]。另一方面，使用其他种类飞秒激光器的双光学频率梳，例如电光调制光学频率梳[179-180]、微环谐振腔光学频率梳[181]、半导体飞秒激光器光学频率梳[182-183]相继实现了物质吸收谱的测量。由于 2~10 μm 波段为常见气体的吸收峰所在波段，使用非线性频率转换、差频技术和相干超连续光谱等技术[184-186]将光学频率梳的输出波长拓展至这一范围后，光谱学的测量变得更加有意义[187-188]。最重要的是，该技术已经不仅仅局限于在实验室中对气体的吸收谱进行测量，已经被投入到实验室之外的应用当中。在文献[189-190]中，均在实验室外实现了实时的、高精度的城市空气中气体吸收谱的测量。另一方面，双飞秒激光光学频率梳还可以被应用到距离测量[191]以及光学成像[192]中。使用 GHz 量级重复频率的微环谐振腔作为光源，距离测量的更新速率会被提高，甚至可以快速扫描高速飞行的子弹的形貌[193]。

使用可以产生双飞秒光学频率梳的单台激光器作为光源，可以在系统中避免复杂的锁相系统，是一种较低成本的双飞秒光学频率梳装置。2016 年，Ideguchi 等人首次在钛宝石飞秒激光器中实现了具有重复频率差的双脉冲序列运转，并用此光源实现了 $Nd:YVO_4$ 晶体吸收谱的测量[194]。光纤激光器由于其丰富的内在锁模机制，更易于实现腔内双脉冲的运转。通常使用的技术手段有如下几种：一、偏振复用法。即在全保偏光纤激光器中，在保偏光纤的快轴和慢轴同时产生飞秒脉冲，利用快慢轴的折射率差异实现两脉冲序列重复频率的不同[195]。二、空间复用法。通过将两台光纤激光器的部分腔结构共享[196-198]、两脉冲序列在腔内双向传输[199-200]或者腔外光纤延迟线的方法[201]产生有重复频率差的两脉冲序列。三、波长复用法。通过在激光器腔内加入一等效的带阻滤波器，使得带阻滤波器两侧的光谱实现锁模。由于光纤中的色散，不同中心波长的脉冲序列有着不同的重复频率。常见的腔内实现等效带阻滤波器的方法有：Lyot 滤波器[202-203]、Sagnac 环滤波器[204]、特定波长的损耗调节技术[205]、腔内拉锥光纤[206]以及模式选择耦合器[207]。在微环谐振腔产生的光学频率梳中，通过双向泵浦微环，也可以实现双飞秒激光光学频率梳的运转[208]。以单台飞秒激光器产生的双光学频率梳为光源，同样也实现了距离测量[209]、光谱学测量[210-211]和太赫兹波频率测量[212]等应用。

## 1.3 相干脉冲合成的发展概述

对于单台飞秒激光器而言，受限于增益介质的增益带宽以及掺杂浓度，很难直接获得超过倍频程的光谱和瓦量级的飞秒脉冲输出。因此，一方面可以将两台及更多不同中心波长的飞秒激光器进行相干合成，扩宽飞秒脉冲的光谱。另一方面，可以将两台相同波长的飞秒激光器进行相干合成，提高飞秒脉冲的功率。在量子噪声及环境噪声等因素的影响下，多台独立的飞秒激光器之间并无相干性，因此，需要对飞秒激光器输出脉冲的重复频率和载波-包络偏移频率进行精确控制，以实现有效的相干脉冲合成。

2001年，Sheldon首次将两台独立运转的钛宝石飞秒激光器输出脉冲的重复频率锁定起来，证明了两台激光器的相干脉冲合成的可能性[89]。一般认为，当两光学脉冲的相对定时抖动的均方根值小于脉冲中心波长对应的光学周期的十分之一时，才可以被认为脉冲已经无法被分辨出是来自于两个独立的激光振荡器[213]。但是，基于光电探测器法的电锁定受限于光电探测器的响应时间，无法实现如此高精度的时间同步。直到2003年，平衡互相关技术[14]的出现将飞秒脉冲的时间同步精度达到阿秒量级，该技术的快速发展使得相干脉冲合成成为了可能。2012年，Cox等人使用腔内压电陶瓷与腔外电光调制器将一台钛宝石飞秒激光器与一台掺铒光纤飞秒激光器的重复频率进行锁定，剩余相对定时抖动为2.2 fs。在此基础上，使用腔外声光调制器对两脉冲的相对载波-包络偏移频率进行锁定，剩余的相对相位噪声为200 mrad。将钛宝石激光器输出的光谱进行展宽后，与掺铒光纤飞秒激光器的脉冲进行合束，得到了超过一个倍频程的光谱和宽度为3.7 fs的时域脉冲[214]。2015年，Fong等人成功地将一台掺镱光纤激光器和一台掺铒光纤激光器进行相干脉冲合成。将被动脉冲同步的方法与光学平衡互相关技术相结合，获得了1.9 MHz的重复频率锁定带宽。两台飞秒激光器的剩余相对定时抖动为0.81 fs。随后，使用声光频移器对两台激光器的载波-包络偏移频率进行前馈锁定，得到了长时稳定的合成信号[215]。2016年，Tian等人实现了两台掺镱光纤激光器的长时间相干脉冲合成。合成后两台光纤激光器的剩余相对定时抖动和剩余相位噪声分别为380 as和375 mrad[216]。

## 1.4 选题意义、研究内容及主要创新点

本文将围绕飞秒激光光学频率梳的重复频率与载波-包络相位的精密控制进行研究：系统研究了飞秒激光光学频率梳的定时抖动特性，实现了两台飞秒激光光学频率梳的时间同步；系统研究了飞秒激光光学频率梳的载波-包络相位噪声

特性，实现了飞秒激光光学频率梳的载波-包络相位噪声的高精度测量；系统研究了两台独立的掺镱光纤激光器的长期稳定相干脉冲合成；最后，系统研究了掺铒光学频率梳中 $f_{rep}$、$f_{ceo}$ 和 $\nu_n$ 的频率噪声功率谱，实验验证了 $f_{rep}$、$f_{ceo}$ 和 $\nu_n$ 的锁相环环路之间的串扰。这些成果从理论和实验方面为低噪声的光学频率梳运转提供了基础。本论文的主要内容和创新点如下：

1、首次实现了两台独立的掺镱光纤激光器的连续 5 天（120 小时）的高精度时间同步。实现时间同步后，环内光学互相关系统输出的剩余时间误差信号的均方根值为 103 as，环外光学互相关系统输出的剩余时间误差信号的均方根值为 733 as。研究了环外定时抖动结果的恶化的主要原因。环内交叠的艾伦方差的 $1.31\times10^5$ s 稳定度为 $8.76\times10^{-22}$。环外交叠的艾伦方差的 $1.31\times10^5$ s 稳定度为 $1.36\times10^{-20}$。研究了环内、环外的剩余定时抖动的噪声特性。研究了光纤引入的额外定时抖动。首次使用光学外差探测法测量了 5 米长的单模光纤所引入的定时抖动。

2、系统研究了飞秒激光光学频率梳中的载波-包络相位噪声的高精度测量及分析。在两台重复频率锁定的掺镱光纤飞秒激光器的基础上，首次使用光学外差探测法实现了超高范围、超高精度的载波-包络相位噪声的功率谱测量。实现了傅立叶频率范围在 5 mHz 至 8 MHz 内的，动态范围高于 270 dB 的载波-包络相位噪声功率谱测量。相位噪声测量基底低于 1 μrad/√Hz。分别使用功率谱分析、Hadamard 方差分析和 Kendall 互相关分析研究了飞秒激光器中的载波-包络相位噪声的特性。首次测量了孤子分子对中的两个光学孤子的相对相位噪声。得到光孤子的相对相位噪声的功率谱密度，相位噪声测量的精度为 $10^{-13}$ $rad^2$/Hz，在 10 Hz 至 10 MHz 傅里叶频率范围内积分，相对相位噪声的积分值为 3.5 mrad。使用 $\beta$-line 分析法，估算两个光学孤子的相对线宽仅为 μHz 量级。

3、系统研究了 $f_{rep}$、$f_{ceo}$ 和 $\nu_n$ 的频率噪声特性。首次使用光纤延迟线系统对掺铒光纤飞秒激光器光学梳齿的频率噪声进行了高精度测量。分析了光学频率梳中定时抖动和所有梳齿的噪声来源，研究了 $n\times f_{rep}$ 噪声与 $f_{ceo}$ 噪声之间的反相关特性。系统验证了 $f_{rep}$、$f_{ceo}$ 和 $\nu_n$ 的锁相环环路之间的串扰。实验发现，如果 $f_{rep}$ 的锁定带宽过高，压电陶瓷会对 $\nu_n$ 引入额外的噪声。反之，$\nu_n$ 的锁定不会对 $f_{rep}$ 的噪声特性产生任何影响。另一方面，在 $f_{ceo}$ 噪声的锁定过程中，泵浦电流调制会对 $n\times f_{rep}$ 和 $\nu_n$ 的噪声功率谱产生明显的影响。

4、首次实现了两台独立的掺镱光纤激光器的长期稳定相干脉冲合成。通过平衡光学互相关系统锁定重复频率，通过腔外声光调制器锁定相对载波-包络相位。合成后，两脉冲的相对剩余定时抖动的积分值为 380 as，剩余的相对载波-包络噪声的积分值为 375 mrad。合成后的光谱干涉的对比度为 58%。60 分钟内，

两脉冲的相对载波-包络相位漂移的均方根值为 165 mrad。

# 第2章 飞秒激光光学频率梳的噪声与稳定性评价的理论基础

本章将对飞秒激光器中噪声的基本理论和稳定性的评价方法进行阐述。通过对飞秒激光器的定时抖动、载波-包络相位以及梳齿噪声机制进行理论建模，可以对飞秒激光器的噪声分析与稳定给予理论性指导。另一方面，通过功率谱密度和多种方差分析，可以从不同角度对稳定后的飞秒激光器的噪声特性进行全面的评估。与此同时，对光电探测器在光电探测时产生的噪声的理论分析可以对噪声探测起到指导性帮助，为实现低噪声光学频率梳运转打下坚实的基础。

## 2.1 飞秒激光光学频率梳噪声的基本理论

### 2.1.1 飞秒激光光学频率梳的定时抖动

飞秒激光器输出脉冲序列的重复频率的噪声在时域上表现为脉冲时域包络的定时抖动。脉冲的定时抖动由多方面噪声源引入，如激光器腔内的放大自发辐射、激光腔的非零色散、脉冲的强度波动以及可饱和吸收体的恢复时间。激光器腔内放大自发辐射通过直接耦合引入的定时抖动的表达式为：

$$S_{\Delta t}^{ASE,QL}(f)=\frac{D_T}{(2\pi f)^2} \tag{2-1}$$

$$D_T=\frac{\pi^2\tau^2}{6E_p}\Theta\frac{2g}{T_{rt}}h\nu_0 \tag{2-2}$$

其中，$f$ 为傅里叶频率；$D_T$ 为放大自发辐射引入群速度的扩散因子；$\tau$ 为脉冲的半高宽；$E_p$ 为脉冲能量；$\Theta$ 为放大自发辐射因子；$g$ 为腔内增益；$T_{rt}$ 为脉冲在腔内的渡越时间；$h\nu_0$ 为单光子能量。激光器腔内放大自发辐射通过腔内非零色散耦合引入的定时抖动的表达式为：

$$S_{\Delta t}^{GH}(f)=\frac{4D^2f_{rep}^2D_\omega}{(2\pi f)^2\left[(2\pi f)^2+\tau_{\omega c}^{-2}\right]} \tag{2-3}$$

$$D_{\omega}=\frac{2}{3E_{p}\tau^{2}}\Theta\frac{2g}{T_{rt}}h\nu_{0} \tag{2-4}$$

$$\frac{1}{\tau_{\omega c}}=\frac{4}{3}\frac{g}{T_{rt}\Delta f_{g}^{2}\tau^{2}} \tag{2-5}$$

其中，$D$ 为激光腔内净色散；$f_{rep}$ 为脉冲的重复频率；$D_{\omega}$ 为脉冲中心频率的扩散因子；$\tau_{\omega c}$ 为频率波动的衰减时间；$\Delta f_{g}$ 为增益带宽。激光器强度抖动通过自陡峭效应引入的定时抖动的表达式为：

$$S_{\Delta t}^{RN-SS}=\left(\frac{\varphi_{NL}}{2\pi^{2}fT_{rt}\nu_{0}}\right)^{2}S_{RIN}(f) \tag{2-6}$$

其中，$S_{RIN}(f)$为脉冲的强度抖动功率谱；$\varphi_{NL}$ 为脉冲在腔内渡越一周所产生的非线性相移；$\nu_{0}$ 为光频率。激光器强度抖动通过 Kramers-Krönig 关系引入的定时抖动的表达式为：

$$S_{\Delta t}^{RIN-KK}=\left(\frac{1}{2\pi\Delta f_{g}}\right)^{2}S_{RIN}(f) \tag{2-7}$$

激光器强度抖动通过腔内慢可饱和吸收体引入的定时抖动的表达式为：

$$S_{\Delta t}^{SA}=\left(\frac{1}{2\pi fT_{rt}}\frac{\partial\Delta t}{\partial s}s\right)^{2}S_{RIN}(f) \tag{2-8}$$

其中 $s$ 为可饱和参量。从以上理论模型中可以看出，降低飞秒激光器输出脉冲的定时抖动主要方法有：降低激光腔的损耗，以降低腔内脉冲的增益，使得放大自发辐射直接耦合产生的定时抖动被减小；减小腔内净色散，以降低放大自发辐射通过腔内色散耦合产生的定时抖动；降低腔内的非线性，减小脉冲的非线性相移，以降低强度抖动通过自陡峭效应引入的定时抖动；降低脉冲的强度波动噪声，以降低强度抖动通过自陡峭效应、Kramers-Krönig 关系和慢可饱和吸收体引入的定时抖动。

### 2.1.2 飞秒激光光学频率梳的梳齿噪声

飞秒激光器输出的脉冲序列经傅里叶变换后，在光频域可以得到梳状的离散光谱。频率梳齿的线宽是衡量光学频率梳噪声特性的一个重要指标。1958 年 A. L. Schawlow 和 C. H. Townes 两人一同提出了在光学谐振腔中振荡的电磁波频率线

宽极限的概念，并将此极限命名为 Schawlow-Townes 线宽极限[217]。Schawlow-Townes 线宽极限的物理来源是激光器中的自发辐射。但是，由于其他众多技术噪声的存在，在大多数激光器中，输出电磁波的线宽均远大于 Schawlow-Townes 线宽极限。该线宽极限的表达式为：

$$S_{vn}^{ASE,ST}(f)=2\frac{f_{rep}^2}{(2\pi)^2}\left(\frac{\left(1+n_{sp}\right)hv_0G}{P_{circ}}\right) \tag{2-9}$$

其中，$n_{sp}$ 是放大自发辐射因子；$G$ 为腔内增益；$P_{circ}$ 为腔内脉冲功率。

飞秒激光器输出脉冲在光频域内，频率梳齿由某单一噪声源引入的技术噪声可以由以下公式描述：

$$S_{vn}^{X}=\left(v_n-v_{fix}^{X}\right)^2S_r^X(f) \tag{2-10}$$

其中 $v_n$ 为光学梳齿的光学频率；$v_{fix}^{X}$ 为该噪声源在整个光学频率梳内产生噪声的频率“固定点”；$S_r^X(f)$ 为该噪声源的功率谱密度。对于泵浦波动、环境引起的腔长波动、环境引起的腔损耗以及放大自发辐射产生的脉冲定时抖动的噪声功率谱分别由以下公式表示：

$$S_r^{pump}(f)=B\frac{1}{1+\left(f/f_{3dB}\right)^2}S_{RINpunp} \tag{2-11}$$

$$S_r^{lengfh}(f)=\left(\frac{f_{rep}}{v_{group}^L}\right)^2S_{length}(f) \tag{2-12}$$

$$S_r^{loss}(f)=B\frac{1}{1+\left(f/f_{3dB}\right)^2}S_{loss}(f) \tag{2-13}$$

$$S_r^{ASE,\text{ timing}}(f)=2f_{rep}^2\left(\frac{\left(1+n_{sp}\right)hv_0G}{P_{circ}}\right)\times\left[t_{rms}^2+\left(\frac{\beta_2}{4D_g\omega_{rms}}\right)\right] \tag{2-14}$$

其中，$S_{RINpump}$ 为泵浦源的强度波动的功率谱；$B$ 为泵浦变化引入的脉冲重复频率变化率；$f_{3dB}$ 为频率响应下降 3 dB 时所对应的傅里叶频率，在掺铒光纤激光器中，通常为 6 kHz；$S_{length}(f)$和 $S_{loss}(f)$分别为腔长波动和腔损耗的功率谱；$v_{group}^L$ 为激光腔中脉冲传播的群速度；$\beta_2$ 为腔内净色散；$D_g$ 为增益色散，在掺铒光纤激光器中，通常为 0.065 ps；$t_{rms}$ 和 $\omega_{rms}$ 分别为腔内脉冲宽度和光谱宽度的均方根值。

在光学频率梳中，第 0 根频率梳齿为光学频率梳的载波-包络偏移频率。将载波-包络偏移频率的频率值带入以上公式中，即可以得到各种不同技术噪声源对载波-包络偏移频率噪声的贡献。通常的，在飞秒激光器中，在低于 1 kHz 的傅里叶频率部分，载波-包络相位噪声主要由环境引起的腔损耗引入；在 1 kHz 至 100 kHz 傅里叶频率范围，载波-包络相位噪声主要由泵浦噪声引入；在高于 100 kHz 傅里叶频率范围，载波-包络相位噪声主要由放大自发辐射产生的量子噪声引入。对于光谱范围内的频率梳齿，其频率噪声的主要来源是激光器腔长的波动。

### 2.1.3 飞秒激光光学频率梳的相对强度噪声

飞秒激光器输出脉冲序列的相对强度噪声是指该光脉冲序列在某个测量时间段内平均功率的波动，该噪声表示了光脉冲序列的平均功率稳定性。脉冲强度噪声可以被量化为相对强度噪声（Relative intensity noise，RIN），其定义为：

$$RIN = \frac{\left\langle \Delta P(t)^2 \right\rangle_T}{\left\langle P(t) \right\rangle_T^2} \tag{2-15}$$

其中，$\left\langle \Delta P(t)^2 \right\rangle_T$ 和 $\left\langle P(t) \right\rangle_T^2$ 分别为一定测量时间 $T$ 内的光功率波动的均方根值和脉冲序列的平均光功率。

相对强度噪声是由量子噪声（例如放大自发辐射引入的噪声和真空波动噪声）和外部技术噪声源（例如泵浦激光器的强度噪声和泵浦激光器驱动器的电源噪声）引起的。激光器强度对微扰的响应是非瞬时的。考虑到增益介质中粒子的有限的寿命，任何小的扰动都会触发粒子瞬态的弛豫振荡，该振荡将以指数形式衰减。超过弛豫振荡频率的部分，频率响应迅速下降。因此，整个激光腔表现出低通滤波特性，将高于弛豫振荡频率的强度波动滤除。在掺铒或掺镱的飞秒光纤激光器中，弛豫振荡频率通常为 kHz 量级。放大自发辐射引入的噪声和泵浦引入噪声仅在弛豫振荡频率以下的部分对激光器的强度噪声产生影响。在激光器中，通常泵浦噪声水平远高于放大自发辐射噪声水平，因此飞秒激光器的相对强度噪声通常由泵浦技术噪声主导。由真空波动引入的激光器输出脉冲的强度噪声的表达式为：

$$S_{RIN}^{vacuum}(f) = \frac{2hv_c}{P_{avg}} \tag{2-16}$$

其中，$h$ 为普朗克常数，$v_c$ 为光学频率，$P_{avg}$ 为脉冲的平均功率。以 1550 nm 波

段的激光器为例，对于输出功率为 1 mW、10 mW 和 100 mW 的激光器，其真空波动引入的强度噪声分别在-156 dBc/Hz、-166 dBc/Hz 和-176 dBc/Hz 量级。

## 2.2 飞秒激光光学频率梳的稳定性分析的基本理论

### 2.2.1 功率谱分析

根据傅立叶分析，任何物理信号都可以分解为多个离散频率或连续范围内的频谱。对于平稳过程的连续信号，功率谱密度（Power spectral density，PSD）描述了信号的功率在频率上如何分布，单位为 W/Hz。对于一个周期信号，其频率稳定性可以通过功率谱密度在频域中进行评价。在周期信号的频率稳定性评价中，功率谱密度描述了相位强度或频率波动随傅立叶频率的变化情况。频谱稳定度评价与潜在的噪声过程直接相关，在描述激光光源的相位噪声时尤其适用。

频率源的随机相位和频率波动的功率谱密度可通过以下形式的幂指数形式表示：

$$S_y(f) = h(\alpha) f^{\alpha} \qquad (2\text{-}17)$$

其中，$S_y(f)$ 为物理量 $y$ 的功率谱密度；$f$ 为傅里叶频率；$h(\alpha)$为强度系数。

通常用于描述频率源的稳定性的功率谱密度通常有以下四种：频率波动的功率谱密度 $S_y(f)$，单位为 $Hz^2/Hz$，对于归一化的频率波动，单位为 1/Hz；相位波动的功率谱密度 $S_\varphi(f)$，单位为 $rad^2/Hz$。时间波动的功率谱密度 $S_x(f)$，单位为 $s^2/Hz$；单边相位噪声功率谱 $\mathcal{L}(f)$，单位为 dBc/Hz。这四种功率谱密度的转换关系为：

$$S_\varphi(f) = (2\pi\nu_0)^2 \bullet S_x(f) = \left(\frac{\nu_0}{f}\right)^2 \bullet S_y(f) \qquad (2\text{-}18)$$

$$\mathcal{L}(f) = 10 \cdot \log\left[\frac{1}{2} \cdot S_y(f)\right] \qquad (2\text{-}19)$$

其中，$v_0$ 为载波的频率，单位为 Hz。

### 2.2.2 方差分析

记录频率源的相位或频率波动随时间变化的序列后，使用特定公式计算不同种类的方差，可以对该频率源进行时域稳定性分析。

**标准方差**：经典的 $N$ 个数据的标准方差计算公式为：

$$s^2 = \frac{1}{N-1}\sum_{i=1}^{N}\left(y_i - \bar{y}\right)^2 \tag{2-20}$$

其中，$y_i$ 为 $N$ 个数据的归一化频率值，$\bar{y}$ 为 $y_i$ 序列的平均值。由于标准方差对于频率源中常见的某些类型的噪声是不收敛的，因此其无法作为频率稳定性的评估标准。

**艾伦方差**：艾伦方差（Allen variance，AVAR）是频率稳定性最常见的时域评估方法。与标准方差相似，它也是关于归一化频率波动的评价，但其具有对大多数类型的时钟噪声收敛的优势。艾伦方差的几种版本均可以提供很好的统计置信度，可以区分白噪声和闪烁相位噪声，并可以描述时间稳定性。艾伦方差的计算公式为：

$$\sigma_y^2(\tau) = \frac{1}{2(M-1)}\sum_{i=1}^{M-1}\left[y_{i+1} - y_i\right]^2 \tag{2-21}$$

其中 $y_i$ 为在门时间 $\tau$ 内，$M$ 次归一化频率测量中第 $i$ 次的频率值。在相位稳定性的分析中，计算公式为：

$$\sigma_y^2(\tau) = \frac{1}{2(N-2)\tau^2}\sum_{i=1}^{N-2}\left[x_{i+2} - 2x_{i+1} + x_i\right]^2 \tag{2-22}$$

其中 $x_i$ 为在门时间 $\tau$ 内，$N$ 次归一化频率测量中第 $i$ 次的相位值。

**交叠艾伦方差**：交叠艾伦方差是艾伦方差的一种形式，通过在每个门时间 $\tau$ 上形成所有可能的重叠样本来最大程度地利用所采集到的数据，以达到提高艾伦方差计算结果置信度的效果，对于频率序列，其计算公式为：

$$\sigma_y^2(\tau) = \frac{1}{2m^2(M-2m+1)}\sum_{j=1}^{M-2m+1}\left\{\sum_{i=j}^{j+m-1}\left[y_{i+m} - y_i\right]\right\}^2 \tag{2-23}$$

对于相位序列，计算公式为：

$$\sigma_y^2(\tau)=\frac{1}{2(N-2m)\tau^2}\sum_{i=1}^{N-2m}\left[x_{i+2m}-2x_{i+m}+x_i\right]^2 \tag{2-24}$$

**修正艾伦方差：**修正的艾伦方差（Modified Allan variation，MVAR）是艾伦方差的另一种形式，通过对相位的一次额外的平均计算，使其具有能够区分白相位噪声和闪烁相位噪声的能力。对于频率序列，其计算公式为：

$$Mod\sigma_y^2(\tau)=\frac{1}{2m^4(M-3m+2)}\sum_{j=1}^{M-3m+2}\left\{\sum_{i=j}^{j+m-1}\left(\sum_{k=i}^{i+m-1}\left[y_{k+m}-y_k\right]\right)\right\}^2 \tag{2-25}$$

对于相位序列，计算公式为：

$$Mod\sigma_y^2(\tau)=\frac{1}{2m^2\tau^2(N-3m+1)}\sum_{j=1}^{N-3m+1}\left\{\sum_{i=j}^{j+m-1}\left[x_{i+2m}-2x_{i+m}+x_i\right]\right\}^2 \tag{2-26}$$

**时间方差：**时间方差（Time variance，TVAR）对修正的艾伦方差乘以门时间的平方，并对√3 归一化。对于白相位噪声调制，时间方差和标准方差的结果是相同的。该方差主要用于对时间频率传递链路的稳定性评价，计算公式如下：

$$\sigma_x^2(\tau)=\left(\frac{\tau^2}{3}\right)\cdot Mod\sigma_y^2(\tau) \tag{2-27}$$

**Hadamard 方差：**Hadamard 方差（Hadamard variance，HVAR）基于 Hadamard 变换。由于该方差在计算过程中提取的是频率的二阶导数、相位的三阶导数，因此在时域频率稳定性评价中，该方差最重要的优点是其对线性频率漂移不敏感，特别适用于分析 Rb 原子钟的稳定性。对于频率序列，其计算公式为：

$$H\sigma_y^2(\tau)=\frac{1}{6(M-2)}\sum_{i=1}^{M-2}\left[y_{i+2}-2y_{i+1}+y_i\right]^2 \tag{2-28}$$

对于相位序列，计算公式为：

$$H\sigma_y^2(\tau)=\frac{1}{6\tau^2(N-3m)}\sum_{i=1}^{N-3}\left[x_{i+3}-3x_{i+2}+3x_{i+1}-x_i\right]^2 \tag{2-29}$$

**交叠 Hadamard 方差：**将数据交叠用于 Hadamard 方差，即可得到交叠 Hadamard 方差的计算公式。对于频率序列，其计算公式为：

$$H\sigma_y^2(\tau)=\frac{1}{6m^2(M-3m+1)\tau^2}\sum_{j=1}^{M-3m+1}\left\{\sum_{i=j}^{j+m-1}\left[y_{i+2m}-2y_{i+m}+y_i\right]\right\}^2 \tag{2-30}$$

对于相位序列，计算公式为：

$$H\sigma_y^2(\tau)=\frac{1}{6(N-3m)\tau^2}\sum_{i=1}^{N-3m}\left[x_{i+3m}-3x_{i+2m}+3x_{i+m}-x_i\right]^2 \quad (2\text{-}31)$$

以上介绍的方差的特点及其适用范围总结如下表：

表 2-1　不同种类方差的特点

| 方差种类 | 特点 |
|---|---|
| 标准方差 | 对于原子钟的噪声不收敛，基本不使用 |
| 艾伦方差 | 经典方法，仅在需要时使用，置信度很低 |
| 交叠艾伦方差 | 首选方法，用途最为广泛的方差 |
| 修正艾伦方差 | 用于分辨白相位噪声和闪烁相位噪声调制 |
| 时间方差 | 基于修正艾伦方差，用于时间频率链路的稳定性评估 |
| Hadamard 方差 | 用于抑制频率漂移，并分析发散的噪声 |
| 交叠 Hadamard 方差 | 比 Hadamard 方差有着更好的置信度 |

## 2.2.3 功率谱与艾伦方差的转换

与功率谱密度分析相同，在艾伦方差中，方差随门时间变化的不同斜率也表征了不同的噪声特性。表 2-2 和图 2-1 中详细描绘了同种噪声特性在相位噪声功率谱、频率噪声功率谱和艾伦方差中的对应关系。

表 2-2　不同噪声在相位噪声功率谱、频率噪声功率谱和艾伦方差中的对应关系

| 噪声种类 | 相位噪声功率谱密度 | 艾伦方差 |
|---|---|---|
| 白相位调制噪声（White phase modulation） | $b_0$ | $h_2 f^2$ |
| 闪烁相位调制噪声（Flicker phase modulation） | $b_{-1} f^{-1}$ | $h_1 f$ |
| 白频率调制噪声（White frequency modulation） | $b_{-2} f^{-2}$ | $h_0$ |
| 闪烁频率调制噪声（Flicker frequency modulation） | $b_{-3} f^{-3}$ | $h_{-1} f^{-1}$ |
| 随机游走频率调制噪声（Random walk frequency modulation） | $b_{-4} f^{-4}$ | $h_{-2} f^{-2}$ |

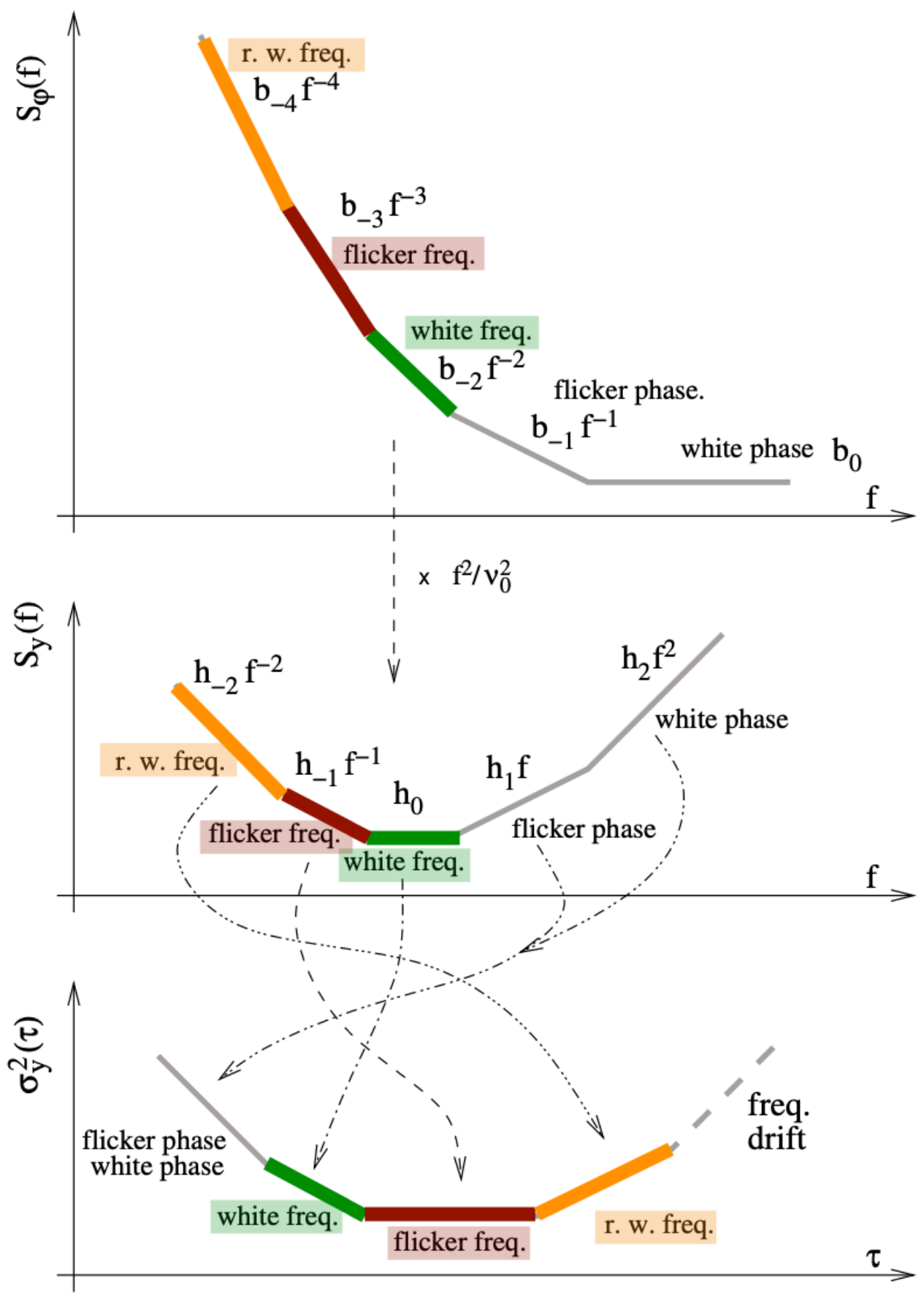

图 2-1　不同噪声在相位噪声功率谱、频率噪声功率谱和艾伦方差中的对应关系

并且，时域的频率稳定性与功率谱密度图可以通过如下公式进行转换：

$$\sigma^2(\tau)=\int_0^\infty S_y(f)\cdot|H(f)|^2\cdot df \tag{2-32}$$

其中，$H(f)$为传递函数。艾伦方差与功率谱密度的传递函数为：

$$|H(f)|^2=2\left[\frac{\sin^4(\pi\tau f)}{(\pi\tau f)^2}\right] \tag{2-33}$$

## 2.3 光电探测的基本理论

### 2.3.1 光电探测器的热噪声

光电探测器中的热噪声为负载电阻的热噪声。热噪声（Johnson 噪声，Nyquist 噪声或 Johnson-Nyquist 噪声）是在平衡状态下电导体内部的电荷载流子（通常

是电子）受到热搅动而产生的电子噪声。无论施加何种电压，电阻都会产生这种噪声。热噪声存在于所有电路中，并且在敏感的电子设备（如无线电接收器）中可能会将微弱的电信号淹没，进而成为限制电子测量仪器灵敏度的重要因素。热噪声随温度增加。一些敏感的电子设备，例如射频望远镜接收器，被冷却至低温，以降低其电路中的热噪声。理想电阻中的热噪声的噪声特性为白噪声，这意味着功率频谱密度在整个频谱中几乎是恒定的。当限制为有限带宽时，热噪声具有接近高斯的幅度分布。

热噪声的单边功率谱密度由下式给出：

$$\overline{i_n^2} = 4k_B T R_L \tag{2-34}$$

其中 $k_B$ 是玻尔兹曼常数，为 1.38×10$^{-23}$ J/K；$T$ 是电阻的绝对温度，以 $K$ 为单位；$R_L$ 是电阻器的值，单位为欧姆。由该公式计算得到的功率谱密度的单位为 $V^2$/Hz。例如，300 K 室温下，当光电探测器的负载为 50 欧姆时，其热噪声量级为：$(4\times1.38\times10^{-23}\times300\times50)^{1/2}$=0.9 nV/√Hz

## 2.3.2 光电探测器的散粒噪声

散粒噪声符合泊松分布，因此又被称为泊松噪声。在电子设备中，散粒噪声源自电荷的离散性质。在光学探测器与光子计数器中，散粒噪声源自探测器吸收光子的随机性。散粒噪声描述了由于彼此独立发生而被探测到的光子数量的波动，其单边功率谱密度可以由以下公式表示：

$$\overline{i_s^2} = 2qI \tag{2-35}$$

其中 $q$ 表示电子电荷量，为 1.6×10$^{-19}$ C；$I$ 为探测器产生的电流。由该公式计算得到的功率谱密度的单位为 $A^2$/Hz。例如，当光电探测器输出电流为 1 mA 时，其散粒噪声量级为：$(2\times1.6\times10^{-19}\times0.001)^{1/2}$=0.0179 nA/√Hz。

## 2.3.3 不同光功率下光电探测器中的噪声

本节讨论光功率对光电探测器中的噪声特性的影响。假设光电探测器的入射功率为 1 mW，对于 1550 nm 波段的电磁波，基于铟镓砷的光电探测器的响应度为 1 A/W。即每 1 mW 的光会使得光电探测器产生 1 mA 的电流，对于 50 欧姆负载的光电探测器，负载两端的电压值为 50 mV。根据上两节中的计算，室温下光电探测器的热噪声为 0.9 nV/√Hz，那么，热噪声基底相对于信号的量级为 $10\times\log[(0.9\ \text{nV}/50\ \text{mV})^2]$=-154 dBc/Hz。这表明，信号功率高于热噪声基底 154 dB。

同样地，1 mW 入射光时，散粒噪声基底相对于信号的量级为 10×log[(0.0179 nA/1 mA)$^2$]=-155 dBc/Hz。这表明，信号功率高于散粒噪声基底 155 dB。根据此计算方法，本文对不同入射光功率下，光电探测器的热噪声基底和散粒噪声基底进行了计算，结果见图 2-2。

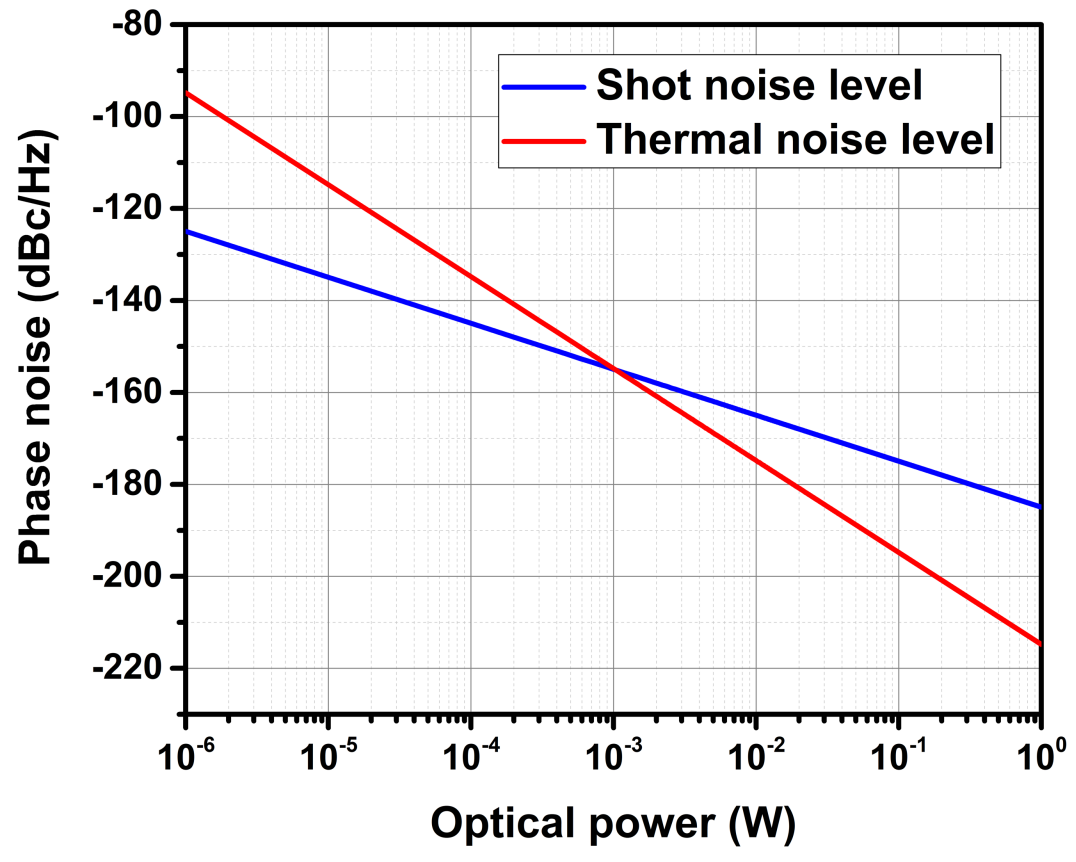

图 2-2　光电探测器的热噪声基底和散粒噪声基底随入射光功率变化曲线

由图可知，当入射光功率为 1 mW 时，热噪声基底与散粒噪声基底基本相同；当入射光功率小于 1 mW 时，光电探测受限于热噪声基底。并且对着光功率的升高，探测基底随 20 dB/W 斜率下降。当功率大于 1 mW 时，光电探测受限于散粒噪声基底。并且对着光功率的升高，探测基底随 10 dB/W 斜率下降。这说明，当光电探测开始受限于散粒噪声时，通过提升光功率带来的探测信噪比提升开始变得缓慢。

## 2.4 本章小结

本章内容主要包括以下三部分：首先，通过对飞秒激光光学频率梳输出脉冲的定时抖动、载波-包络相位噪声和梳齿噪声的理论建模，分析了放大自发辐射、泵浦功率波动、环境引起的腔长波动和腔损耗等因素对光学频率梳中不同种类噪声的影响。这是实现稳定光学频率梳运转的重要理论基础。其次，本文介绍了功率谱分析法和方差分析法，两种方法分别从频域和时域的角度对相位锁定后的光学频率梳进行稳定性评价。特别地，方差分析法包括标准方差、艾伦方差、交叠艾伦方差、改进的艾伦方差、时间方差和 Hadamard 方差等，不同的方差适用于不同种类的实验系统。功率谱分析和方差分析从不同角度反映出不同类型的噪声，例如白相位调制噪声、闪烁相位调制噪声、白频率调制噪声等。最后，分析了在

光电探测过程中，光电探测器产生的热噪声和散粒噪声对噪声探测的信噪比产生的影响。本章的理论分析均为实现低噪声光学频率梳运转打下了坚实的基础。

# 第3章 飞秒激光光学频率梳定时抖动的测量及时间同步

本章将介绍两台掺镱光纤飞秒激光器的定时抖动测量、长时间稳定同步以及光纤的定时抖动测量。对飞秒激光器定时抖动的高精度测量是实现激光器输出脉冲包络稳定的重要前提。低定时抖动的激光振荡器可作为高精度的时间基准，被应用到诸多领域中，包括实现阿秒精度的光学脉冲同步和高精度时间基准的远距离传输。在 X 射线自由电子激光器的时间分辨泵浦探针实验中，通常需要长期不间断的日期采集，以记录完整的分子电影[218]。因此，长期稳定的时间同步是实现该技术的重要前提。与此同时，掺镱光纤激光源由于其平均功率易提升的特点，而在此类 X 射线自由电子激光器系统中的作用与日俱增。

但是，独立激光器的脉冲同步对激光器本身锁模状态的不稳定和环境扰动的影响均非常敏感。特别是在建立极高精度的时间同步时，任何轻微的泵浦功率波动或声学噪声都很容易破坏锁相环路。出于此目的，本章工作中，使用光学平衡互相关技术对两台掺镱光纤激光器的定时抖动进行高精度探测与时间同步。在使用铝制盒子、隔音海绵对环境噪声进行隔离的条件下，高精度的时间同步持续了 5 天（120 小时），是现有文献报道的同步时长的 3 倍左右。使用环外光学互相关系统对同步后的时间误差信号进行监测，得到的环外剩余定时抖动的均方根值为 733 as。通过重叠的艾伦方差对时间同步的长期稳定性进行了评估，在 $1.31\times10^{5}$ s 门时间内，稳定度达到 $1.36\times10^{-20}$。并且发现，对环外互相关器采取精确温度控制可以进一步提高时间同步的指标。同时，使用光纤对时间基准进行传输时，光纤链路由于受到环境温度的变化，也会引入定时抖动。本章工作的另一部分对脉冲在光纤中传输过程中引入的定时抖动进行测量。使用光学外差探测法和其一系列改进的外差探测法，测得了光纤引入脉冲的定时抖动的功率谱图。脉冲在光纤中传输过程中引入的额外的定时抖动在 10 Hz 至 10 MHz 频率范围内的均方根值为 64.7 as。

## 3.1 飞秒激光光学频率梳定时抖动的理论模型

飞秒激光器中，定时抖动主要源于两部分：一、放大自发辐射直接引入的定时抖动；二、激光器中心波长的改变通过激光器腔内色散间接引入的 Gorden-Haus 定时抖动。从孤子扰动方程中可以得到的量子极限的定时抖动功率谱的表达式如

下：

$$S_{\Delta T}^{ASE,soliton}(f)=\frac{D_{T,sol}}{(2\pi f)^2}+\frac{4D^2\cdot f_{rep}^2\cdot D_{\omega_c,sol}}{(2\pi f)^2[(2\pi f)^2+\tau_{\omega_c}^{-2}]} \tag{3-1}$$

其中，第一项为放大自发辐射直接引入的定时抖动。第二项为 Gorden-Haus 定时抖动。此外，还有激光器强度噪声通过自陡峭效应引入的定时抖动，强度噪声通过 Kramers-Krönig 效应引入的定时抖动以及慢可饱和吸收体引入的定时抖动。该部分内容的详细介绍请见第二章。

## 3.2 飞秒激光光学频率梳定时抖动的测量

实验中对两台掺镱光纤飞秒激光器的定时抖动进行测量，激光器的结构如图 3-1。两台激光器均为 NPR 锁模激光器，振荡级 1 为线性腔结构，泵浦源为中心波长为 976 nm 的低噪声二极管激光器，增益光纤为长度 25 厘米的高掺杂掺镱增益光纤（Yb 214，CorActive）。腔内光栅对用于提供负材料色散以调节整个激光腔的净色散。激光器工作在近零色散点，以获得最低的 Gordon-Haus 噪声。激光器重复频率为 157 MHz。与振荡级 1 不同的是，振荡级 2 为$\sigma$腔，一个直径为 2 mm 的 1040 nm 波段高反射镜粘在 2 mm 长的压电陶瓷上，达到小范围高带宽调节振荡级 2 的重复频率的目的。并且，在一个光纤准直器的支架下方有一个电控平移台，用于大范围低带宽调节振荡级 2 的重复频率。压电陶瓷与电控平移台一同作用，可以同时实现两台激光器的大范围，高带宽的重复频率锁定。

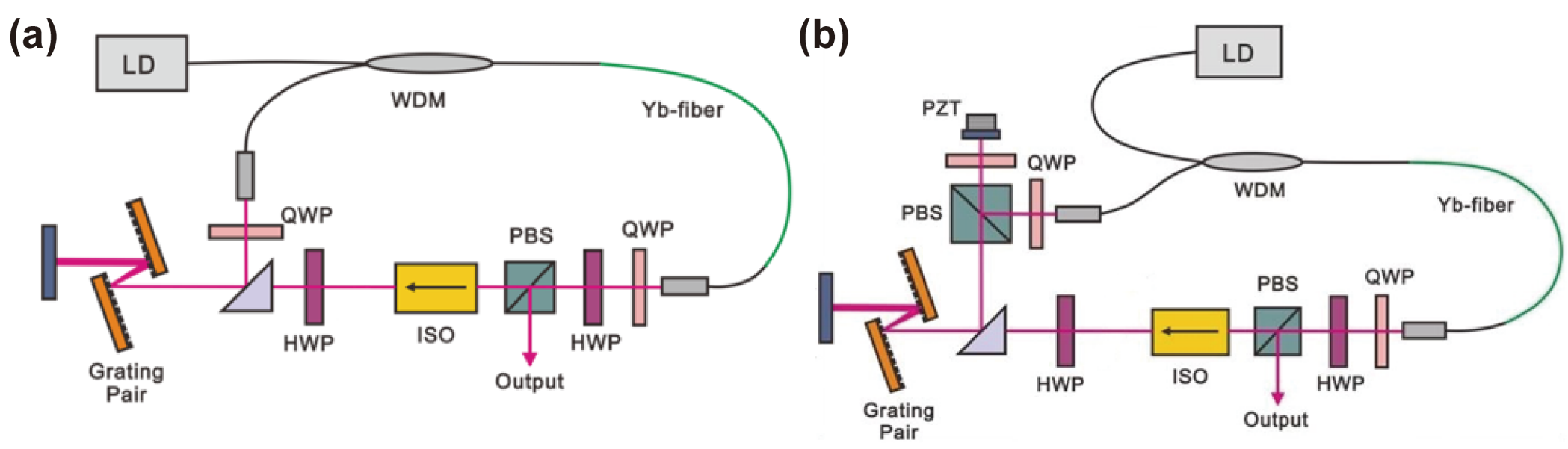

图 3-1 掺镱光纤飞秒激光器示意图。（a） 振荡级 1；（b） 振荡级 2。LD，980 nm 半导体二极管激光器；WDM，波分复用器；PZT，压电陶瓷；QWP，四分之一波片；HWP，半波片；PBS，偏振分束器；ISO，隔离器；Grating Pair，光栅对

### 3.2.1 定时抖动测量实验装置

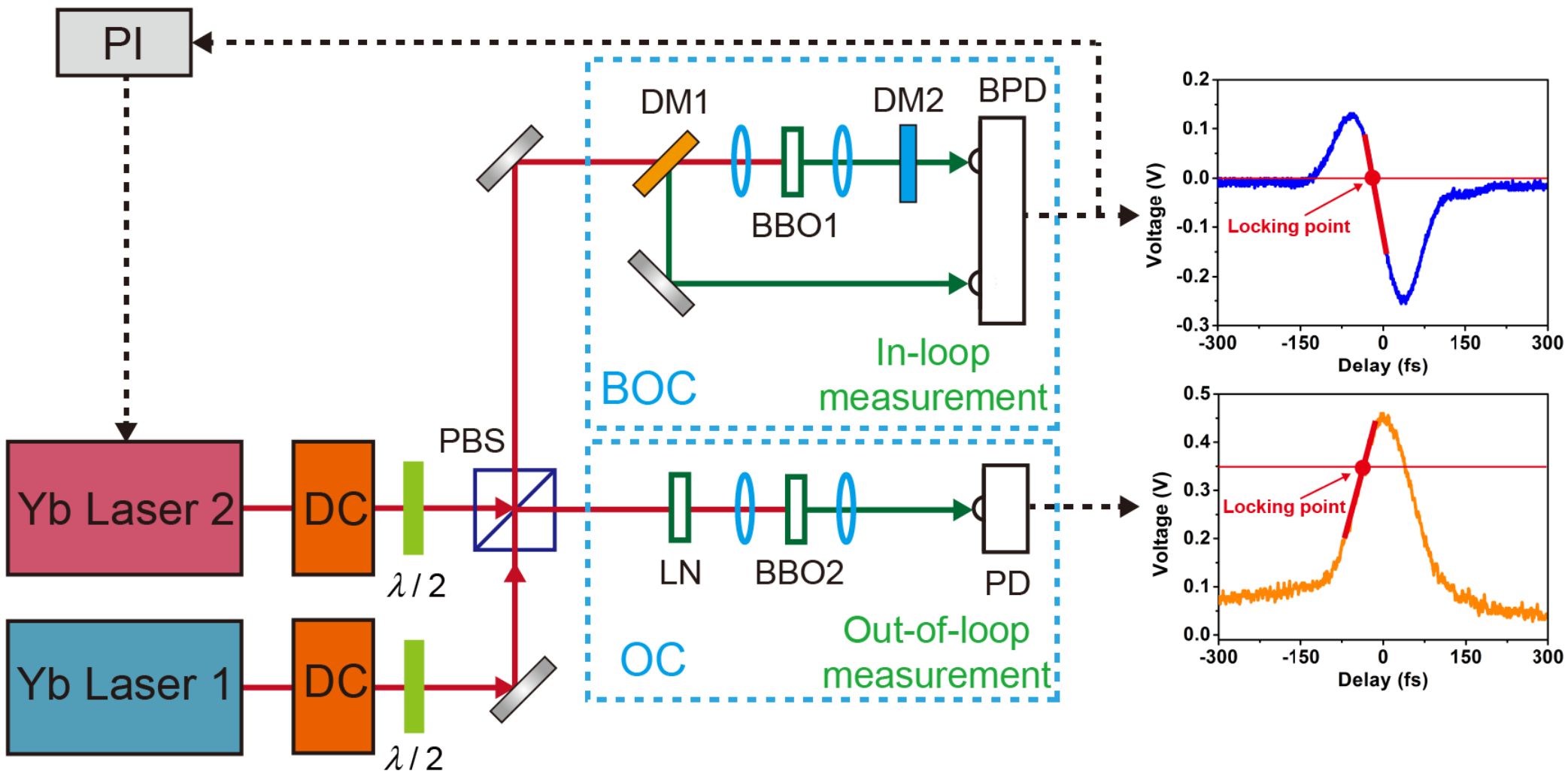

图 3-2 定时抖动测量实验装置示意图。BBO，*β*-偏硼酸钡晶体；BPD，高速平衡探测器；BOC，平衡光学互相关；DC，色散补偿器件；DM，双色镜；OC，光学互相关；LN，铌酸锂晶体；PBS，偏振分束器；PD 高速光电探测器；PI，比例积分伺服控制器

两台激光器输出的脉冲分别经过两半波片后，在一个偏振分束器处进行合束。将近一半功率的脉冲进入平衡光学互相关，用于激光器环内的重复频率锁定与定时抖动测量。平衡互相关由 BBO 晶体、双色镜、聚焦透镜、以及高速平衡探测器（BPD，Newfocus，1807，80MHz 带宽）组成。由于 DM1 对 1040 nm 波段的光高透，对 520 nm 波段的光高反。两台激光器输出的脉冲首先透过双色镜 DM1，随后被聚焦透镜聚焦至 BBO1 上，产生中心波长在 520nm 的和频信号。BBO1 为 0.65 mm 厚的 II 类 *β*-偏硼酸钡晶体。随后基频光与和频光一同被第二个透镜准直。DM2 为对 1040 nm 波段的光高反、对 520 nm 波段的光高透的双色镜。和频光透过 DM2 入射进入平衡探测器上，基频光被 DM2 反射，再次聚焦至 BBO1，产生和频信号。第二次产生的和频信号被 DM1 和一个银镜反射进入平衡探测器。由于 BBO 晶体的双折射效应，0.65 mm 厚的 BBO 晶体会使得两个偏振态互相垂直的两脉冲产生~100 fs 的延迟，进而会导致两次产生的和频信号在时域上有~100 fs 的延迟。两个有延迟和频信号在平衡探测器上被探测后相减，就可以得到与两脉冲时间延迟相关的、S 形状的误差信号，如图 3-2 中插图所示。和频过程有效地将两脉冲的定时抖动转化为和频信号功率的强弱，并且由于使用平衡探测，脉冲的强度抖动会被平衡探测器减掉，这样就有效地消除了激光器强度噪声对定时抖动测量的影响。使用该 S 曲线的线性部分进行相位鉴别，可以实现高精

度、阿秒量级的定时抖动的探测。该实验中，鉴相曲线的相位鉴别斜率为 6.0 mV/fs。将该误差信号输入进入一个高速比例积分伺服控制器（Newfocus，LB1005），产生的反馈信号经过低噪声电放大器进行放大后，加载到腔内的压电陶瓷上，最终实现两台激光器重复频率的高带宽锁定。

与此同时，从偏振合束器透过的脉冲入射进入一个独立的互相关系统进行定时抖动的环外测量。如图 3-2 所示，在环外测量系统中，两脉冲聚焦在一块 1 mm 厚的 II 类 BBO 晶体上。产生的和频互相关信号被一个光电探测器探测，如图 3-2 中插图。鉴相曲线的相位鉴别斜率为 4.0 mV/fs。一块厚度为 0.4 mm 的铌酸锂晶体被置于聚焦透镜之前。利用铌酸锂晶体的双折射效应对两个偏振态脉冲的时域延迟进行调节，使得两脉冲在时域重叠部分约为 1/2 个脉冲，使得环内使用的锁定点在时域上同时也为环外的鉴相曲线斜率最大的点附近，实现最大灵敏的定时抖动测量。与文献[34]中的基于干涉仪的环外互相关系统对比，由于舍弃了空间光学延迟线，该共路设计可以更好地抑制环境噪声对定时抖动测量的影响，达到更准确地测量低频定时抖动的目的。

在定时抖动测量系统的搭建过程中，应注意如下几点：

一、环内光学互相关系统的搭建：首先，将合束后的两路激光光束调节至空间完全重合，并且合束后的光路水平方向上与光学平台平行。合束后的光束经过 45 度角放置的双色镜后，聚焦至 BBO 晶体上。尽量保证光斑通过聚焦透镜中心。使用光电探测器找到互相关信号后，调节双色镜前的反射镜底部的线性平移台，使得两光斑在空间上平行地分开，两光斑距离大约在 2 mm 左右。微调聚焦透镜位置，尽量保证两光斑的中心与聚焦透镜中心重合。微调某一路光束的角度，寻找互相关信号。利用此步骤进行和频系统的搭建，可以保证最大的非共线相位匹配的和频效率。

二、两激光器重复频率锁定的调节：该实验中使用 Newfocus 公司生产的比例积分私服控制器（Newfocus，LB1005）进行锁相环的锁定。首先将锁相环的比例增益调节至零，调节伺服控制器的输出偏置，使得重复频率的误差信号的频率被尽量地“抻开”。将伺服控制器的积分拐点设置在 3 kHz 或者 10 kHz 档位，伺服控制器的开关拨至低频增益限制（Low frequency gain limit，LFGL）档位，随后逐渐增加锁相环的比例增益。在此过程中，不断地调节伺服控制器的输出偏置电压，使得误差信号一直在零电压附近。当增益增加到一定程度时，可以发现误差信号开始产生调制，并且随着增益的加大，调制会越来越深。这时将伺服控制器的开关推至 Lock on 档位，逐渐降低增益，直到调制消除后，即可实现锁相环的锁定。

三、环外互相关系统中脉冲零延迟的调节：可以更换不同厚度的铌酸锂晶体，

并且适当旋转铌酸锂晶体的角度和 BBO 晶体的角度来调整两脉冲的时间延迟，确保使得两脉冲在时域重叠部分约为 1/2 个脉冲，使得环内使用的锁定点在时域上的同时也为环外的鉴相曲线斜率最大的点附近。

### 3.2.2 定时抖动测量结果

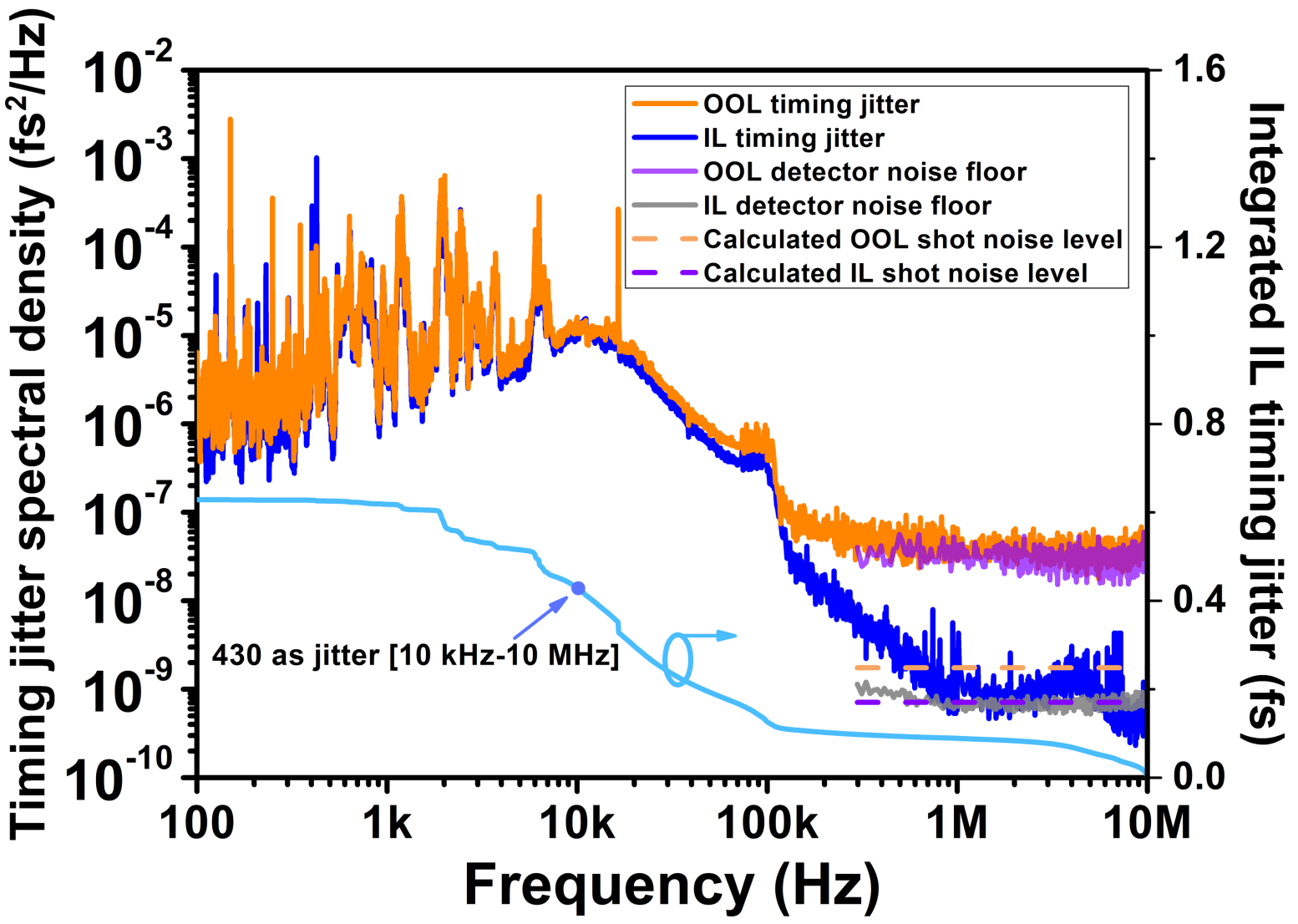

图 3-3　定时抖动的功率谱密度图

成功地将两台激光器的重复频率锁定后，使用快速傅里叶变换分析仪（Stanford Research Systems, SR770）和频谱分析仪（Agilent，8560EC）分别测量环内（In loop，IL）和环外（Out of loop，OOL）的定时抖动噪声功率谱密度图，结果见图 3-3。图中蓝色曲线对应环内测量的定时抖动功率谱密度。从功率谱密度图中可以看出，重复频率的锁定带宽在 11 kHz 左右。对于高于锁定带宽的频率范围（>11 kHz），功率谱密度遵循 20 dB/decade 的斜率。这表明在此频率范围内，定时抖动主要表现为由增益光纤中的自发辐射引入的随机游走噪声。从 10 MHz 偏频积分至 10 kHz 偏频，功率谱密度的积分值为 430 as，见图中淡蓝色积分曲线。在环内定时抖动功率谱密度的测量中，散粒噪声与二极管的热噪声基本在同一量级，见图中紫色虚线与灰色曲线，这表明该测量主要受限于散粒噪声和热噪声。

图中橙色曲线对应环外测量的定时抖动功率谱密度。可以看出，对于< 100 kHz 的频率范围内，环内功率谱密度与环外功率谱密度图有着较好的一致性。环内与环外噪声测量的一致性验证了环内测量结果的正确性。不同地是，由于环外

测量过程中，分配给互相关系统的光功率较少，因此信号的信噪比相比于环内测量较低。环外测量的散粒噪声远低于二极管的热噪声，见图中橙色虚线与紫色曲线。这表明，该环外测量主要受限于二极管的热噪声。

## 3.3 飞秒激光光学频率梳的长时间稳定时间同步

在实现了飞秒激光器定时抖动的功率谱密度测量后，本小节研究了两台飞秒激光器的长时间稳定的时间同步。为了实现时间长度为天量级甚至更长的时间同步，需要对整个系统进行严格的环境噪声隔离。具体做法为：将整个系统的光学部分，包括激光器，环内平衡互相关系统，环外互相关系统置于防震光学平台上。将两台飞秒激光器搭建在光学面包板上，面包板底部依次垫有隔音海绵—橡胶—铅皮—橡胶—铅皮—橡胶。其中，橡胶厚度为 3 mm，铅板厚度为 1 mm。如下图所示。激光器置于一个铝盒子内，铝板厚度为 11 mm，盒子内部粘有隔音海绵。这样可以很好地隔离环境中声学噪声对两台激光器的影响。与此同时，环内与环外互相关系统同样地也搭建在光学面包板上，置于一个厚度为 10 mm 的聚合材料组成的大盒子内。整个实验室的空调持续工作，保证环境温度在 24.1±0.3 C。这样，测量系统自身受温度漂移的影响也会最大程度地得到改善。

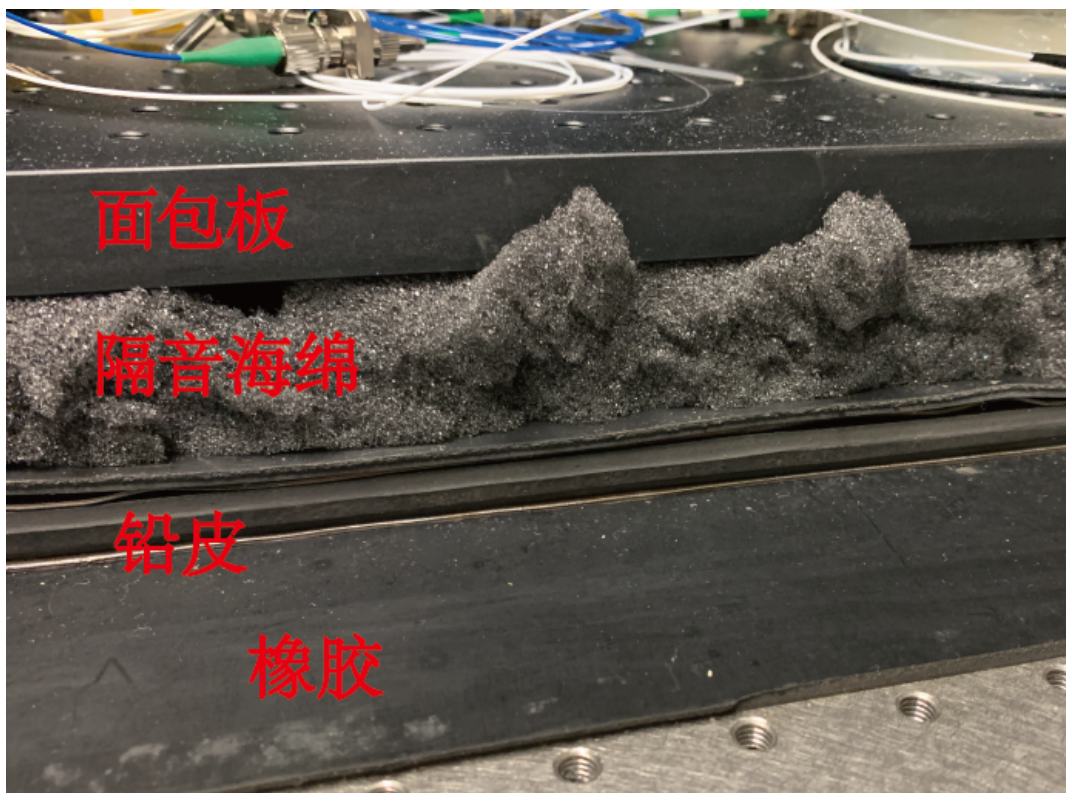

图 3-4　激光器底部隔振装置设计

### 3.3.1 时域误差信号

本工作实现了两台激光器进行了时间长达 5 天的时间同步。需要指出的是，在激光器 2 的一个准直器底部，一个电动位移平台可对激光器的重复频率较大范围调节，实现重复频率的慢锁定。在 5 天的锁定过程中，实时监测比例积分伺服控制器的输出信号，保证腔内压电陶瓷在其中心位置附近进行补偿。当压电陶瓷偏离中心位置过远时，慢锁相环路会小幅度调节电控平移台的位置，使得压电陶

瓷回到中心位置附近。使用数据采集卡（National Instruments，NI USB-6008）对同步后的误差信号进行数据采集，采样率为 1 kHz。同时，对实验室的温度和相对湿度进行实时监控。温度与湿度的测量精度分别为 0.1 °C 与 0.1%。图 3-5（a）给出了 5 天时间内，环内（蓝色）与环外（橙色）时域误差信号的变化曲线。在实现了时间同步后，环内误差号在 120 小时内的波动的均方根值为 103 as，环外误差信号在 120 小时内波动的均方根值为 733 as。特别地，环外信号近似以 24 小时为周期，以 1.5 fs 的幅度振荡。这说明了，尽管整个光学系统已经做了相应的环境隔离与温度控制，但是环外误差信号仍旧会受到环境温度及湿度变化的影响。

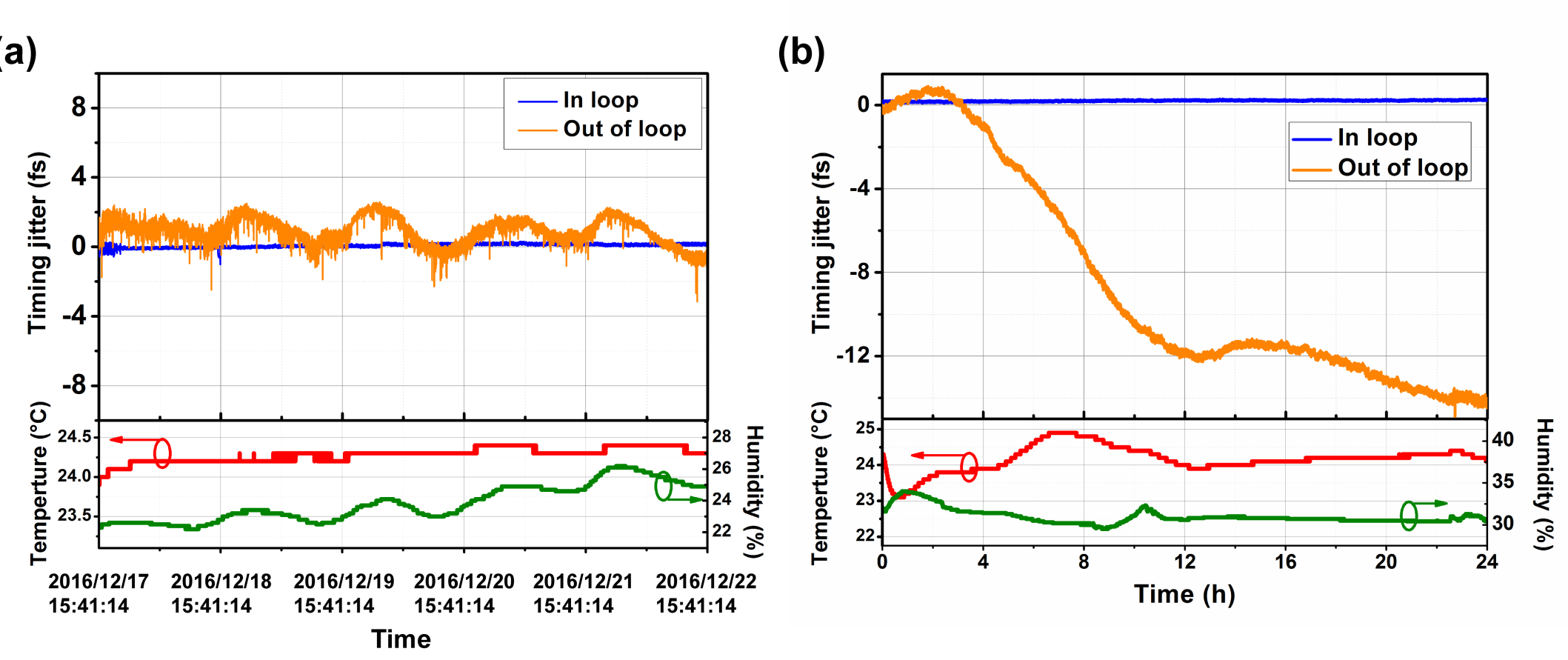

图 3-5 时间同步的时域误差信号。（a）在温度控制条件下，时间长度为 5 天的时间同步结果。（b）在无温度控制条件下，时间长度为 1 天的时间同步结果

为了探究环外误差信号长期漂移的原因，本文做了对比实验。在关闭了空调的情况下，对两台激光器时间同步进行了 24 小时的测量。时域误差信号见图 3-5(b)。可以看出，在第 0~1 小时内，室内温度有所下降，然后在随后的 1~7 小时，室内温度有所回升。随后的 7~24 小时内，室内温度先缓慢下降后缓慢上升。对应的，环外误差信号随着温度变化呈现相似的波动趋势。不同的是，总体而言，环外误差信号的整体趋势一直在向一个方向漂移，漂移幅度在 24 小时内达到了 15 fs。同时也说明了，随着温度的改变，环外误差信号会随之漂移，但是，如果温度变化量过大，即使温度变回初始温度，环外误差信号却无法回归初始值。温度变化过大导致的系统的失谐是不可逆的。

接下来对环外信号的长期周期性漂移做出定性分析。由于环外测量系统是共路的，其产生长期漂移的可能原因有两种：一、环境因素改变导致的具有双折射效应的光学晶体的折射率改变，偏振态相互垂直的两脉冲所经历的光程发生改变，进而两脉冲的时域位置发生改变；二、和频效率改变导致误差信号强度的改变。

首先，本文分别计算了 BBO 晶体和铌酸锂晶体中，寻常光和异常光的折射率随温度的变化量。对于 BBO 晶体，寻常光的折射率随温度的变化量为 $\mathrm{d}n_o/\mathrm{d}T$ = -9.3×10$^{-6}$/°C，异常光为 $\mathrm{d}n_e/\mathrm{d}T$ = -16.6×10$^{-6}$/°C。因此对于 1 mm 厚的 BBO 晶体，1040 nm 波段的光在温度变化 1 °C 时，寻常光与异常光的时间延迟改变 2.43×10$^{-2}$ fs。如需产生 2 fs 的延迟量，对应的温度变化为 82 °C。类似地，对于 0.4 mm 厚的铌酸锂晶体，产生 2 fs 的延迟量所对应的温度变化为 15 °C。这两种情况均远远超出了该实验中记录到的温度变化量，该原因可排除。另一个可能原因是非线性晶体的相位失配导致的和频效率的改变。相位匹配参量与相位匹配角度的关系，如图 3-6 所示。2 fs 的定时抖动对应的和频信号的强度变化为 1.63%，对应相位匹配角度的变化仅为 0.00693 度（4.158’）。这说明和频信号的效率对于非线性晶体的角度非常敏感，任何轻微的晶体支架角度的改变都会导致和频效率的改变。虽然实验室的空调一直在运转，但是实验室的温度和湿度均在以 24 小时为周期变化，这可能是由于外部环境的昼夜交替变化超出了实验室空调的最大控制分辨率。对于环内而言，由于锁相环的作用，温度和湿度的昼夜交替变化导致的光路失谐会被锁相环消除，因此环内信号一直保持着良好的被锁定状态。然而，对于环外测量而言，没有锁相环的作用，误差信号会受到温度和湿度变化导致的光路失谐的影响，例如晶体支架角度的微量改变，进而导致和频光效率的周期性改变。综上所述，本文对环外信号的周期性改变归因于此。

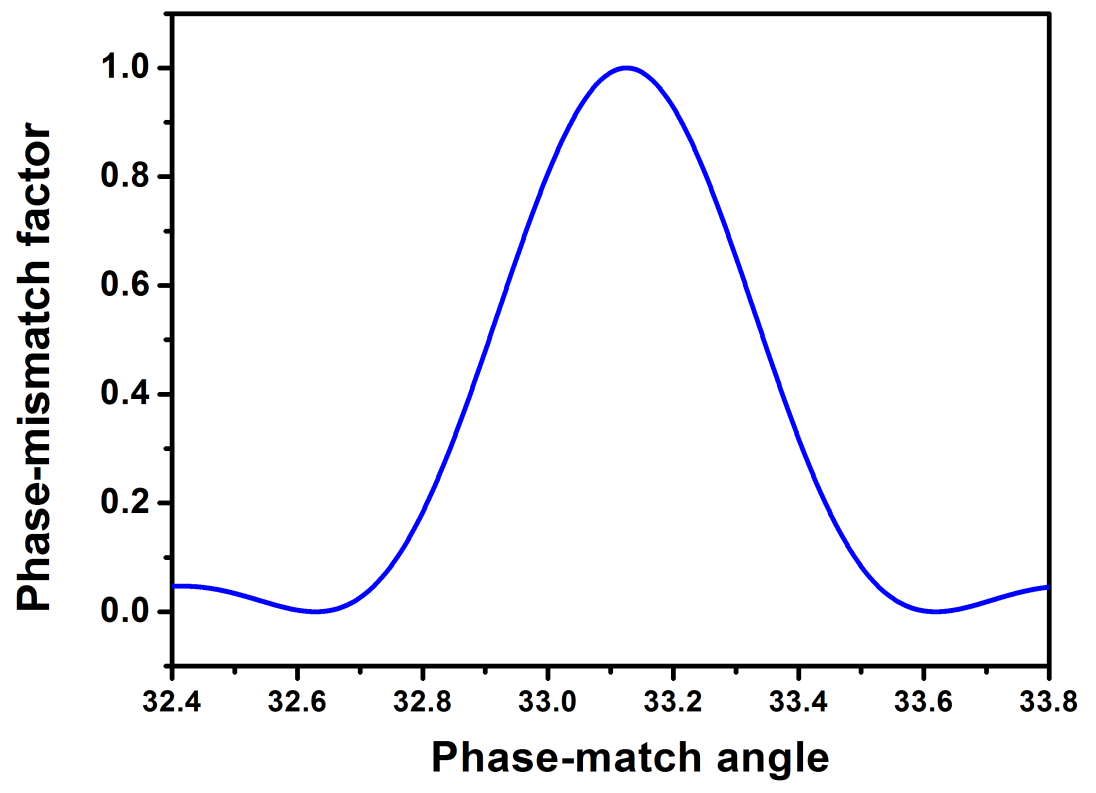

图 3-6 相位失配参量随相位匹配角度的改变

### 3.3.2 方差分析

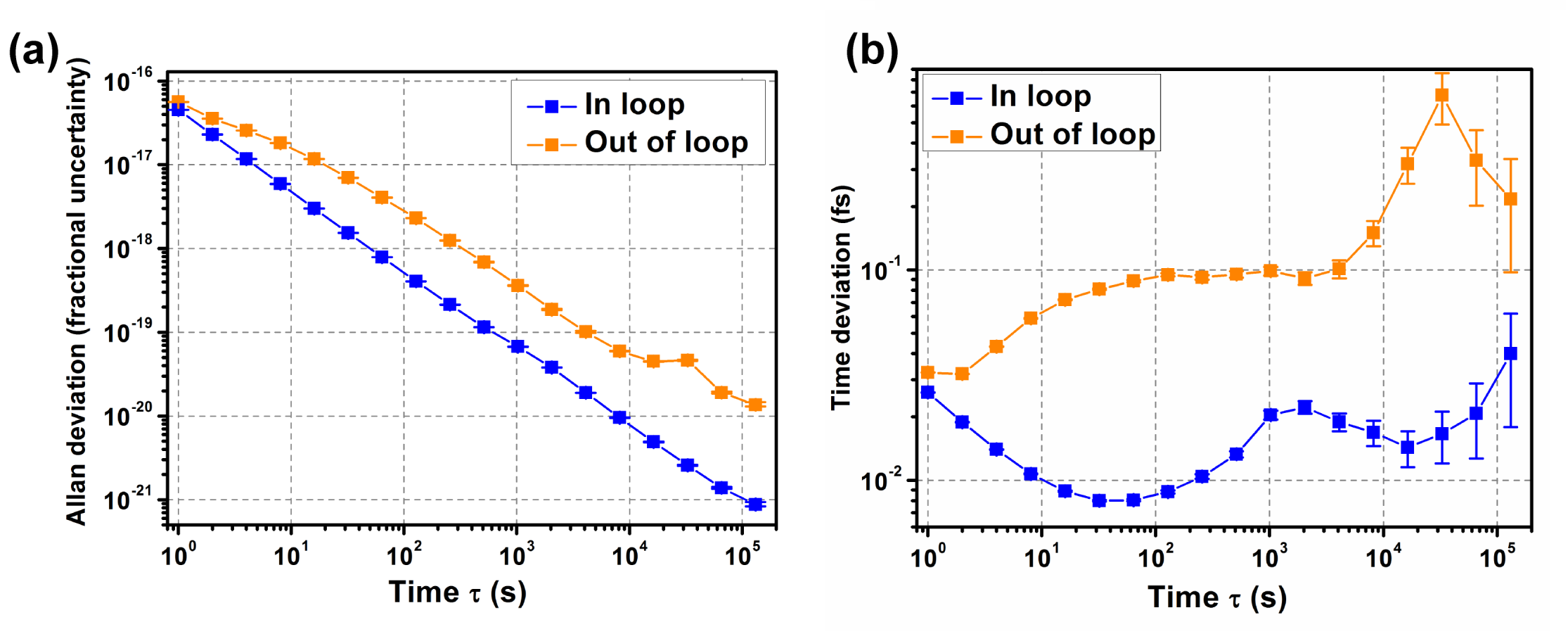

图 3-7 时间同步的时域稳定性评价。(a) 环内（蓝色）与环外（橙色）的交叠艾伦方差。(b) 环内（蓝色）与环外（橙色）的时间艾伦方差

接下来，本文对两台激光器的时间同步的稳定性进行了方差分析，分别使用了交叠艾伦方差和时间艾伦方差对时间同步的稳定性做出了评价。交叠艾伦方差是一种非常常用的时间频率稳定性的评价标准，是艾伦方差的一种常见的形式。通过将采样时间内的数据进行交叠处理，交叠艾伦方差最大程度地利用了所有的采样数据。相比于普通的艾伦方差，方差结果可以延伸到更长的门时间，并且有着更小的置信区间和数据可靠度。图 3-7 (a) 展示了从时域误差信号计算得到的环内（蓝色）与环外（橙色）的交叠艾伦方差。对于环内的交叠艾伦方差，从 1 s 到 $10^5$ s 门时间范围内内，方差曲线以 1/τ 的斜率下降，最终 $1.31\times10^5$ s 的稳定度为 $8.76\times10^{-22}$。然而，由于环境因素的影响，环外测量的交叠艾伦方差比环内大约高 10 dB。环外测量的交叠艾伦方差的 $1.31\times10^5$ s 的稳定度为 $1.36\times10^{-20}$。

另一方面，本文使用时间艾伦方差对时域误差信号进行了分析。时间艾伦方差是一种基于修正艾伦方差的，适用于评价分布网络时间稳定性的评价方法。图 3-7 (b) 展示了环内（蓝色）与环外（橙色）的时间艾伦方差。从 1 s 到 32 s 门时间，环内的时间方差曲线以 $\tau^{-2}$ 的斜率下降。这表明了在低于 0.03 Hz 的傅里叶频率范围，误差信号的功率谱密度的斜率为 $1/f$ 的白相位噪声。对于环外时间艾伦方差，并没有一个确定的斜率。这说明在低于 1 Hz 的傅里叶频率范围，环外测量由于受到环境因素的影响，是无规律的长期漂移而非白噪声。

### 3.3.3 功率谱分析

将同步后的时域误差信号做傅里叶变换，即可得到在至 2 μHz 至 1 Hz 范围

内环内与环外测量的定时抖动的功率谱密度图。同 3.2.2 节中高频部分的定时抖动功率谱密度图拼接在一起，可以得到时间同步后的，2 μHz 至 10 MHz 范围的剩余定时抖动功率谱密度图，如图 3-8。可以看出，由于环境温度的改变，环外剩余噪声的功率谱也明显高于环内。

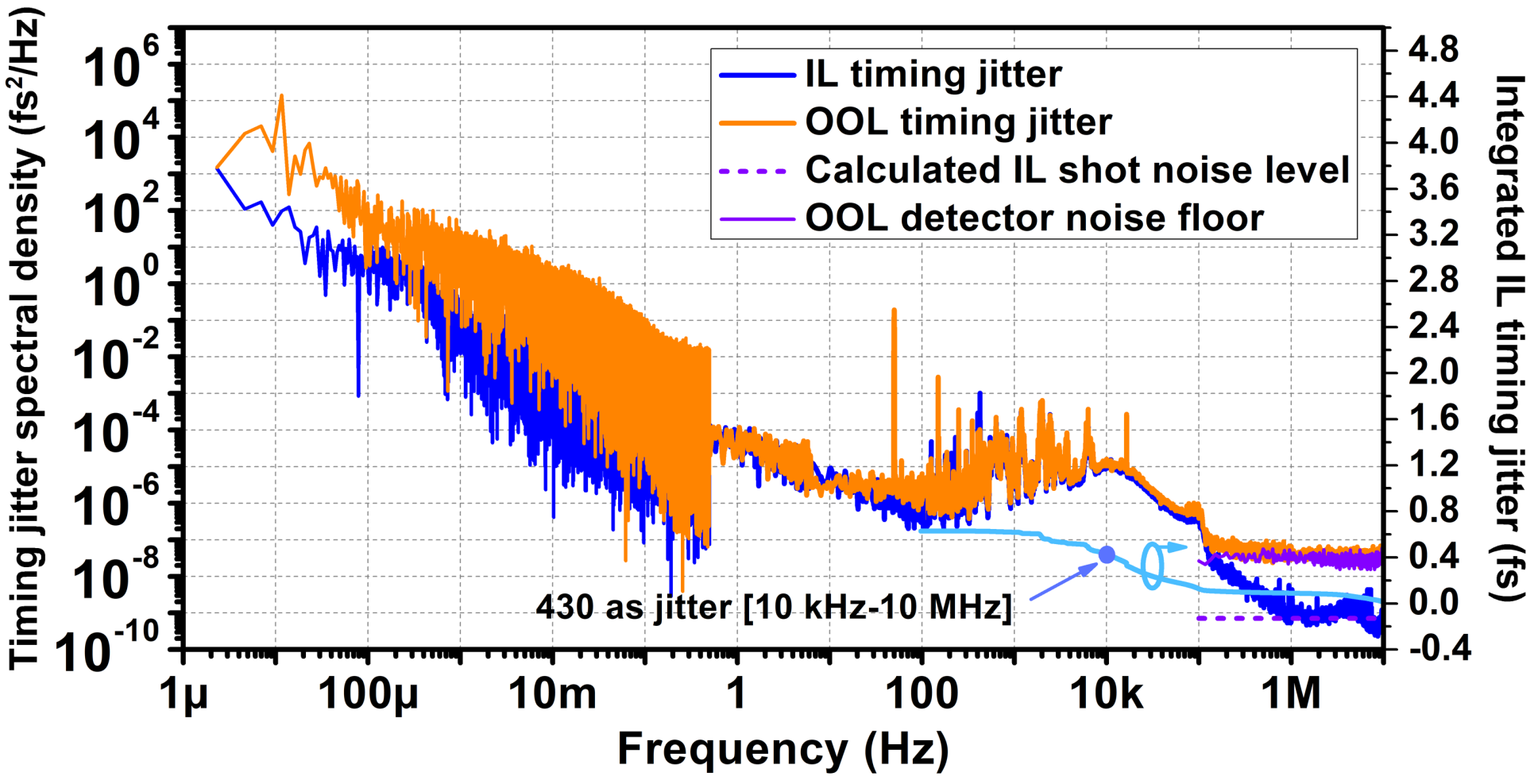

图 3-8　时间同步时域误差信号功率谱密度图

## 3.4 光纤引入的定时抖动测量

得益于低定时抖动的飞秒激光器和光学频率梳技术的蓬勃发展，世界各地的许多高校、研究所均拥有了低噪声时间基准或光学频率基准。那么伴随而来的问题是如何对处于不同地点的时间、频率基准进行比对。光学脉冲在自由空间中传输很容易受到大气湍流扰动的影响。相比之下，光纤链路可以作为一种更理想的时间、频率基准传输介质。德国联邦物理技术研究院首次使用光纤实现了百公里长度的光学频率基准传输系统[163]。2014 年，韩国高等技术研究院（KAIST）也实现了公里长度的时间基准传输[46]。光学脉冲在光纤中传输时，光纤受到温度、湿度、应力等众多环境因素的影响，会对光学脉冲引入额外的定时抖动。因此在这些时间、频率基准传输系统中，对光纤引入额外的噪声的测量和分析，是实现低噪声、高精度时间、频率基准传输的重要环节。

### 3.4.1 基于光学外差法的定时抖动测量原理

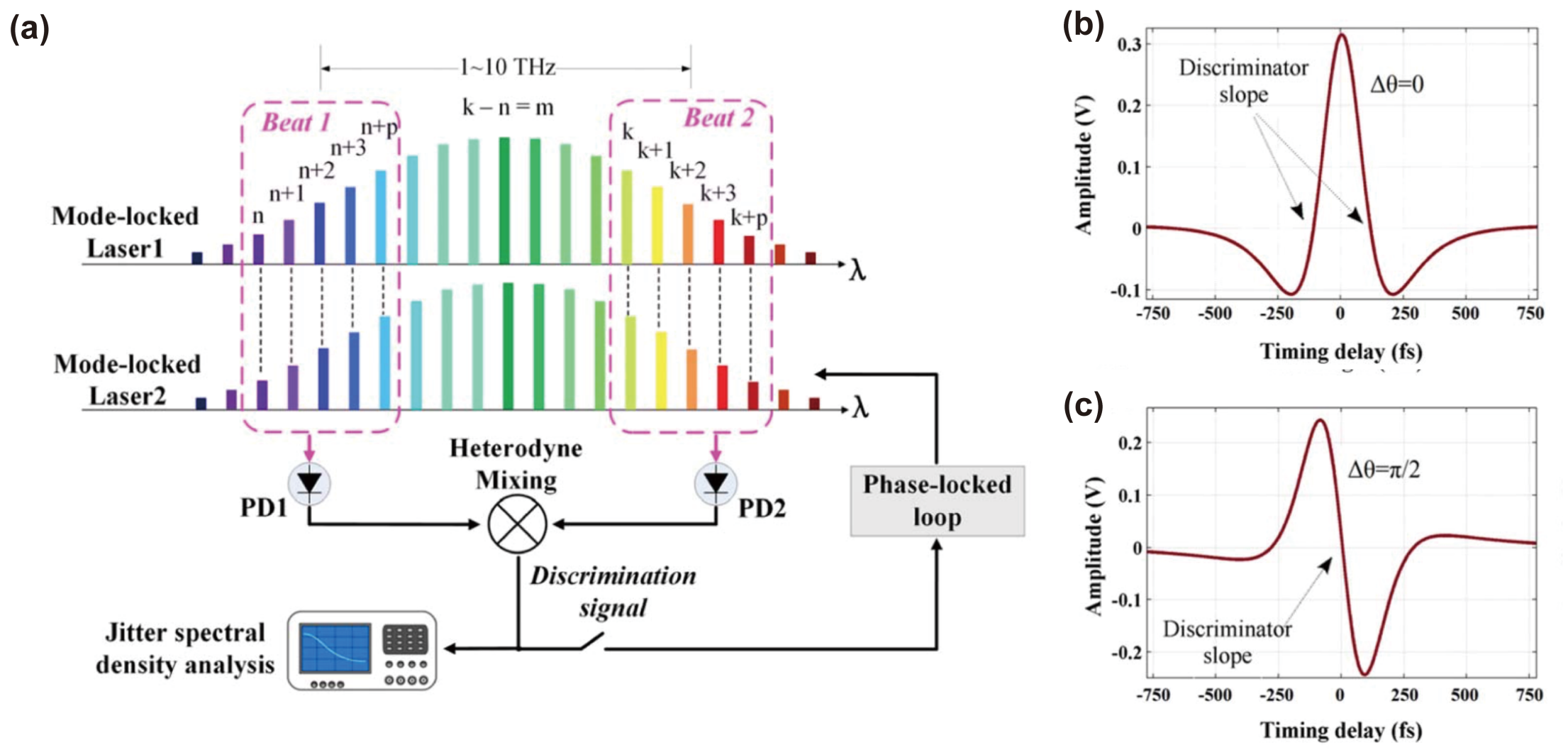

图 3-9　光学外差探测法的定时抖动测量原理示意图

图 3-9（a）为基于光学外差探测法测量两台激光器定时抖动的测量原理示意图。将两台激光器输出光谱的短波部分和长波部分分别做拍，得到拍频信号 1 与拍频信号 2。两个拍频信号的频率大小为两台激光器的相对载波-包络偏移频率。将拍频信号 1 与拍频信号 2 在正交方向上进行混频，即对其中之一加以 π/4 的相位偏移，可以得到一个高相位鉴别度的，与两台激光器定时抖动相关的相位鉴别信号。使用该相位鉴别信号作为误差信号，并用比例积分伺服控制系统将两台激光器的重复频率进行低带宽的锁定后，通过频谱分析仪或相位噪声分析仪测量锁定后的误差信号，即可得到能够反映激光器自由运转的定时抖动特性的功率谱密度图。

假设两台飞秒激光器的重复频率差为 $\Delta f_{rep}$，相对载波-包络相位偏移频率为 $\Delta f_{ceo}$。拍频信号 1 是两台激光器的第 $n$ 个纵模至第 $n+p$ 个纵模的外差拍频的结果。拍频信号 2 是两台激光器的第 $k$ 个纵模至第 $k+p$ 个纵模的外差拍频的结果。光电探测器 1 产生的信号为：

$$I_{d1}(t) \propto \mathrm{Re}\left[\sum_{j=n}^{n+p} A_{1,j} A_{2,j}^{*} \cos(2\pi(j\Delta f_{rep} + \Delta f_{ceo})t)\right] \tag{3-2}$$

其中 $A_{1,j}$ 和 $A_{2,j}$ 分别为两台激光器的复振幅项。在实验中，使用一个截止频率在 10 MHz 左右的低通滤波器对误差信号进行滤波。因此，在以上公式中，激光器的重复频率项被省去了。同样地，光电探测器 2 产生的信号也可以用类似的表达式表示，累加范围从 $k$ 到 $k+p$。混频器将两个误差信号混频后，相对载波-

包络偏移频率的项被消除，得到的相位鉴别信号可以近似被表达为：

$$V_\theta(t) \propto \sum_{\alpha=-p}^{p} G_\alpha \cos[2\pi(k-n+\alpha)\Delta f_{rep} t] \tag{3-3}$$

其中，$G_\alpha$ 为与复振幅相关的归一化参数项。从以上公式反映了，误差信号 $V_\theta$ 由一系列中心频率在 $m\Delta f_{rep}$，间隔为 $\Delta f_{rep}$ 的信号叠加而成。该信号的中心频率 $m\Delta f_{rep}$ 与模式数 $m$ 成正比，$m$ 范围在 $10^3$ 至 $10^6$ 量级，$m\Delta f_{rep}$ 则在 $10^{12}$ 量级。这是由激光器输出光谱宽度决定的。$m\Delta f_{rep}$ 越大，相位鉴别的灵敏度越高。相比于直接用光电探测器进行重复频率噪声测量的方法，由于光电探测法只能探测重复频率的若干次谐波，因此 $m$ 只能为一个小于 100 的整数，远低于 $10^3$ 至 $10^6$ 量级。所以，采用该光学外差探测法，能够有更灵敏的探测精度。当增大拍频 1 和拍频 2 的光谱范围至最大，即两个光谱中间没有光谱间隙，鉴相信号为图 3-9（b）中所示形状。在长波部分加入了一个 1/4 波片后，两个拍频信号会产生 π/4 的相位差，鉴相信号会变成如图 3-9（c）所示形状。利用该鉴相信号的中间直线部分，可以实现高精度的定时抖动相位鉴别。

### 3.4.2 高重复频率掺铒固体激光器

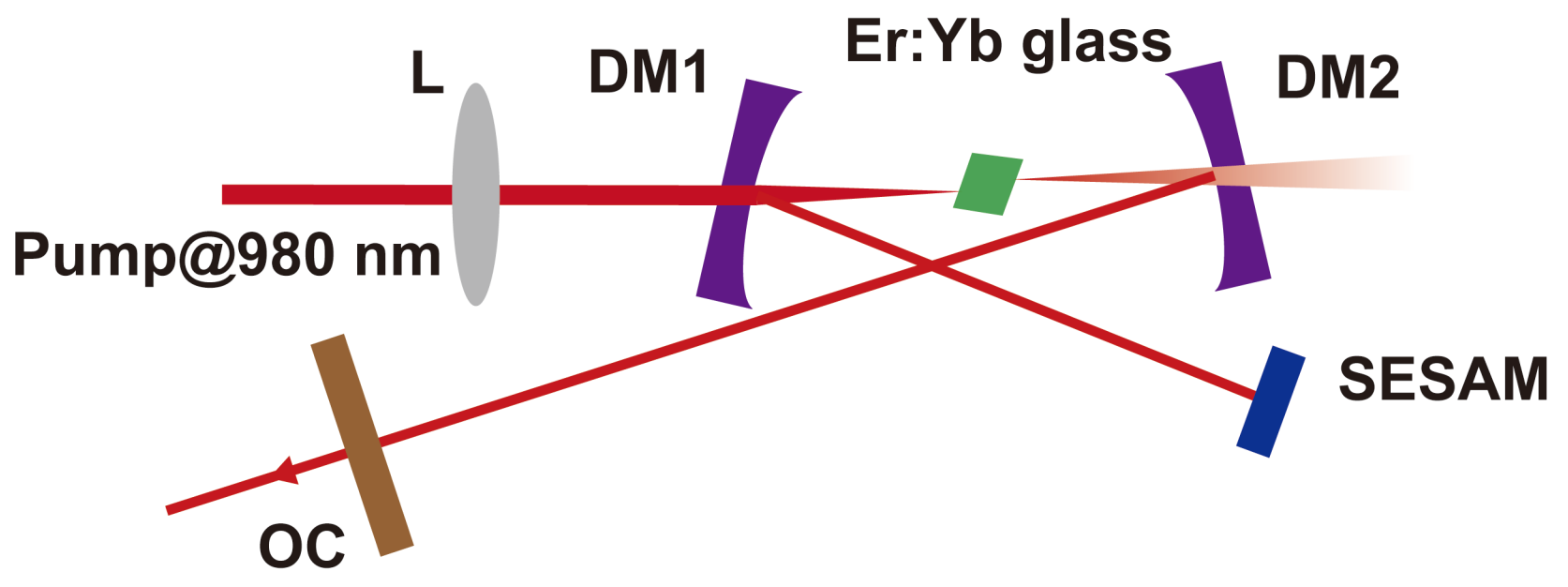

图 3-10 高重频掺铒固体激光器示意图。

该实验使用重复频率为 500 MHz 的 1550 nm 波段的固体激光器作为光源。激光器的结构图如图 3-10 所示。使用铒镱共掺的光学玻璃晶体作为增益介质，晶体厚度约为 1.5 mm。使用 980 nm 的激光二极管作为泵浦源。泵浦光经焦距为 75 mm 的透镜 L 聚焦到增益介质上。焦距为-100 mm 的两双色镜 DM1 与 DM2 置于增益介质两侧，用于收集以及准直增益介质发射出来的荧光。激光腔振荡产生激光后，DM1 反射的激光随后入射到一片特殊定制的半导体可饱和吸收镜（Semiconductor Saturable Absorber Mirror，SESAM）上，被 SESAM 原路返回。

DM2 反射的激光入射到一个输出率为 0.5%的输出镜 OC 上，0.5%的激光被输出，99.5%的激光返回腔内继续振荡。特别的是，为了尽量减小激光器的 Gorden-Haus 色散，激光腔的净色散需要尽量接近于零色散。为此，双色镜 DM1 与 DM2 的镀膜为在 1560 nm 波段特殊设计的二阶色散/三阶色散补偿膜系。输出镜 OC 也在 1560 nm 波段色散平坦。该激光器中的各个器件的距离为：DM1—glass：46 mm; glass—DM2：27 mm；DM1—SESAM：54 mm；DM2—OC：170 mm。

在搭建该激光器时，应注意以下问题：

一、关于聚焦透镜 L 的焦距的选择:首先用 reZonator 软件模拟整个激光腔，计算得到腔内振荡的激光在增益介质上的光斑半径约为 26 μm。对应的，泵浦光经过透镜 L 和双色镜 DM1 后，光斑半径也应该为 26 μm 左右。用该软件进行计算，当透镜 L 的焦距为 75 mm，L 与 DM1 的距离为 50 mm 时，泵浦被聚焦后的半径符合该条件。

二、关于象散补偿角：该激光器中，1.5 mm 增益介质对应的象散补偿角为 10°。但在实际搭建过程中，DM1 的镜架边缘很容易切光，因此可以适当地加大该角度，实验证明，对激光器的运转影响不大。

三、关于晶体的布儒斯特角：晶体表面与入射光的角度应满足晶体的布儒斯特角条件，这样晶体表面对于泵浦光和腔内激光的反射才会达到最低，这一点尤为重要。为了实现超低定时抖动的飞秒激光器运转，腔内的输出镜仅为 0.5%，这已经是一个非常低的输出率，晶体的放置角度稍有不当，晶体表面的反射率极易超过 0.5%，该激光器的定时抖动会大大增加。晶体角度在搭建激光器过程中需要反复仔细优化，确保在最优的角度下，激光腔达到最大的输出效率。

四、关于 SESAM 的位置：DM2 与 SESAM 的距离可以在 52 mm 至 55 mm 之间调整，不同的距离对应着激光在 SESAM 上有不同大小的光斑。如果 SESAM 上光斑太大，光功率密度不够，则无法使 SESAM 饱和，无法实现锁模运转。如果光斑太小，光功率太高有可能会打坏 SESAM。

五、在优化激光腔效率过程中，激光器的振荡阈值是一个非常重要的指标。通常会将泵浦电流降低至激光器刚刚开始振荡，随后优化腔效率至最大，接着继续降低泵浦电流至激光器刚刚开始振荡。如此反复，一般地，激光的振荡阈值在 90~100 mA 时，为较理想情况。通常，当泵浦电流为 200~300 mA 左右时，激光器会进入调 Q 状态，继续增加泵浦电流至 400~500 mA 时，激光器会实现锁模运转，在 700 mA 泵浦电流下，激光器可达到最理想状态的锁模运转。

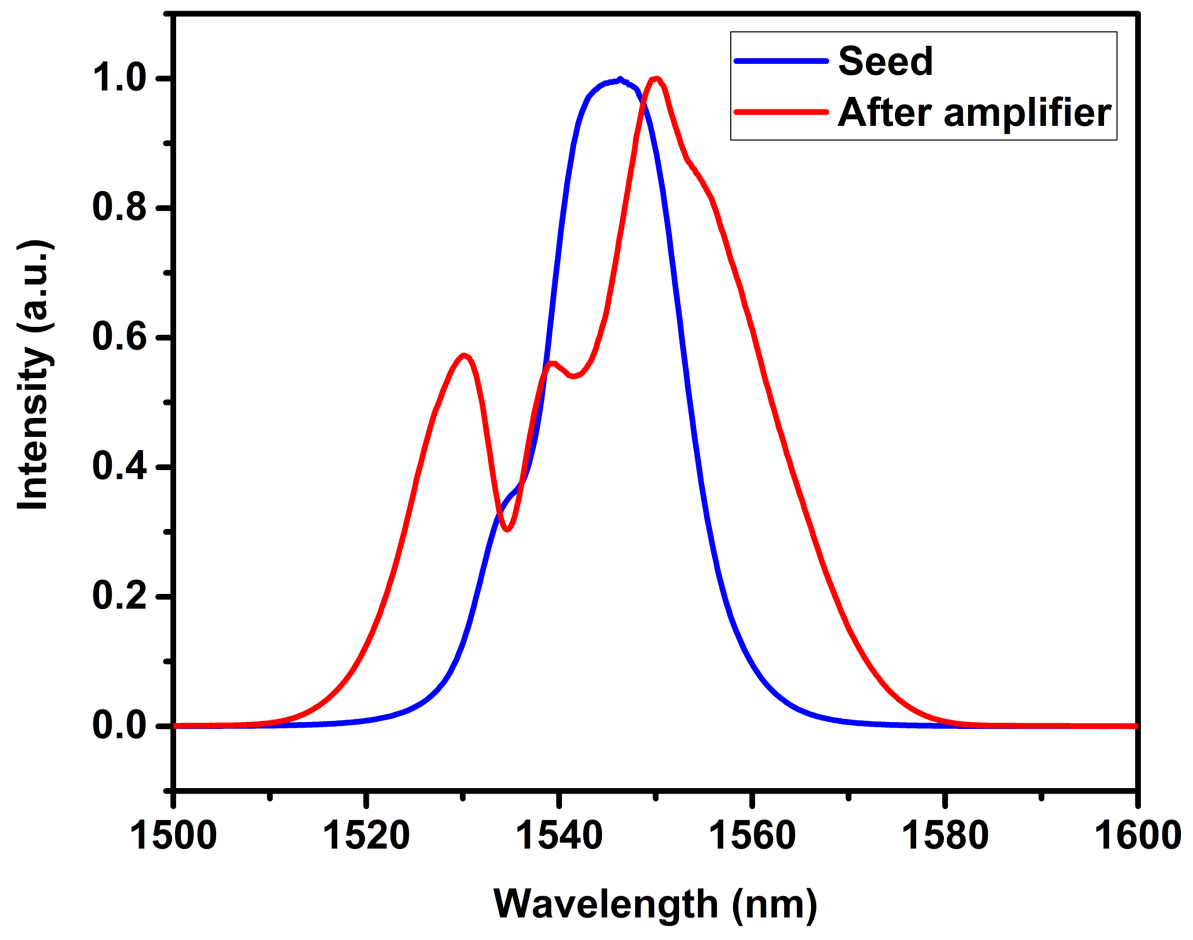

图 3-11　激光器直接输出的光谱（蓝色曲线）和经过光纤放大后的光谱（红色曲线）

激光器锁模后，脉冲的重复频率约为 500 MHz，输出功率为 48 mW，脉宽为 120 fs 左右，光谱半高全宽为 15 nm，激光器输出光谱见图 3-11 蓝色曲线。随后，为了获得更大的激光功率，使用全保偏结构的光纤放大器对振荡级输出的脉冲进行放大。光纤放大器的结构非常简单，采用了后向泵浦的放大技术。激光器输出的脉冲作为放大器的种子光，依次通过长度为 80 cm 的保偏 Er-80-4/125 增益光纤和一个 WDM 后输出。当放大级的泵浦电流为 1.4 A（对应大约 500 mW 的泵浦功率）时，放大级的输出功率在 150 mW 左右。放大后的光谱见图 3-11 红色曲线。

### 3.4.3 基于光学外差探测的光纤定时抖动测量

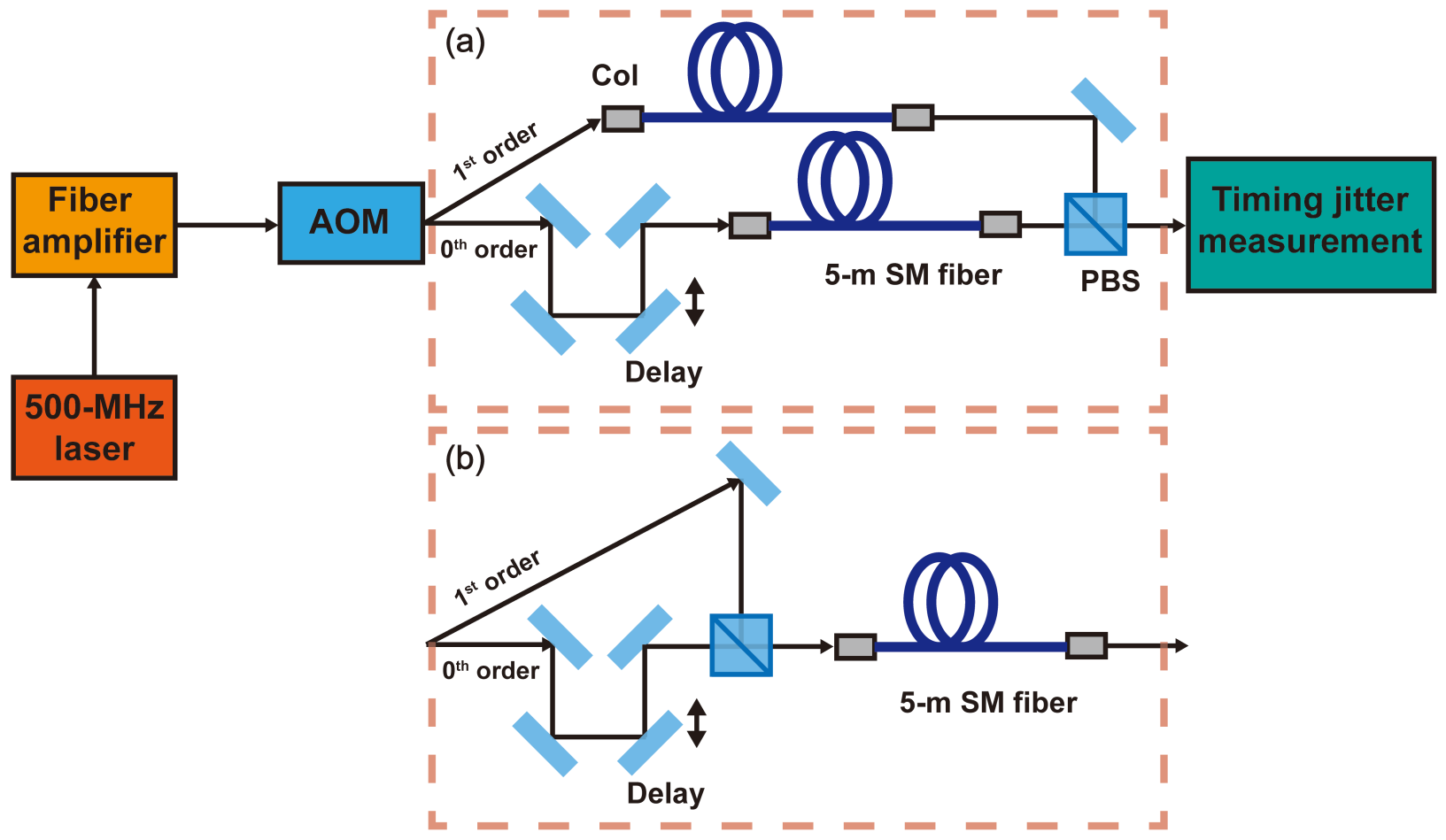

图 3-12　光纤定时抖动测量系统结构示意图。（a）两段长度为 5 m 的单模光纤的定时抖动的测量。（b）对比实验

图 3-12（a）为基于光学外差探测法的光纤定时抖动测量系统示意图。将放大后的脉冲经过一个工作频率为 80 MHz 的声光频移器后，零级衍射光的光学频率不会被改变，1 级衍射光的频率会被频移 80 MHz。零级衍射光经过一个由电动平移台控制的可调延迟线之后，耦合进入长度为 5 m 的单模光纤。1 级衍射光直接耦合进入 5 m 单模光纤。最后，零级衍射光和 1 级衍射光经偏振合束器合束后进入光学外差探测系统。由此可见，与文献[28]中不同的是，该实验用声光频移器产生的 1 级与零级衍射光作为待测对象，替代文献[28]中的两台飞秒激光器。图 3-12（b）为对比实验结构示意图，在该对比实验中，移除了一根 5 m 长的单模光纤，将零级衍射光与 1 级衍射光直接合束后进入 5 m 单模光纤。该对比实验的目的是测量干涉仪对光纤时间抖动测量结果的影响，对比实验的结果详见 3.4.5 小节。

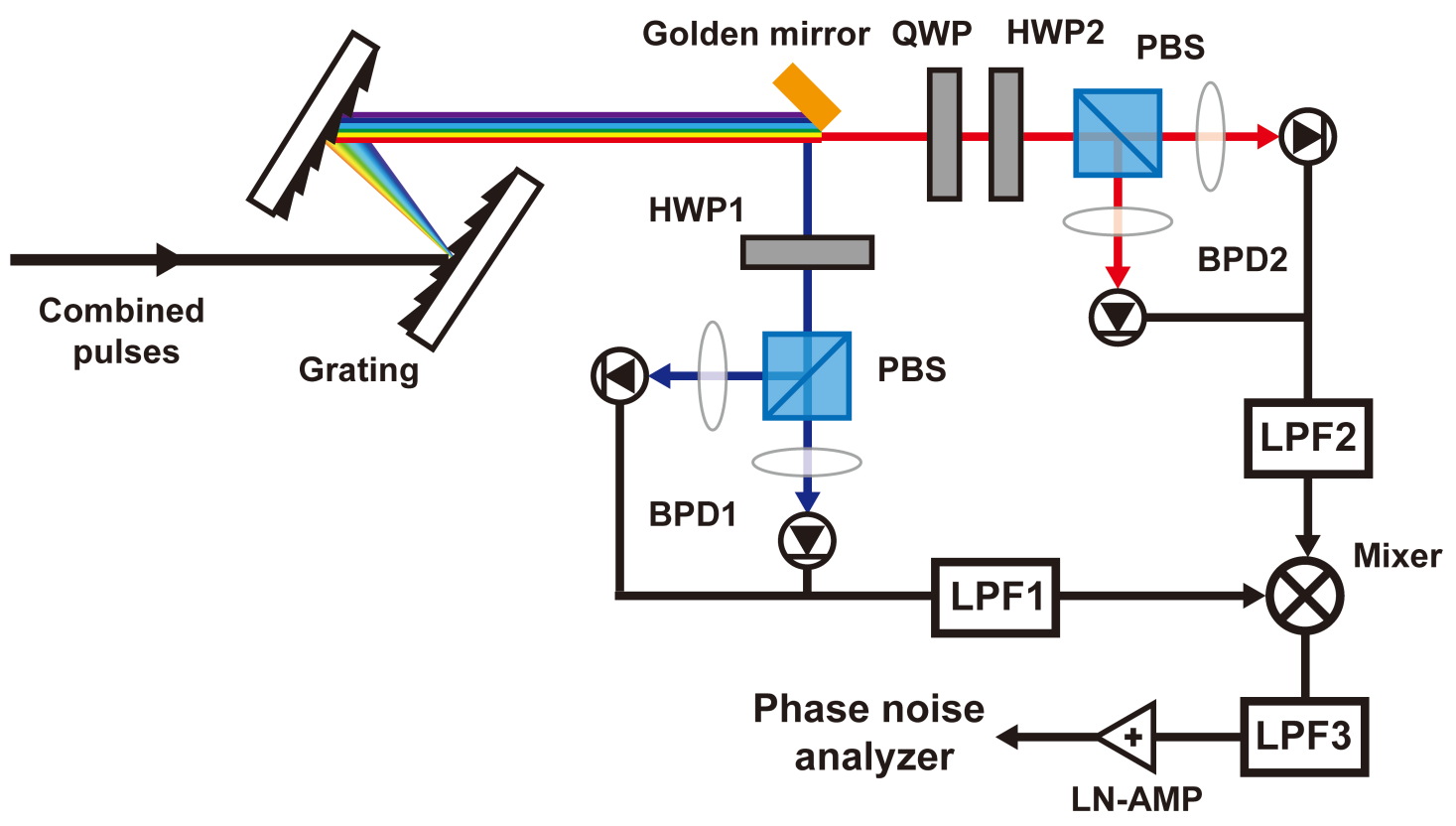

图 3-13　光学外差探测系统结构示意图。HWP，半波片；QWP，四分之一波片；PBS，偏振合束器；BPD，平衡探测器；LPF，低通滤波器；LN-AMP，低噪声电放大器

图 3-13 为光学外差探测系统的结构示意图。合束后的脉冲经过一个闪耀波长在 1550 nm 波段的闪耀光栅后，不同光谱成分在空间上散开。随后第二个相同的光栅将具有空间色散的光谱准直。短波部分的光谱成分被一块镀金膜的反射镜反射，经过一个半波片 HWP1 后，被偏振分束器分束为能量相等的两部分，分别经过聚焦后入射进入平衡探测器 BPD1 的两个输入端。长波部分的光谱成分继续向前传输。经过一个四分之一波片 QWP 和半波片 HWP2 后，同样地，被偏振分束器分束为能量相等的两部分，分别经过聚焦后进入平衡探测器 BPD2 的两个输入端。平衡探测器 BPD1 由两个低噪声的铟镓砷光电探测器组成（Hamamatsu，G12180-003A），该探测器的感光面直径为 300 um，等效电容为 5 pF，等效噪声功率（Noise equivalent power，NEP）为 $4.2\times10^{-5}$ W/Hz$^{-1/2}$。由于两个脉冲相差的光学频率为 80 MHz，因此，每个探测器上都可以探测到频率为 80 MHz 的拍频

信号。该拍频信号包含了光纤的定时抖动的信息。两个探测器串联构成一个平衡探测器，平衡探测器的输出的拍频信号经过一个滤波电容（1~100 nF）将直流成分滤除掉。随后经过一个 3：1 的变压器（Minicircuits，T3-1+）提高平衡探测器的输出功率。使用一个截止频率为 120 MHz 的低通滤波器 LPF1 滤除重复频率信号后，进入混频器（Minicircuits，TUF-1+）的一端。平衡探测器 BPD2 为完全相同的结构，输出经过滤波电容、变压器、低通滤波器 LPF2 后，进入混频器的另一端。混频器将两个 80 MHz 的拍频信号进行混频再进行低通滤波器 LPF3（Minicircuits，DC-1.9 MHz）后，经过一个自行搭建的低噪声放大器进行信号放大，就可以得到与定时抖动相关的在基带附近的噪声信号。对此信号使用相位噪声分析仪（Keysight，PXA Signal Analyzer N9030A）进行相位噪声功率谱测量，即可得到光纤引入的定时抖动的噪声功率谱。

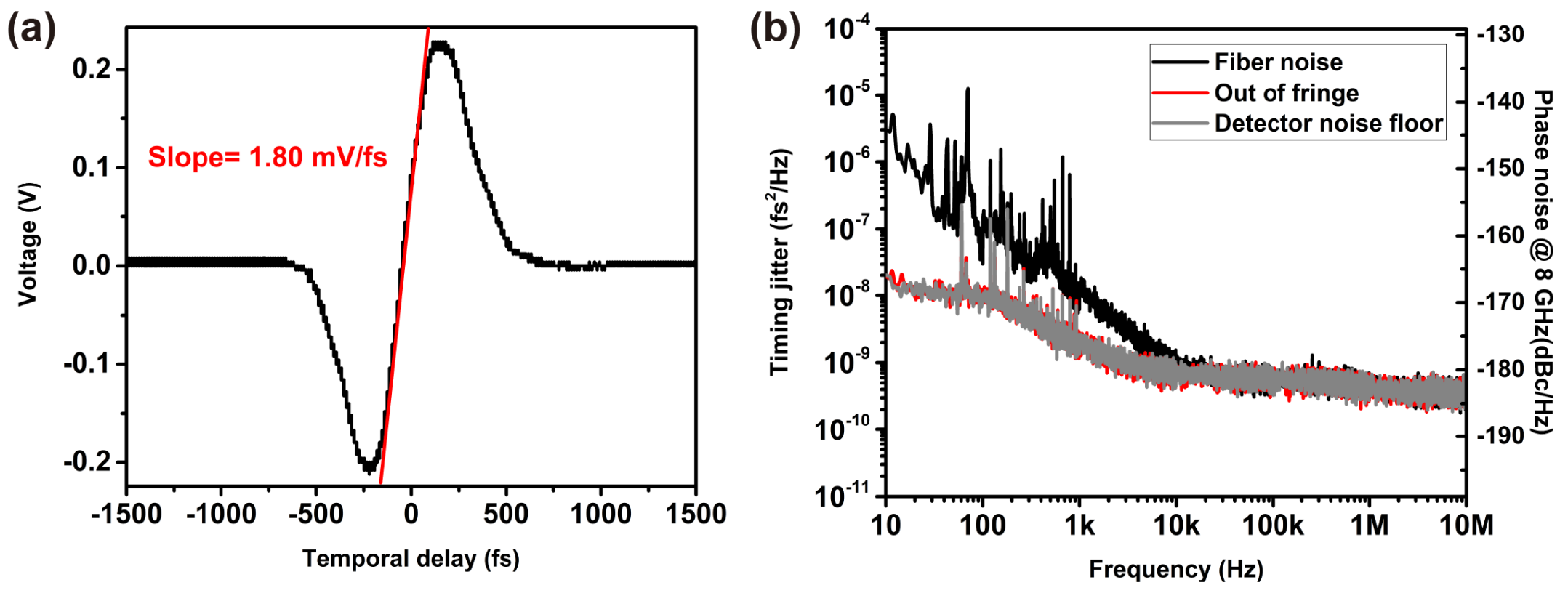

图 3-14 基于光学外差法的光纤定时抖动测量结果。（a）误差信号。（b）定时抖动功率谱密度

数据测量过程中，首先将电控延迟线中的电动位移平台移动至干涉信号最强的位置。随后让电动位移平台在 50 μm 范围内进行循环扫描。用示波器探测混频器输出的误差信号，可以在示波器上看到 S 型的误差信号曲线。如图 3-14（a）所示。误差信号的鉴相斜率为 1.8 mV/fs。随后进行三组数据测量：一、将两个平衡探测器用感光卡遮挡，测量误差信号的功率谱。该步骤是为了测量探测器与低噪声放大器的噪声基底，对应于图 3-14（b）中灰色曲线。该噪声基底决定了整个测量系统的分辨率；二、移动平移台至远离 S 曲线中心的点，测量误差信号的功率谱，对应于图 3-14（b）中红色曲线。该步骤是为了验证测量的准确性。如果散粒噪声低于探测器和放大器的噪声基底，那么该测量得到的功率谱应当完全与系统的噪声基底一致。如果散粒噪声高于探测器和放大器的噪声基底，那么该测量得到的功率谱应当在散粒噪声的量级，可计算散粒噪声的量级以验证。三、

移动平移台至 S 曲线中心处，测量误差信号的功率谱，对应于图 3-14（b）中黑色曲线。该结果对应光纤的定时抖动功率谱。最后，使用误差信号的鉴相斜率将测量到的三组电压的功率谱转化为定时抖动的功率谱，结果请见图 3-14（b）。可以看出，光纤的定时抖动功率谱在 10 Hz 至 10 kHz 频率范围内，一直以 10 dB/decade 斜率下降。10 kHz 偏频处，测量开始受限于光电探测器和放大器的噪声基底。光纤引入的定时抖动在 10 Hz 至 10 kHz 频率范围内的积分值为 64.7 as。

该实验在系统搭建与数据测量过程中，需要注意如下事项：

一、平衡探测的目的是为了消除激光器强度噪声对定时抖动测量带来的干扰。在数据测量前应当测量自行搭建的平衡探测器的强度噪声抑制比。具体做法为：使用一个声光调制器对进入平衡探测器之前的光进行一个固定强度、固定频率的振幅调制。使用频谱分析仪分别探测单个探测器输出的和平衡探测器输出的调制信号的强度，计算该信号的抑制比。一般来讲，平衡探测器的强度噪声抑制比应当大于 30 dB。

二、为了保证长波与短波光谱分量的功率相同，需要调节金镜的位置，使被金镜反射的光功率与没有被反射的光功率大致相等。

三、为了保证激光被光电探测器全部接收，需分别优化每个光电探测器的耦合效率。在这个过程中，可以通过测量探测器中的 100 欧姆负载电阻两端的电压来判断是否达到理想耦合效率。例如，如果光电探测器前的光有 1 mW 的平均功率，那么对应地光电探测器应产生 1 mA 左右的电流，负载电阻的电压应为 100 mV。最后，分别调节两个偏振分束器前的半波片，使得平衡探测器输出的平均电压为 0 V。

四、需要特别注意的是，在放置 HWP2 和 QWP 之前，偏振分束器透射的光应全部来自于一段单模光纤，反射的光应全部来自于另一段单模光纤。这时首先加入四分之一波片，调整波片的角度，保证偏振分束器的分束与加四分之一波片之前一致，此时的四分之一波片的角度为正确的角度，在随后实验过程中不应再改变。

五、该实验的测量分辨率受限于混频器的效率。混频器的 LO 端口输入信号的电功率需要> +7 dBm 才能保证混频器的损耗在 3 dB 以内。然而，从平衡探测器输出的信号即使经过变压器后，仍旧很难达到这一数量级。如果增加电学放大器，又会引入额外的非线性，使得测量到的功率谱产生畸变。

### 3.4.4 基于电学降频的光学外差探测法

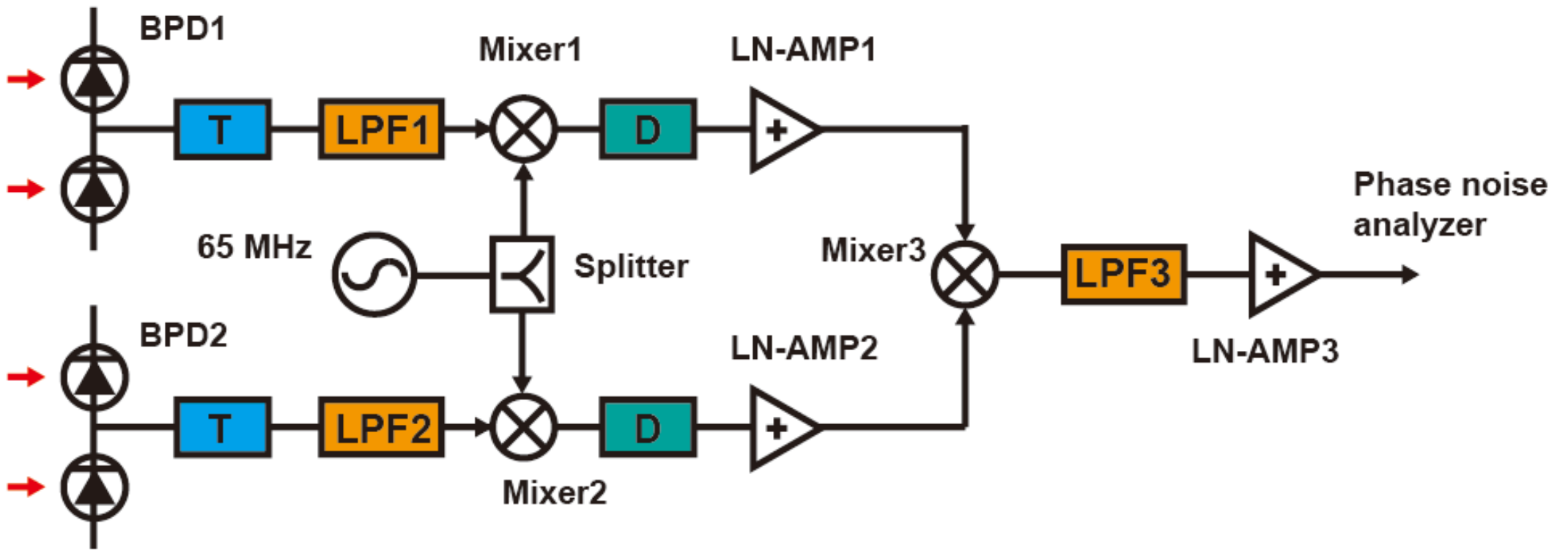

图 3-15　基于电学降频的光学外差探测法电学装置示意图。BPD，平衡光电探测器；T，变压器；LPF，低通滤波器；Mixer，混频器；Splitter，功率分束器；D，双工器；LN-AMP，低噪声放大器

为了克服混频器 LO 端口输入信号功率不足导致的混频器效率下降的问题，本工作提出了基于电学降频的光外差探测法。该方法的实验转置中，光学部分与上一节的光学外差探测法装置相同，本工作对电学部分进行了改进。电学部分的实验装置示意图如图 3-15。平衡探测器的输出的 80 MHz 拍频信号经过一个滤波电容（1~100 nF）将直流成分滤除掉。随后经过一个 8：1 的变压器（Minicircuits，T8-1T+）提高平衡探测器的输出功率。使用一个截止频率为 270 MHz 的低通滤波器 LPF1 滤除重复频率信号后，进入混频器 1（Minicircuits，TFM-3H+）的 RF 端口。使用一个低噪声信号发生器输出频率为 65 MHz，功率为+20 dBm 的参考信号。该信号经过一个功率分束器（Minicircuits，SYPS-2-1+）后，功率为+17 dBm 的 65 MHz 信号输入至混频器 1 的 LO 端口。该混频器为高功率，高线性度的混频器，能够提供更大范围的线性相位鉴别。并且+17 dBm 的 LO 端口输入保证了混频器的工作效率。65 MHz 的参考信号与 80 MHz 拍频信号混频后，产生 15 MHz 的拍频信号。15 MHz 的拍频信号通过一个双工器（Minicircuits，RDP-2150+）将混频器残留的高频信号滤除后，被一个低噪声电放大器（Minicircuits，PHA-13LN+）放大。放大后的信号进入混频器 3（Minicircuits，TUF-1+）的一端。平衡探测器 BPD2 为完全相同的结构，输出经过滤波电容、变压器、低通滤波器 LPF2 后，与 65 MHz 参考信号进行混频。混频得到的 15 MHz 信号经过双工器和低噪声电放大器后，进入混频器 3 的另一端。混频器将两个 15 MHz 的拍频信号混频再通过低通滤波器 LPF3（Minicircuits，SLP-1.9+）后，经过一个自行搭建的低噪声放大器进行信号放大，就可以得到与定时抖动相关的在基带附近

的噪声信号。同样地，对此信号使用相位噪声分析仪（Keysight，PXA Signal Analyzer N9030A）进行相位噪声功率谱测量，即可得到光纤引入的定时抖动的噪声功率谱。总得来说，电学降频法的基本原理是，由于频率为 80 MHz 的高频率拍频信号的放大很困难，因此先用一个 65 MHz 的参考信号与之混频，得到一个低频率的拍频信号，然后对该低频的拍频信号进行放大后进入混频器。这样有效的克服了混频器 3 的效率下降的问题，实现更高精度的定时抖动测量。

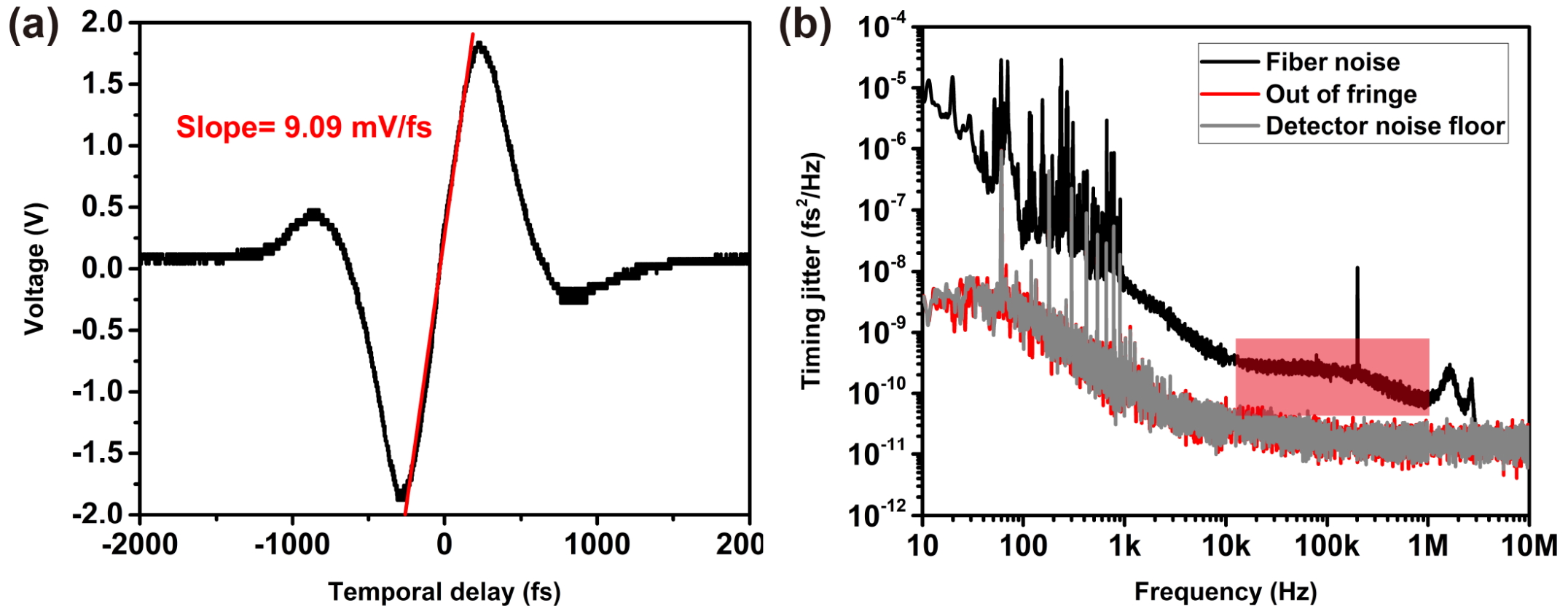

图 3-16　基于电学降频的光学外差法光纤定时抖动测量结果。（a）误差信号 (b)定时抖动功率谱密度

同样地，让电动位移平台在 50 μm 范围内进行循环扫描。用示波器探测混频器输出的误差信号，在示波器上记录 S 型的误差信号曲线。如图 3-16（a）所示。误差信号的鉴相斜率为 9.09 mV/fs。与未改进的光学外差探测法相同，测量的系统的噪声基底、远离干涉点的噪声和光纤的噪声。使用误差信号的鉴相斜率将测量到的电压功率谱转化为定时抖动的功率谱，结果请见图 3-16（b）。可以看出，光纤的定时抖动功率谱在 10 Hz 至 10 kHz 频率范围内，一直以 10 dB/decade 斜率下降。通过对比图 3-14（b）与 3-16（b）可得，该频率范围内，定时抖动测量结果与未改进的光学外差探测法的测量结果一致。然而，高于 10 kHz 频率部分的功率谱存在一个有一定程度隆起的噪声形状，见图 3-16（b）中红色部分。该部分噪声并非光纤引入的定时抖动。本工作推断该噪声是由信号发生器产生的 65 MHz 参考信号的噪声通过系统耦合进入定时抖动引起的。如果使用更低噪声的信号发生器作为参考信号的信号源，该情况可能会得到改善。

### 3.4.5 基于互相关算法的光学外差探测法

为了提高定时抖动测量的精度，本文再次对光学外差探测法进行了改进。改进方法是，采用两套结构相同但是独立的外差探测系统进行探测，将两个系统得

到的功率谱使用互相关算法进行处理，这样可以部分消除两套系统中非相关的噪声成分，例如光电探测器的热噪声、光电探测器的非线性以及散粒噪声。达到提高测量分辨率的目的。改进的光学外差探测法系统的实验装置见图 3-17。

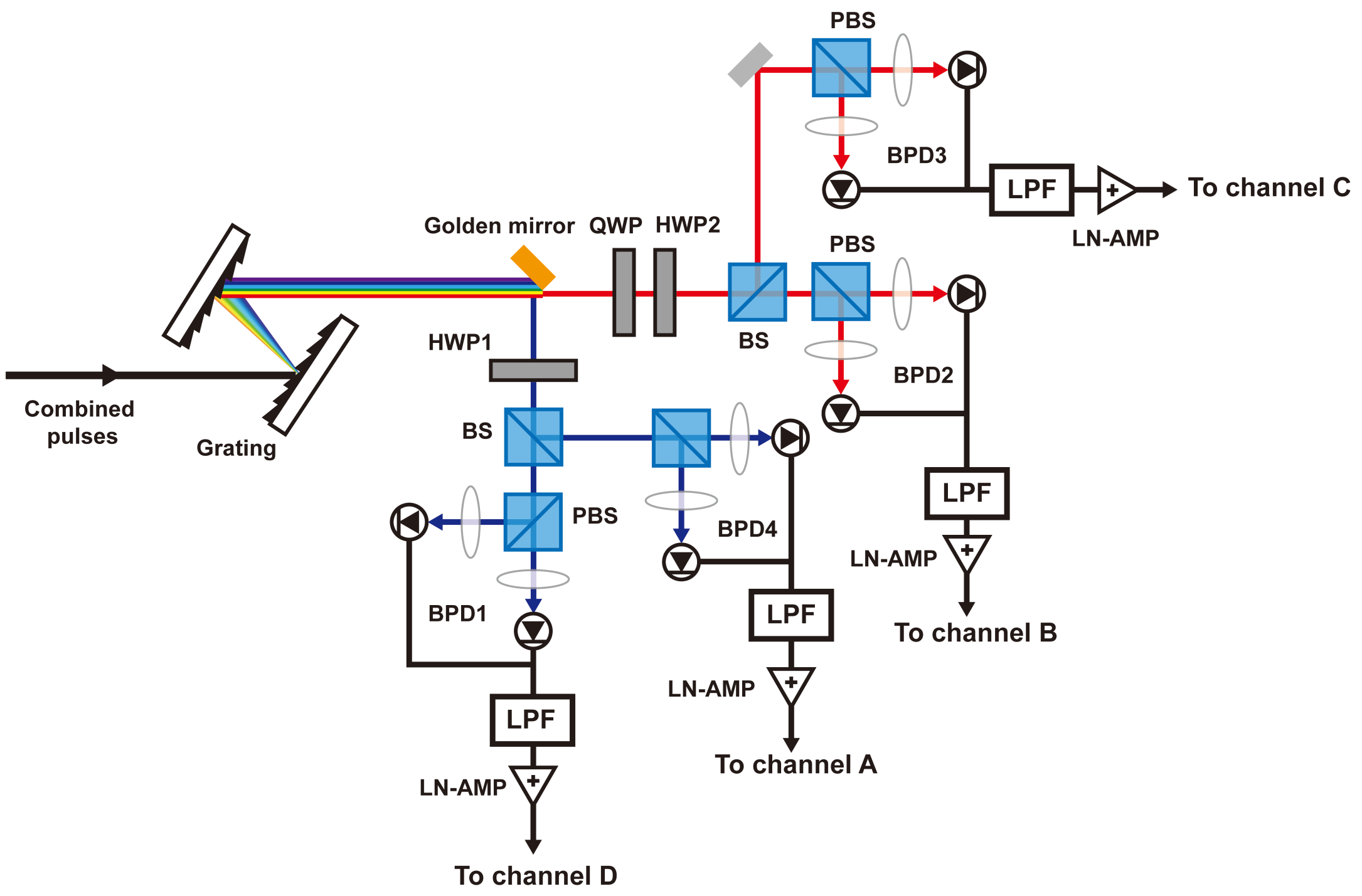

图 3-17　改进的光学外差探测系统示意图。BS，50:50 分束器

可以看出，相比于 3.4.3 节中介绍的方法，改进的光学外差探测法在两个半波片后方分别增加了两个 50:50 分束器，分别将一半的光功率入射进入平衡探测器 BPD3 和平衡探测器 BPD4 中。在该实验中，4 个平衡探测器中的光电二极管由低噪声的铟镓砷光电探测器组成（Hamamatsu，G11193-02R）。在 4 个平衡探测器之后，使用 4 个低通滤波器（Minicircuits，SLP-100+）分别滤除重复频率信号。最后，使用一个 4 通道的相位噪声分析仪进行相位噪声的测量和数据的处理。由于该相位噪声分析仪每个通道对于输入信号的功率要求在 0 dBm 以上，需要使用 4 个低噪声电放大器（Minicircuits，PHA-13LN+）将误差信号从-15 dBm 左右放大至+5 dBm 左右。

该实验中使用的相位噪声分析仪为日本国家计量科学研究院（NMIJ/AIST）自行开发的相噪分析系统。该相位噪声分析仪包含 4 个基于 FPGA 的高速数据采集通道。每个通道可以对待测信号进行时域采集。采集后，通过自行开发的特殊算法，对周期信号的上升沿的过零点进行高精度的鉴别。通过对信号过零点的处理，可以得到待测信号的频率噪声，经过转换可以得到待测信号的相位噪声功率

谱密度。该相位噪声分析仪有着高精度、高集成化、鉴相范围宽等特点。

在该实验中，光纤的定时抖动信息均被调制到了 80 MHz 的电学载波上。并且，1550 nm 波段电磁波 1 rad 的相位改变量对应于 80 MHz 拍频信号的 1 rad 相位改变量。使用相位噪声分析仪的 4 个通道分别对 4 个平衡探测器输出的 4 个 80 MHz 拍频信号进行采样。将通道 A 与通道 B 的采集信号在电脑中进行相位鉴别后，即可得到与光纤定时抖动相关的时域信号。同样地，采用相同手段可以从通道 C 与通道 D 得到另一组定时抖动相关的时域信号。将两组时域信号使用互相关算法处理后，进行傅里叶变换，即可得到光纤的定时抖动功率谱密度如下图。

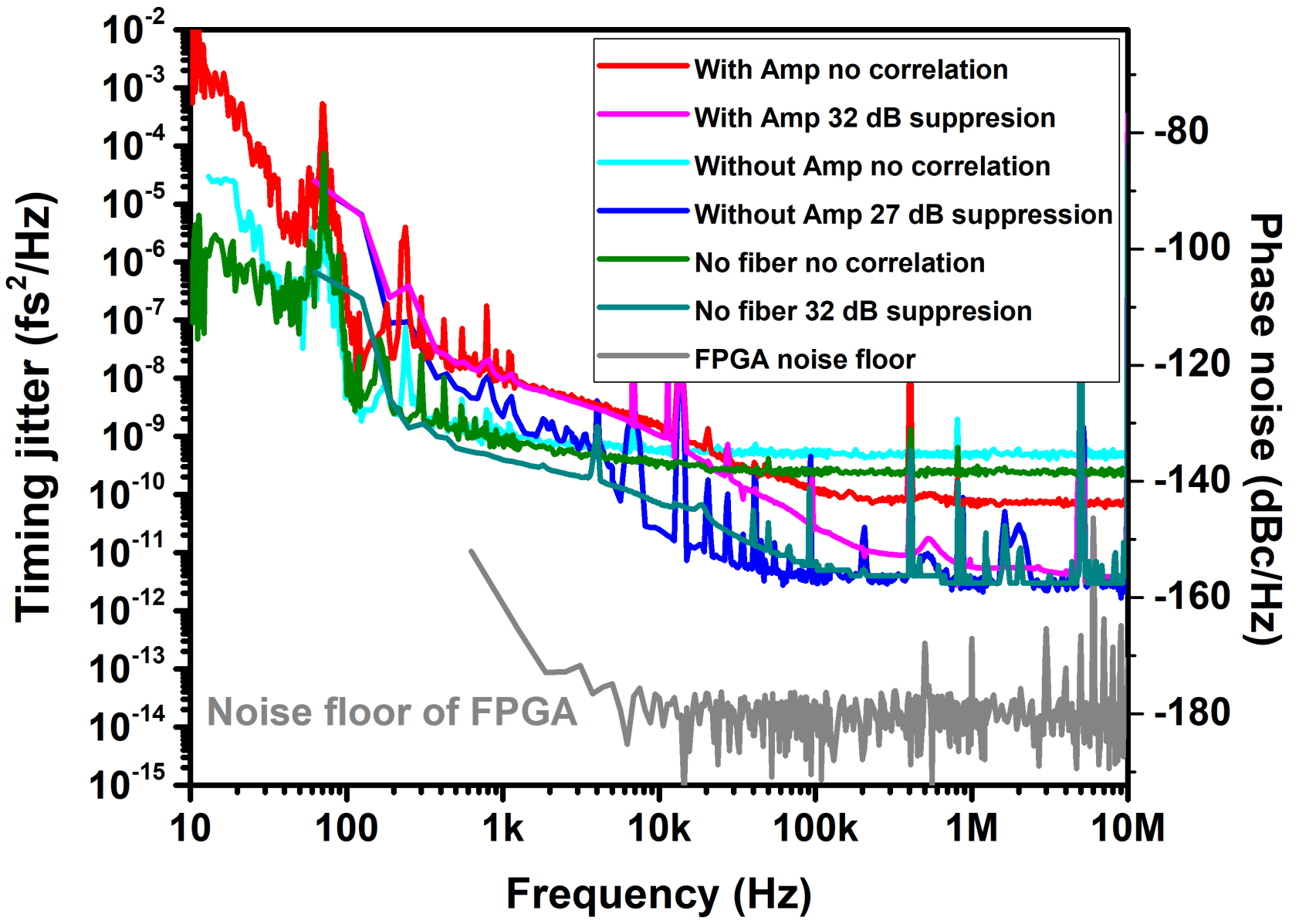

图 3-18　改进的光学外差探测法测量的定时抖动功率谱密度图

图 3-18 中，红色曲线为未经过互相关算法处理的光纤定时抖动功率谱密度图。可以看出，低于 1 kHz 偏频部分，定时抖动功率谱与上一节中的测量结果基本吻合。对于高于 100 kHz 的偏频，测量开始受限于系统相噪分析仪中模拟-数字转换电路的噪声基底。图中粉色曲线为经过 32 dB 抑制的互相关算法后光纤定时抖动功率谱密度图。相比于未经过互相关算法的功率谱，从 10 kHz 开始，粉色曲线开始渐渐低于红色曲线。互相关算法成功消除掉了两套光学外差探测系统中的非相关噪声，如 8 个光电探测器的热噪声、8 个光电探测器的散粒噪声和 4 个模拟-数字转换电路的噪声。对于紫色曲线，在 1 kHz 至 100 kHz 频率范围内，功率谱并不是以 10 dB/decade 斜率下降的直线，而是有一个凸起。对此，本文做了一组对比实验。在去除了 4 个低噪声电放大器后，对定时抖动功率谱再次进行了测量。未经互相关算法处理的功率谱图为图中浅蓝色曲线，经过 32 dB 抑制的互相关算法后的功率谱为图中深蓝色曲线。由于去除了电放大，噪声信号的信

噪比大大降低，浅蓝色曲线在 1 kHz 偏频处便开始受限于模拟-数字转换器的噪声基底。并且，在去除了 4 个电放大器后，凸起很大程度地消失了。因此猜想该功率谱的畸变是由于放大器中的非线性引入的。

需要注意的是，理论上来讲，进行 32 dB 抑制的互相关算法处理后，高频部分的噪声基底应当下降 32 dB。该实验中，不管是紫色曲线还是深蓝色曲线，均只下降至 $4\times10^{-14}$ $fs^2/Hz$ 量级就不再下降。图中深灰色曲线为经过 32 dB 抑制的互相关算法后，模拟-数字转换器的噪声基底。深灰色曲线远远低于 $4\times10^{-14}$ $fs^2/Hz$ 量级。4 个模拟-数字转换器是独立的，它们的噪声基底是非相关的，可以被互相关算法消除。因此该量级的噪声源于 4 个拍频信号采样过程中无法被互相关算法消除的相关噪声。为了探究该噪声的来源，本文进行了另一个对比实验。移除了一根 5 m 长的单模光纤，仅测量一个简单干涉仪引入的定时抖动。实现装置图见图 3-12（b）。未经过互相关算法的定时抖动功率谱见图 3-18 中浅蓝色曲线，经过 27 dB 抑制的互相关算法后的功率谱为图中深蓝色曲线。$4\times10^{-14}$ $fs^2/Hz$ 量级的某些非相关噪声仍旧存在。该噪声并不是来源于待测量的光纤。对此，猜想该噪声的来源为驱动声光调制器的 80 MHz 的信号源的噪声。由于每个 80 MHz 拍频信号都来源于声光调制器零级衍射和 1 级衍射光的拍频，该信号源驱动声光调制器会对 4 个拍频信号同时直接产生作用。因此，该噪声对于 4 个拍频信号是相关的，无法用互相关算法消除。这也是使用光学外差探测技术进行光纤的定时抖动测量的一个难以克服的技术瓶颈。如果使用该技术进行两台独立的飞秒激光器的定时抖动进行测量，便不会受到该技术问题的困扰。

## 3.5 本章小结

本章工作主要分为两部分，第一部分是在优化了激光器量子噪声引入的定时抖动特性后，使用光学平衡互相关技术首次实现了两台独立的掺镱光纤激光器的连续 5 天（120 小时）的阿秒量级的精密时间同步。在优化激光器自身噪声特性的同时，高速的电学锁相装置也是维持长期时间同步的重要因素。实现时间同步后，环内光学互相关系统输出的剩余时间误差信号的均方根值为 103 as，环外光学互相关系统输出的剩余时间误差信号的均方根值为 733 as。经分析，环外定时抖动结果的恶化主要是由于温度和湿度变化导致的光路失谐，如晶体支架的角度的微量改变，进而导致和频光效率的周期性改变，环外时间误差信号也呈现周期性变化的趋势。本工作首次以平均约 $10^5$ s 的时间评估了锁模激光器之间的时间同步的长期稳定性。环内交叠艾伦方差的 $1.31\times10^5$ s 稳定度为 $8.76\times10^{-22}$。环外

交叠艾伦方差的 $1.31\times10^5$ s 稳定度为 $1.36\times10^{-20}$。从方差分析可得知，环内的剩余定时抖动主要为白相位噪声，环外的剩余定时抖动受到环境因素的影响，是无规律的长期漂移。环内噪声功率谱在 10 kHz 至 10 MHz 频率范围的积分值为 430 as。这一结果是迈向新兴的亚周期光脉冲合成、光学频率基准的传输和基于 X 射线自由电子激光器的高时间分辨技术的关键一步。

第二部分内容为使用光学外差法对飞秒脉冲在光纤中传播过程中，光纤引入的额外定时抖动的测量。使用重复频率为 500 MHz 的 1550 nm 波段的低噪声固体激光器作为光源，放大后的脉冲经过两段 5 米长的单模光纤，用光学外差探测技术对单模光纤引入的定时抖动进行了测量。得到了积分值为 64.7 as 的定时抖动结果。分别使用电学降频法和互相关算法对光学外差探测法进行改进，改进后定时抖动的测量仍受限于系统中仪器的噪声。对光纤引入的定时抖动的高精度测量是实现阿秒量级时间、光学频率基准远距离传出的重要理论和实验基础。

# 第4章 飞秒激光光学频率梳的载波-包络相位噪声测量及分析

本章将介绍飞秒激光光学频率梳中的载波-包络相位噪声的高精度测量及分析。本工作提出了一种无需 *f*-2*f* 干涉仪的载波-包络相位噪声功率谱密度测量的方法。该方法可以实现掺镱光纤飞秒激光器中，傅立叶频率范围在 5 mHz 至 8 MHz 内的，动态范围高于 270 dB 的载波-包络相位噪声功率谱测量。相位噪声测量基底低于 1μrad/√Hz。这是在报道过的文献中最高灵敏度的载波-包络相位噪声频谱测量。然后，本工作使用 Kendall 互相关分析对不同飞秒激光器的载波-包络相位噪声进行了分析。最后，本工作对孤子分子对中的两个光学孤子的相对相位噪声进行了测量，相对相位噪声的积分值为 3.5 mrad（从 10 Hz 至 10 MHz 积分）。相位噪声测量的精度为 $10^{-13}$ $rad^2/Hz$。使用 $\beta$-line 分析法，估算两个光学孤子的相对线宽仅为 μHz 量级。

## 4.1 载波-包络相位噪声的基本理论

在飞秒激光器中，由于腔色散的作用，输出脉冲的相速度与群速度失配会导致脉冲载波峰值与包络峰值之间的周期性不重合。载波相位与包络相位之差定义为载波-包络相位 $\varphi_{CE}$。包含该相位的脉冲电场的表达式为：

$$E(t) = A(t)\cos(\omega_C t + \varphi_{CE}) \tag{4-1}$$

飞秒激光器的载波-包络相位噪声的主要来源有环境噪声、泵浦波动引入的噪声以及放大自发辐射引入的量子噪声。载波-包络相位噪声来源在第二章中已经详细讨论。

## 4.2 大范围高精度载波-包络相位噪声探测

### 4.2.1 实验原理

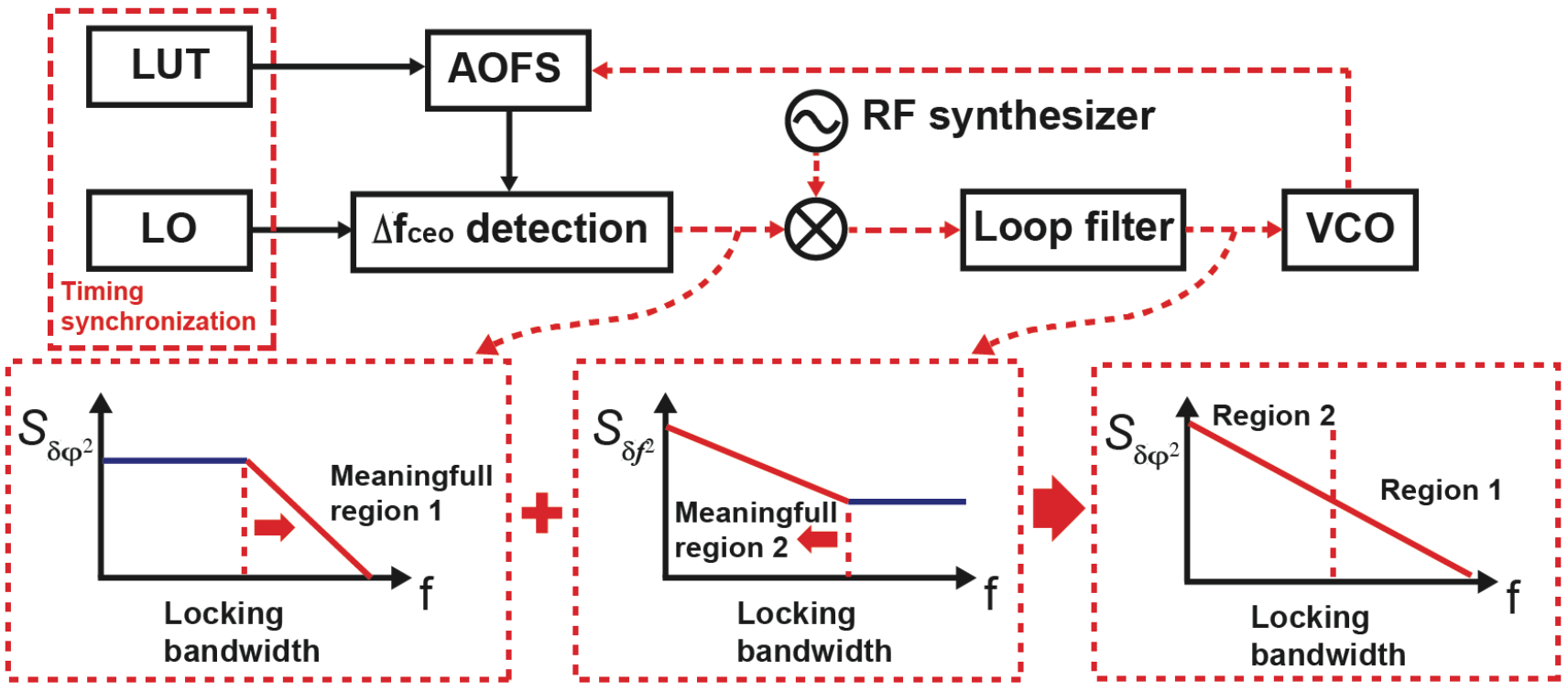

图 4-1　高精度载波-包络相位噪声探测原理。LUT，待测激光器；LO，参考激光器；AOFS，声光频移器；Loop filter，比例积分伺服控制器；VCO，压控振荡器

本工作的提出的高精度载波-包络相位噪声的测量原理见图 4-1。该相位噪声测量借鉴了文献[26]中的相位噪声测量方法，是一种被广泛适用于测量射频振荡器相位噪声的方法。该实验中，测量对象为两台独立运转的掺镱光纤飞秒激光器的载波-包络相位噪声。其中，一台激光器作为参考激光器，另一台激光器作为待测激光器。使用光学平衡互相关技术将两台激光器的重复频率进行高带宽的锁定后，通过将两个光学频率梳齿在频域上做拍，就可以得到在射频波段的两台激光器相对载波-包络偏移频率信号 $\Delta f_{ceo}$。由于两台激光器各自运转，互不相关，这个相对载波-包络偏移频率信号为两台激光器的载波-包络偏移频率信号的噪声之和。对该信号的噪声进行测量后，将测量得到的功率谱密度除以 2，即可得到单台激光器的载波-包络相位噪声。2010 年，Jung 等人通过分析测量锁相环输出的纠正信号的噪声，实现了宽范围的自由运转激光器的定时抖动功率谱测量。本工作将该测量方法应用到激光器载波-包络相位噪声测量中。使用一个声光频移器将 $\Delta f_{ceo}$ 锁定至一个频率为 3 MHz 的射频基准上。在实现了该信号的相位锁定后，提取加载到声光频移器的控制器（压控振荡器）上的电压信号，使用控制器的电压-频率转换系数，将该电压信号的功率谱转化为 $\Delta f_{ceo}$ 的频率噪声的功率谱。通过公式

$$S_{\varphi}(f)=S_{v}(f)/f^{2} \tag{4-2}$$

将频率噪声谱 $S_\nu(f)$ 转化为相位噪声谱 $S_\varphi(f)$。从而实现锁相带宽内，自由运转的两台激光器 $\Delta f_{ceo}$ 噪声功率谱的测量。另一方面，通过测量相位锁定后的误差信号，可以得到锁相带宽之外的 $\Delta f_{ceo}$ 的噪声功率谱。将两个实验得到的噪声功率谱进行连接，便可以实现宽傅里叶频率范围内的、激光器相对载波-包络相位噪声分析。

### 4.2.2 实验装置

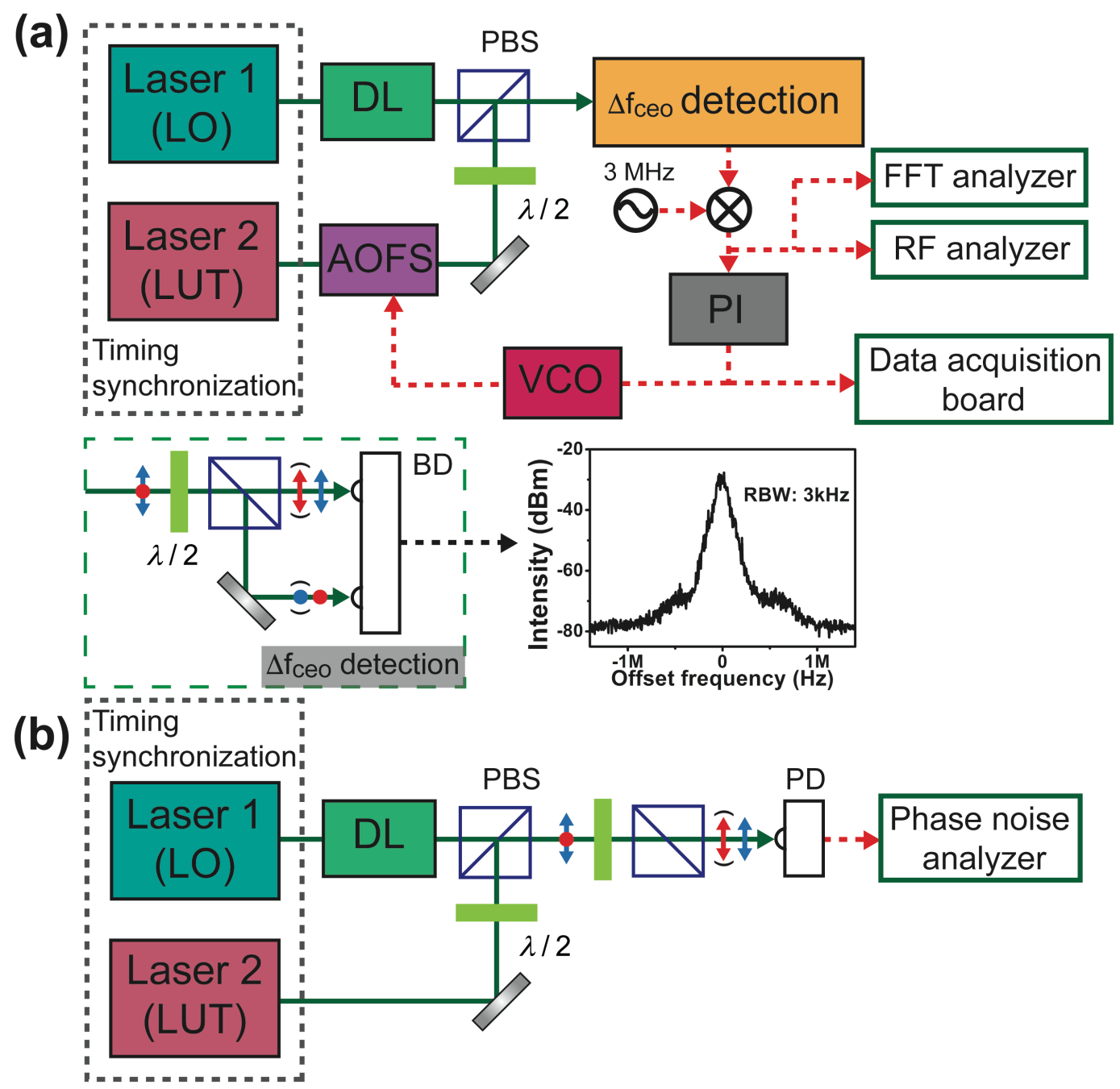

图 4-2 光学外差探测法测量载波-包络相位噪声的实验装置图。(a) 光学外差探测法。(b) 相位分析仪探测法。AOFS，声光频移器；BD，平衡光电探测器；DL，延迟线；PBS，偏振分束器；PD，光电探测器；PI，比例积分伺服控制器；VCO，压控振荡器

本工作提出的高精度载波-包络相位噪声测量的实验装置见图 4-2。实验中使用的激光器为两台相似的掺镱光纤飞秒激光器。两激光器工作在呼吸孤子域。两台激光器平均功率分别为 85 mW 和 65 mW，对应单脉冲能量分别为 0.54 nJ 和 0.43 nJ。脉冲宽度分别为 77 fs 和 69 fs。通过光学平衡互相关技术探测两台激光器的重复频率差，一台激光器腔内的压电陶瓷调节激光器的腔长，使用比例积分伺服控制器实现两台激光器的时间同步，重复频率锁定带宽为 10 kHz，剩余定时抖动均方根值为 380 as（从 200 Hz 积分至 10 MHz）。关于时间同步的具体细节在第三章中已经做了详细介绍。

在实现了高精度、长时间的时间同步后，通过半波片将两台激光器输出脉冲

的偏振态调节至互相垂直，随后用一个偏振合束器进行合束。在偏振分束器合束之前，有一个空间延迟光路进行两脉冲的时域调节，以确保两台激光器重复频率相等的前提下，脉冲在时域上重合。这样，在保证了两脉冲在各自相干长度内的前提下，才能在频谱分析仪上探测到 $\Delta f_{ceo}$ 的拍频信号。两脉冲的偏振态被一个1/2 波片旋转 45 度后，再次经过一个偏振分束器分束。来源于两台激光器的两脉冲分别被投影到水平偏振和垂直偏振方向上，产生 $\Delta f_{ceo}$ 的拍频信号。透射端和反射端分别被平衡探测器（Thorlabs，PDB420A）的两个光电探测器探测。使用平衡探测的目的是为了消除激光器强度抖动的影响。图 4-2（a）中的插图为 $\Delta f_{ceo}$ 信号，该信号在 3 kHz 分辨率带宽下的信噪比为 50 dB。该信号的 3 dB 带宽为 68 kHz。由于两台激光器的腔色散均为近零色散，这个带宽已经是光纤激光器能都达到的较为低噪声的载波-包络偏移频率信号。对于固体激光器，例如钛宝石飞秒激光器，$f_{ceo}$ 信号的带宽会更窄。特别地，在文献[72]中，经过特别仔细的腔设计和腔内色散优化后，自由运转的 $f_{ceo}$ 信号的带宽可以降低至 10 Hz（100 kHz 测量分辨率带宽下）。接下来，该工作对 $\Delta f_{ceo}$ 信号进行锁定和噪声测量。通常的，会将待测信号锁定至零频率，然后测量基带附近的剩余噪声。但是这种方法会不可避免地受到光电探测器闪烁噪声的影响。因此，本工作选择了将 $\Delta f_{ceo}$ 信号锁定至 3 MHz 的频率基准。频率基准由信号发生器（Stanford Research Systems，SG382）提供。随后，使用一个声光频移器（Gooch & Housego，33080-16-.7-I-TB）对待测激光器的输出光进行频移。该声光频移器的驱动器为一个中心频率为 80 MHz 的压控振荡器。压控振荡器的调节范围为 20 MHz。声光频移器将 $\Delta f_{ceo}$ 信号锁定至 3 MHz，锁定带宽为 25 kHz。需要说明的是，在该实验中，特意地降低了 $\Delta f_{ceo}$ 的锁定带宽，以避免高带宽锁定过程中，声光调制器可能对激光频率产生过量调制，导致剩余噪声功率谱在锁定带宽附近会出现一个由锁相环引入的鼓包，对功率谱的测量产生干扰。另外，相比于泵浦调制锁定载波-包络偏移频率，使用声光频移器的优势在于重复频率的锁相环路不会被干扰。

该实验系统搭建过程中，应注意以下几点：

一、在对 $\Delta f_{ceo}$ 信号进行平衡探测时，首先调节偏振合束器前的半波片的角度，使得 $\Delta f_{ceo}$ 信号的信噪比最高。随后可以在平衡探测器中的一个光电探测器前放置一个可调中性密度衰减片，调节衰减片角度，使两个光电探测器输出的直流电压相等，平衡探测器的输出在零电压附近。

二、相对载波-包络频率偏移信号锁定的调节方法：该实验中使用的比例积分伺服系统为 Newfocus 公司的 LB1005。首先，在比例积分伺服系统中，将锁相环的比例增益调节至零，调节伺服控制器的输出偏置，使得 $\Delta f_{ceo}$ 信号的频率在 3 MHz 附近。将伺服控制器的积分拐点设置在 30 kHz 或者 100 kHz 档位，伺服控

制器的开关拨至 LFGL 档位，随后逐渐增加锁相环的比例增益。在此过程中，不断地调节伺服控制器的输出偏置电压，使得 $\Delta f_{ceo}$ 信号一直在 3 MHz 附近。当增益增加到一定程度时，可以发现 $\Delta f_{ceo}$ 信号在主动地向 3 MHz 靠拢，并且 3 MHz 频率附近会产生若干对称的“相干峰”。这时将伺服控制器的开关推至“Lock on”档位，稍微降低比例增益直至“相干峰”消失，即可实现锁相环的锁定。如果无法实现锁相，可以尝试将输入信号从 A 通道换至-B 通道，将输入信号反相，再尝试锁定。

## 4.2.3 功率谱分析

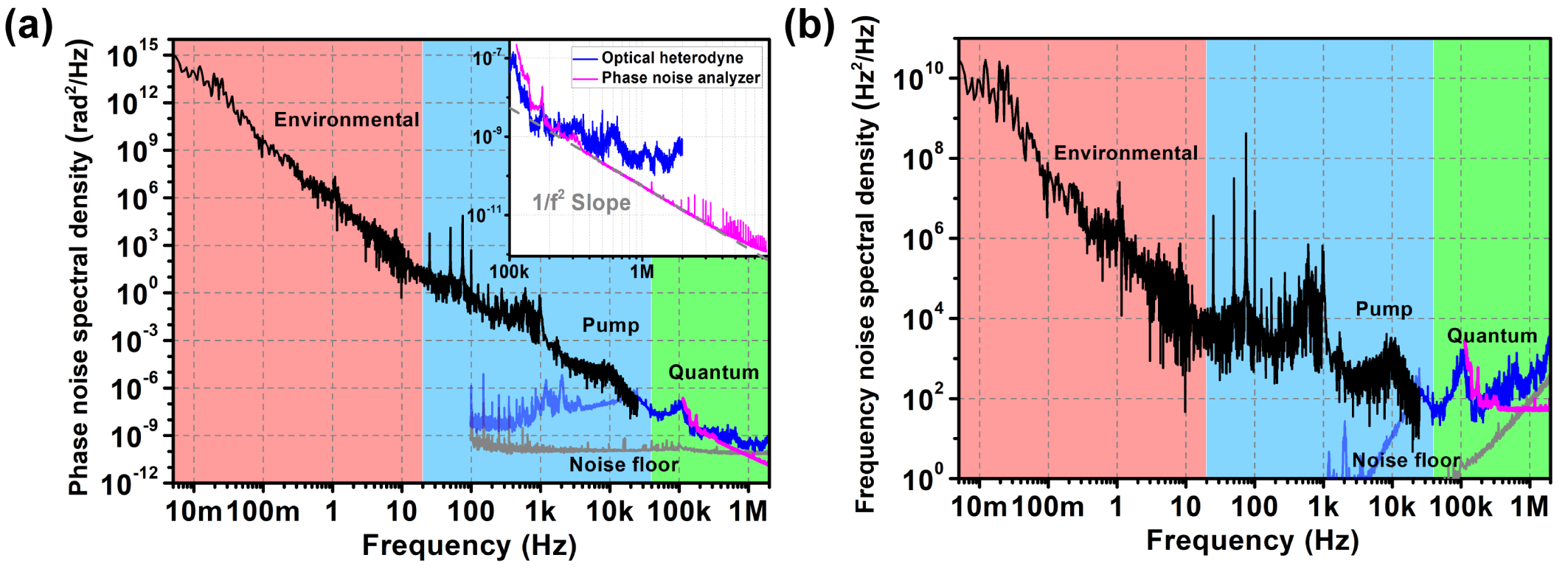

图 4-3 载波-包络相位噪声测量结果。（a）相位噪声功率谱。（b）频率噪声功率谱。黑色曲线为从压控振荡器的输入端提取的噪声功率谱。蓝色曲线为平衡探测器输出的剩余误差信号功率谱。灰色曲线为平衡探测器的热噪声基底。紫色曲线为相噪分析仪测量的噪声功率谱

在实现了 $\Delta f_{ceo}$ 的锁定后，对平衡探测器输出的误差信号和压控振荡器的输入信号进行了噪声功率谱测量。使用频谱分析仪（Agilent，8560EC）和快速傅里叶变换分析仪（Standford Research Systems，SR770）分别对平衡探测器输出的误差信号的 100 kHz 至 2 MHz 频率范围和 100 Hz 至 100 kHz 频率范围的功率谱进行了测量，得到误差电压的功率谱密度图。为了将电压功率谱转化成相位噪声的功率谱，需要用示波器记录未锁定的 $\Delta f_{ceo}$ 信号，利用该正弦信号的线性部分作为相位鉴别，实现电压功率谱到相位噪声功率谱的转换。该实验中，鉴相斜率为 5.43 mV/rad。得到的数据为图 4-3（a）和（b）中的蓝色曲线。其中，25 kHz 以下部分为锁定带宽内的剩余噪声，25 kHz 以上的部分为锁定带宽外的剩余噪声。另一方面，使用一个数据采集卡（National Instruments，PXIe-5122）对输入压控振荡器的信号进行数据采集。采集得到的数据经过傅里叶变化后，得到傅里

叶频率在 5 mHz 至 25 kHz 范围的功率谱。同样地，使用压控振荡器的增益 $K$，实现电压功率谱到频率噪声功率谱的转换。在该实验中，$K = 5.91$ MHz/V。使用公式（4-2）将频率噪声功率谱转化为相位噪声功率谱，与平衡探测器输出的功率谱一同绘制在图 4-3 中，为图中黑色曲线。最后，将黑色曲线与蓝色曲线连接起来，就可以得到自由运转的 5 mHz 至 2 MHz 范围内的 $\Delta f_{ceo}$ 相位噪声功率谱和频率噪声功率谱。考虑到两台激光器的结构以及输出脉冲的参数基本相似，两台激光器独立运转，两激光器的 $f_{ceo}$ 相位噪声功率谱是非相关的，将测量到的功率谱除以 2，便可得到单台激光器的 $f_{ceo}$ 相位噪声的功率谱。另一方面，也可以使用一台载波-包络相位稳定的商业激光器系统作为参考源，这样的话，测量到的噪声结果就主要为待测激光器的噪声。图 4-3 中，灰色曲线为平衡探测器的热噪声基底。在 100 kHz 偏频处，信号发生器产生的 3 MHz 射频信号的噪声基底大约在-140 dBc/Hz 量级。压控振荡器的噪声基底在 1 MHz 偏频为-130 dBc/Hz。这两个噪声基底均低于待测 $f_{ceo}$ 信号的噪声（-100 dBc/Hz @ 1MHz）,因此，$f_{ceo}$ 信号的噪声测量不会受到这两台仪器的噪声基底的影响。

在高于 1 MHz 频率范围，测量受限于平衡探测器的热噪声基底。为了实现更高精度的载波-包络相位噪声测量，本工作使用相位噪声分析仪直接对自由运转的 $\Delta f_{ceo}$ 的相位噪声进行了测量，实验装置见图 4-2（b）。该实验中，使用热噪声基底更低的高速 PIN 探测器（Menlo Systems，FPD510）和相位噪声分析仪（Rohde & Schwarz，R&S FSWP26）对 $\Delta f_{ceo}$ 的高频部分进行相位噪声的测量。100 kHz 至 8 MHz 范围内的相位噪声功率谱为图 4-3（a）中紫色曲线。由于自由运转的 $\Delta f_{ceo}$ 漂移范围非常大，在若干 MHz 量级，已经超出了该相位噪声分析仪的测量能力，因此该相位噪声分析仪无法对低频部分的功率谱进行测量。

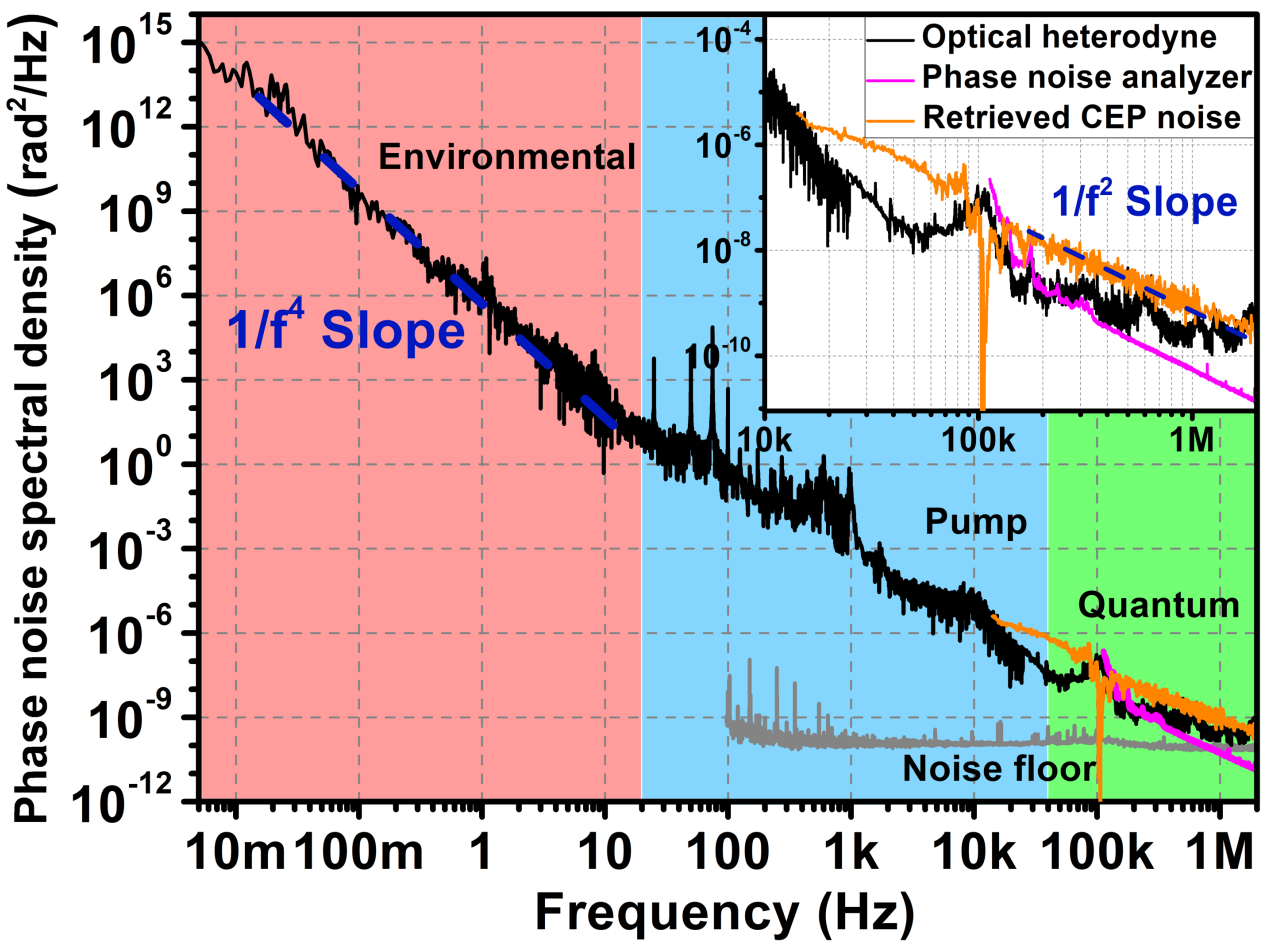

图 4-4　载波-包络相位噪声功率谱密度。橙色曲线为使用公式计算得到的功率谱

由于重复频率的锁定带宽为 10 kHz，载波-包络相位噪声谱高于 10 kHz 的部分会受到自由运转的重复频率噪声的影响。也就是说，高于 10 kHz 部分的噪声谱实际上为光学频率梳梳齿的噪声，而不是简单的载波-包络相位噪声。接下来对测量到的载波-包络相位噪声谱的高频部分进行校对。根据文献[219]中的理论，飞秒激光器中的重复频率噪声和载波-包络相位噪声呈负相关的关系，其噪声谱的关系式如下：

$$S_{\Delta\nu_n\Delta\nu_n} = S_{\Delta f_{ceo}\Delta f_{ceo}} + S_{N\cdot\Delta f_{rep}N\cdot\Delta f_{rep}} + \Gamma_\Delta(\omega)\times\sqrt{S_{\Delta f_{ceo}\Delta f_{ceo}}\cdot S_{N\cdot\Delta f_{rep}N\cdot\Delta f_{rep}}} \qquad (4\text{-}3)$$

其中 $S_{\Delta\nu_n\Delta\nu_n}$、$S_{N\cdot\Delta f_{rep}N\cdot\Delta f_{rep}}$ 和 $S_{\Delta f_{ceo}\Delta f_{ceo}}$ 分别为梳齿频率、重复频率和载波-包络偏移频率的相位噪声，$\Gamma_\Delta(\omega)$ 为相关系数。使用第三章中测量得到的激光器定时抖动的数据结果，通过该公式可以计算得到高于 10 kHz 部分的 $f_{ceo}$ 的相位噪声，如图 4-4 中的橙色曲线所示。

接下来对载波-包络相位噪声的噪声特性进行分析。从图 4-4 中，可以看出，在 2 mHz 至 20 Hz 范围内，载波-包络相位的抖动主要是由环境因素引起的。在相位噪声谱中，该频率范围功率谱的斜率为 $1/f^4$。这表明了载波-包络相位在低频范围内为频率漂移特性。在 20 Hz 至 35 kHz 范围内，载波-包络相位的抖动主要是由激光器泵浦波动引入的振幅调制-相位调制噪声。在 35 kHz 至 8 MHz 范围，$f_{ceo}$ 相位噪声的斜率为 20 dB/decade，主要是由放大自发辐射引入的量子噪声。这一现象与文献[100]中的理论模型相符。该实验中的相位噪声探测灵敏度为 $10^{-10}$ rad$^2$/Hz。

## 4.2.4 方差分析

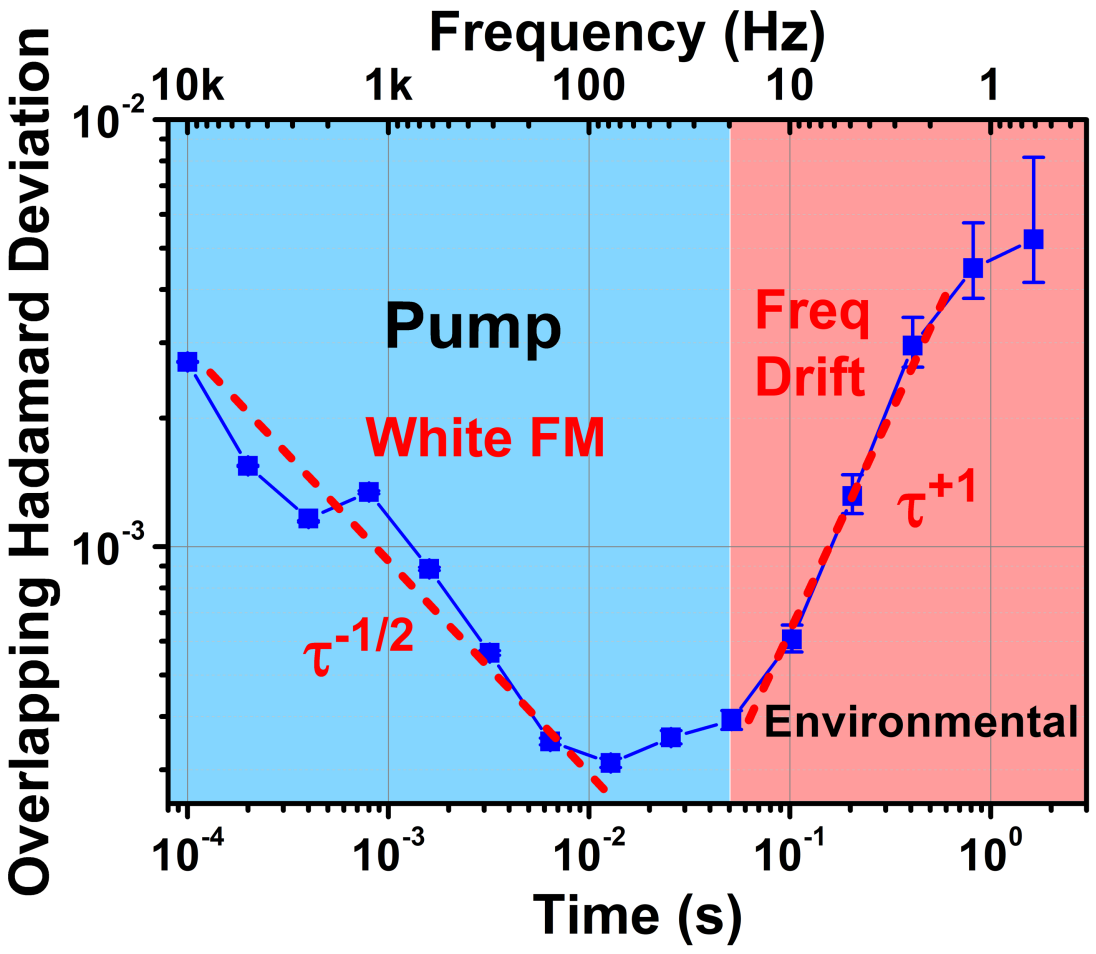

图 4-5　载波-包络偏移频率的交叠 Hadamard 方差分析

本小节对自由运转的 $\Delta f_{ceo}$ 噪声进行方差分析。这里，本文使用了 Hadamard 方差作为 $\Delta f_{ceo}$ 频率噪声的分析工具。Hadamard 方差作为艾伦方差的一种改进形式，已经被广泛应用于有长时间漂移的频率源的稳定性分析中。这恰好符合自由运转的 $\Delta f_{ceo}$ 信号的特点。Hadamard 方差的具体理论已经在第二章中做了详细介绍。使用频率计数器（Agilent，53220A）对自由运转的 $\Delta f_{ceo}$ 信号的频率进行测量，设定门时间为 100 ms，采样时间为 10 s。使用公式(2-30)计算交叠的 Hadamard 方差。Hadamard 方差随门时间变化的曲线见图 4-5。可以看出，在门时间小于 10 ms 的范围，Hadamard 方差随门时间的增长以 $\tau^{1/2}$ 的斜率下降。这表明在高于 100 Hz 频率范围内，$\Delta f_{ceo}$ 信号的噪声特性为白频率调制噪声。在门时间大于 50 ms 的范围，Hadamard 方差展现了频率漂移的特性，更准确地说，是频率漂移和随机游走频率调制的混合，这主要是由环境扰动引入的。Hadamard 方差中 50 ms（对应傅里叶频率为 20 Hz）的泵浦噪声与环境噪声的分界线与图 4-3 中功率谱分析中的两噪声分界线相吻合。

## 4.3 载波-包络相位噪声的 Kendall 互相关分析

### 4.3.1 Kendall 互相关的基本原理

Kendall 互相关是对两个变量进行互相关计算，相比于其他互相关计算方法，Kendall 互相关算法有着更高的灵敏度，可以揭示两个变量之间微弱的相关性。将时域的 $\Delta f_{ceo}$ 信号 $u(t)$进行傅里叶变换后，滤出±$\Delta f$ 范围内的信号成分。然后将 $\Delta f_{ceo}$ 信号的振幅项 $a_i(t)$和频率项 $f_i(t)$解调出来。将每 $M$ 个时刻的振幅 $a_i(t)$进行平均后，得到一个长度为 $N$ 的向量 $\vec{a}=\left(a_{\zeta+1},a_{\zeta+2},\ldots,a_{\zeta+N}\right)$。同样地，将每 $M$ 个时刻的频率 $f_i(t)$进行平均后，得到一个长度为 $N$ 的向量 $\vec{f}=\left(f_{\zeta+1},f_{\zeta+2},\ldots,f_{\zeta+N}\right)$。对两向量进行 Kendall 互相关计算：

$$\rho(\vec{a},\vec{f})=\mathrm{corr}\left[\left(a_{\zeta+1},a_{\zeta+2},\ldots,a_{\zeta+N}\right),\left(f_{\zeta+1},f_{\zeta+2},\ldots,f_{\zeta+N}\right)\right] \tag{4-4}$$

该互相关算法可以揭示 $\Delta f_{ceo}$ 信号中振幅和频率的相关性。

### 4.3.2 相关分析结果

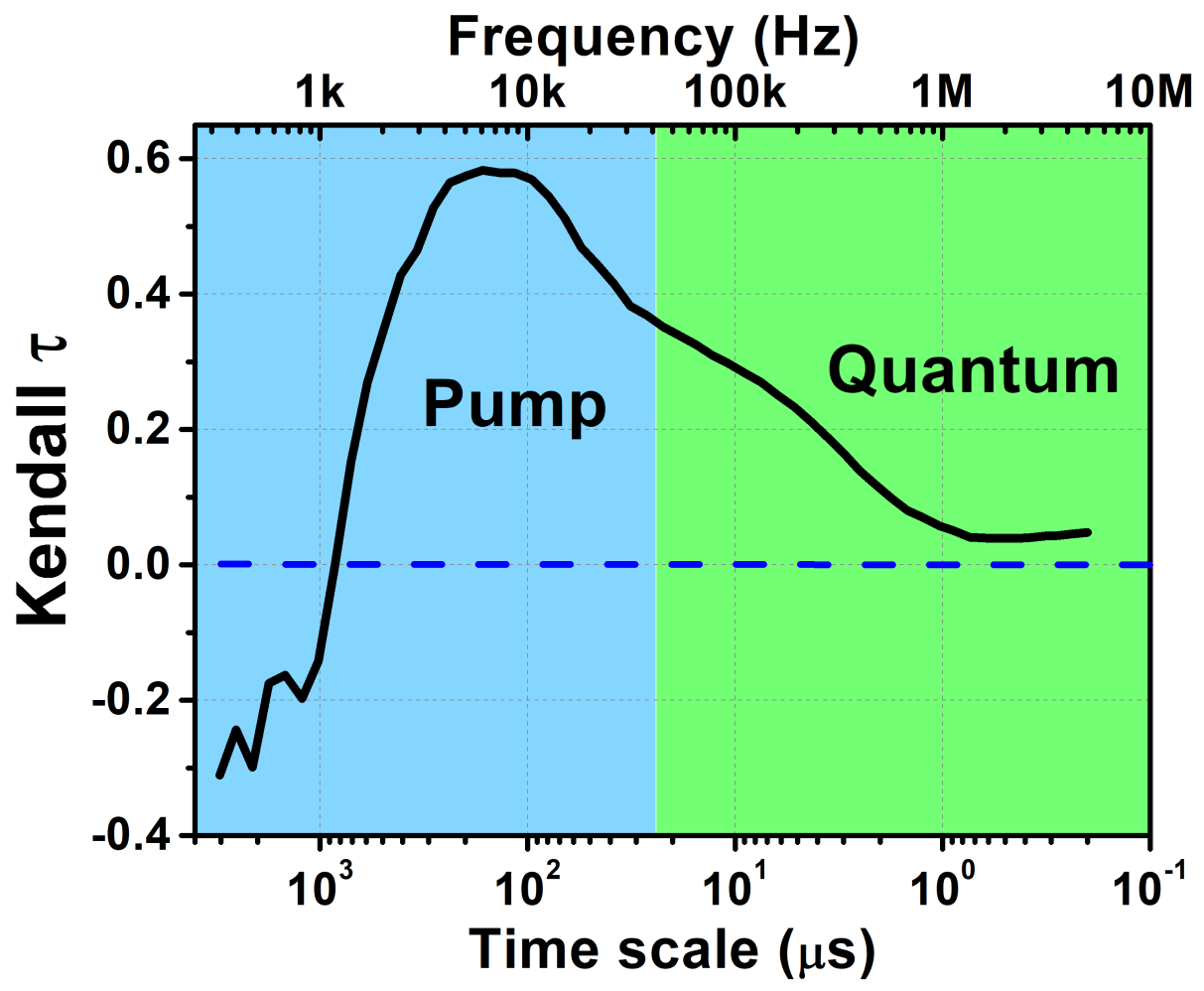

图 4-6　Kendall 相关分析结果

在实验中，本工作首先对掺镱光纤激光器的 $\Delta f_{ceo}$ 信号进行了 Kendall 互相关计算。将 $\Delta f_{ceo}$ 信号的频率调节到 10 MHz，使用数据采集卡对 $\Delta f_{ceo}$ 信号进行数据采集，采样率为 50 MHz，采样时长为 1 s。将 $\Delta f_{ceo}$ 信号的振幅项和频率项解调之后，通过改变两向量的时间延迟和分析的时间窗口，就可以得到掺镱光纤激光器的 $\Delta f_{ceo}$ 信号的 Kendall 互相关系数随分析时间窗口变化的曲线，如图 4-6。可以看出，在大于 0.1 ms 的时间窗口（傅里叶频率小于 10 kHz）时，$\Delta f_{ceo}$ 信号的振幅和频率有着很强的相关性。这是由激光器泵浦波动产生的振幅-频率调制引入的。随着分析时间窗口的减小，振幅和频率的相关强度逐渐下降，这表明，$\Delta f_{ceo}$ 信号的噪声特性由泵浦引入的振幅-相位调制噪声向放大自发辐射引入的量子噪声过渡。这与功率谱分析的结果相吻合。

为了将掺镱光纤激光器的 $\Delta f_{ceo}$ 信号与其他类型的激光器进行比较，本工作对其他 3 台飞秒激光器的 $f_{ceo}$ 信号进行了分析。3 台飞秒激光器分别为：钛宝石飞秒激光器、基于非线性放大环形镜（NALM）锁模的掺铒光纤飞秒激光器和基于可饱和吸收镜（SESAM）锁模的掺铒光纤飞秒激光器。分别使用自行搭建的 $f$-$2f$ 干涉仪探测这 3 台激光器的 $f_{ceo}$ 信号。同样地，将 3 台激光器的 $f_{ceo}$ 信号调节到 10 MHz 频率，用至少 25 MHz 的采样频率进行数据采样，采样时间为 1 s。掺镱光纤激光器和这 3 台飞秒激光器的主要参数见表 4。

表 4 激光器参数对比。KLM，克尔透镜锁模；NPR，非线性偏振旋转锁模；BD，平衡探测法。$\tau_{rad}$，增益介质上能级粒子寿命；$\Delta\nu$，$f_{ceo}$ 信号的 6 dB 带宽。

| 激光器类型 | 钛宝石飞秒激光器 | NPR 掺镱光纤飞秒激光器 | NALM 掺铒飞秒激光器 | SESAM 掺铒飞秒激光器 |
|---|---|---|---|---|
| 重复频率（MHz） | 84 | 157 | 250 | 80 |
| 输出功率（mW） | 520 | 70 | 3 | 1 |
| 脉冲宽度（fs） | 10 | 70 | 72 | 500 |
| $\tau_{rad}$（ms） | 0.004 | 1-2 | 8-10 | 8-10 |
| $f_{ceo}$ 信号探测方法 | $f$-$2f$ 干涉仪 | BD | $f$-$2f$ 干涉仪 | $f$-$2f$ 干涉仪 |
| 锁模机制 | KLM | NPR | NALM | SESAM |
| $\Delta\nu$（kHz） | 50 | 300 | 50 | 1200 |
| $\Delta\nu\times\tau_{rad}$ | 0.2 | 450 | 450 | 11000 |

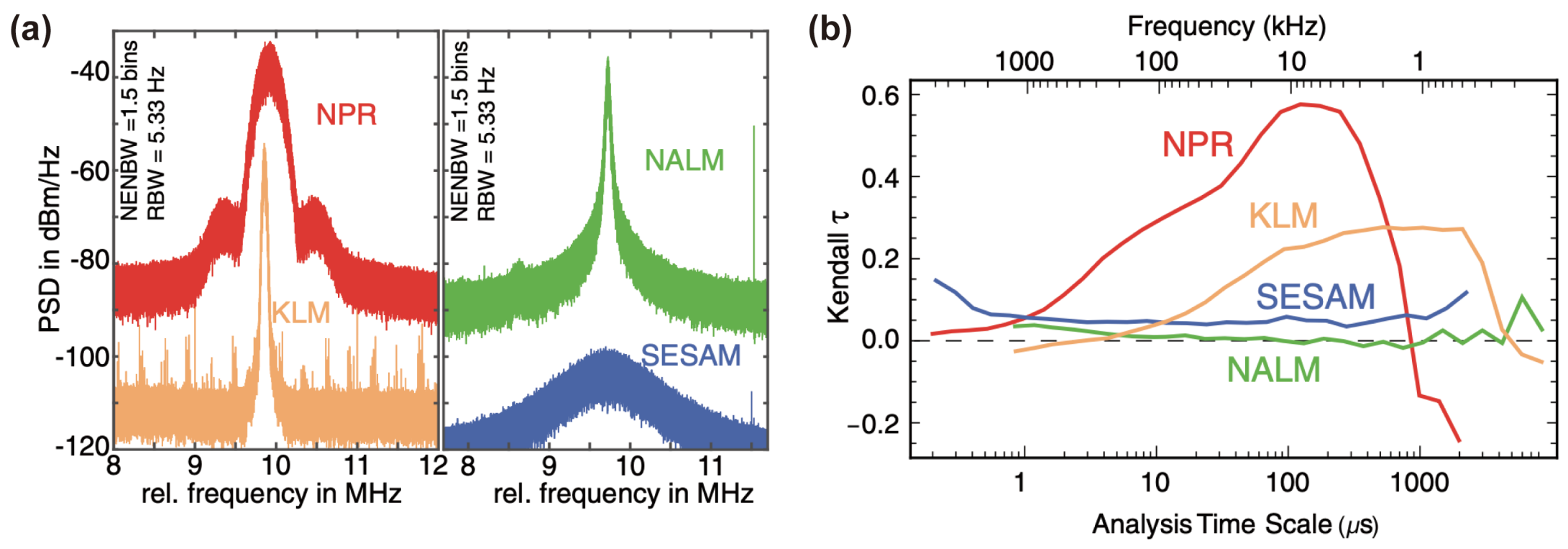

图 4-7 钛宝石飞秒激光器、掺镱光纤飞秒激光器、NALM 掺铒光纤飞秒激光器和 SESAM 锁模的掺铒光纤飞秒激光器的（a）自由运转的 $f_{ceo}$ 信号的频谱（b）Kendall 互相关分析结果。

图 4-7（a）为掺镱光纤飞秒激光器和另外 3 台飞秒激光器自由运转时的 $f_{ceo}$ 信号的频谱图。4 台激光器的 $f_{ceo}$ 信号的带宽各不相同，频谱最窄的为钛宝石飞秒激光器和 NALM 掺铒光纤激光器。频谱最宽的为 SESAM 锁模的掺铒光纤激光器。其中，钛宝石飞秒激光器的 $f_{ceo}$ 信号接近于洛伦兹线型。图 4-7（b）为 4 台激光器的自由运转 $f_{ceo}$ 信号的 Kendall 互相关分析结果。对于 NPR 锁模的掺镱光纤飞秒激光器，$f_{ceo}$ 信号的振幅和频率在 10 kHz 处有着很强的相关性。对于钛宝石飞秒激光器，$f_{ceo}$ 信号的振幅和频率在 1 kHz 处也有一定的强度的相关性。SESAM 锁模的掺铒光纤激光器的互相关分析曲线在 1 us 至 1 ms 时间窗口内较

为平坦，均保持在 5%左右。然而，基于 NALM 锁模的掺铒光纤激光器的振幅和频率在 1 μs 至 1 ms 时间窗口内均未体现出明显的相关性。综合以上实验结果，基于 NALM 锁模的掺铒光纤激光器的 $f_{ceo}$ 信号受到振幅-相位调制噪声影响最小。因此，该类激光器的 $f_{ceo}$ 信号受泵浦强度的波动影响最小。

## 4.4 光学平衡探测法的孤子分子对相位噪声测量

飞秒激光器自从出现以来，已经成为研究大自然中耗散系统非线性动力学过程的实用平台。在自然界中，原子在化学键的作用下，紧密结合起来，形成分子。类似的，当飞秒激光器输出两个紧密束缚起来的脉冲时，这两个脉冲被称作为孤子分子。由于孤子分子在非线性吸引子动力学研究中的重要地位，关于孤子分子的研究已经吸引了众多科学家们的眼球。然而，对于不同类型激光器中光学孤子间相互作用机制的研究需要先进的、高速的、高精度的探测技术。基于脉冲时域展宽的色散傅里叶变换技术（Time-stretch dispersive Fourier-transform，TS-DFT）提供了一个高精度的、快速的、实时的激光器输出脉冲光谱的测量方法[220]。该方法的主要原理为，将激光器输出的脉冲经过公里量级的单模光纤进行时域的展宽。展宽后的时域脉冲的形状对应于脉冲光谱的形状。使用高速采样示波器，便可以采样得到脉冲的光谱信息。该技术的优势在于，传统的光谱仪受到 CCD 积分时间和机械器件的响应速度的限制，无法实现 MHz 量级的光谱测量更新速度。然而在 TS-DFT 技术中，高速采样示波器的采样速率可以达到若干 GHz 量级。因此，可以很容易地实现对于激光器输出脉冲光谱的实时监测，以实现激光器输出脉冲在激光腔内复杂的动力学过程的研究。基于 TS-DFT 技术对激光器中多孤子动力学过程的研究在近几年被广泛报道[221]。

然而，尽管 TS-DFT 技术有诸多优势，拥有 GHz 采样率的高速示波器仍然造价昂贵。本小节提出了一种测量孤子分子相位噪声功率谱密度的测量方法，该方法具有系统结构简单，造价低廉等特点。该相位噪声功率谱测量方法可以被广泛应用于各种类型的激光器中，可以为复杂非线性系统动力学过程的研究、高速光通信、高容量光存储系统以及高分辨率光谱测量奠定更为实用的理论基础。

### 4.4.1 孤子分子对相位噪声测量原理

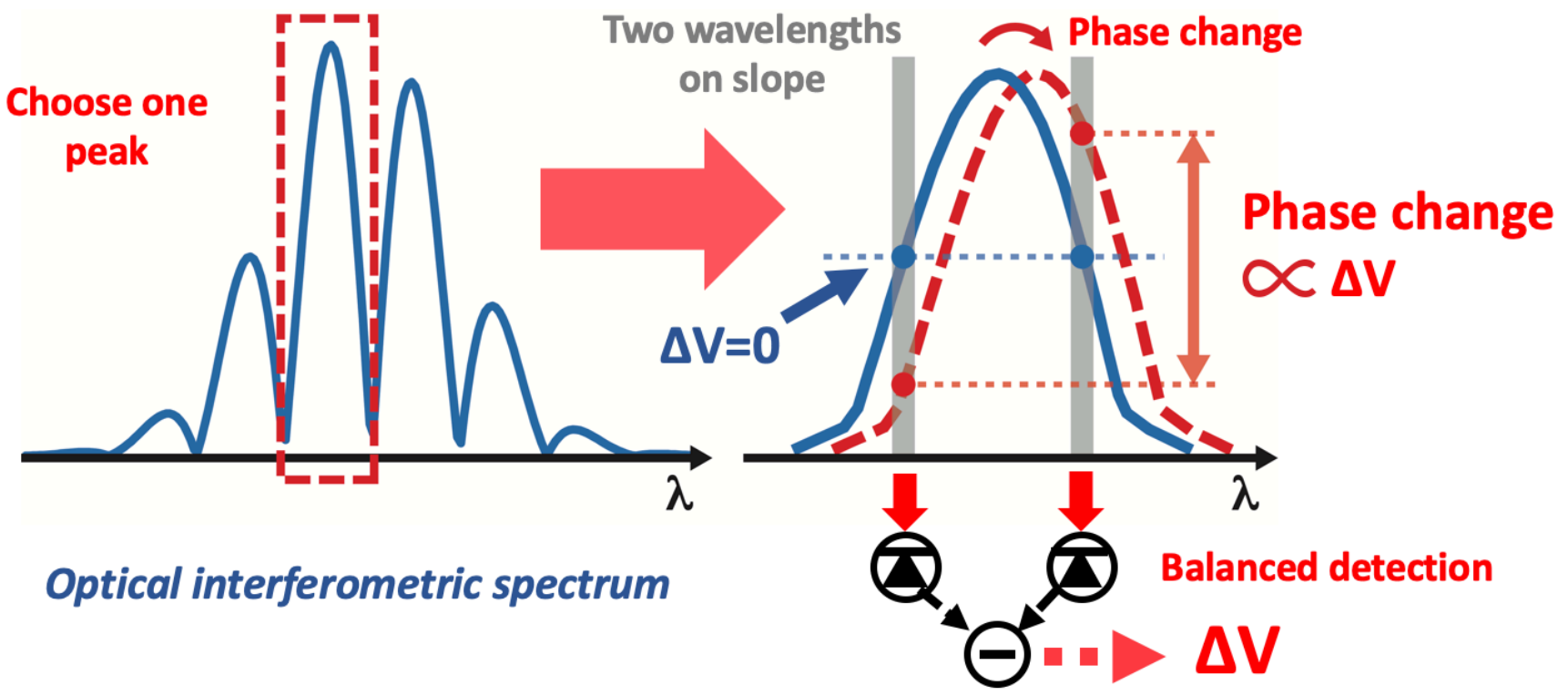

图 4-8　孤子分子对相位噪声测量原理图

图 4-8 为孤子分子相位噪声功率谱测量原理示意图。激光器输出的孤子分子对的光谱在光频域有着明显的周期性干涉条纹。挑选出某一个干涉条纹进行观察。当两个光孤子之间的相位发生改变时，光谱干涉条纹的位置便会产生平移。滤出该光谱干涉峰的两翼的两个波长，两个波长分别进入一个平衡探测器的两个输入端。当两个光学孤子之间的相对相位发生改变时，干涉峰平移，滤出的两个窄带光谱中，一个波长光谱的强度增加，另一个波长光谱的强度降低，平衡探测器的输出电压发生改变。光孤子的相对相位的改变被巧妙地转化为平衡探测器输出电压的改变，通过测量平衡探测器输出电压的功率谱密度，就可以得到光孤子的相对相位噪声的功率谱密度。

### 4.4.2 孤子分子对相位噪声测量装置

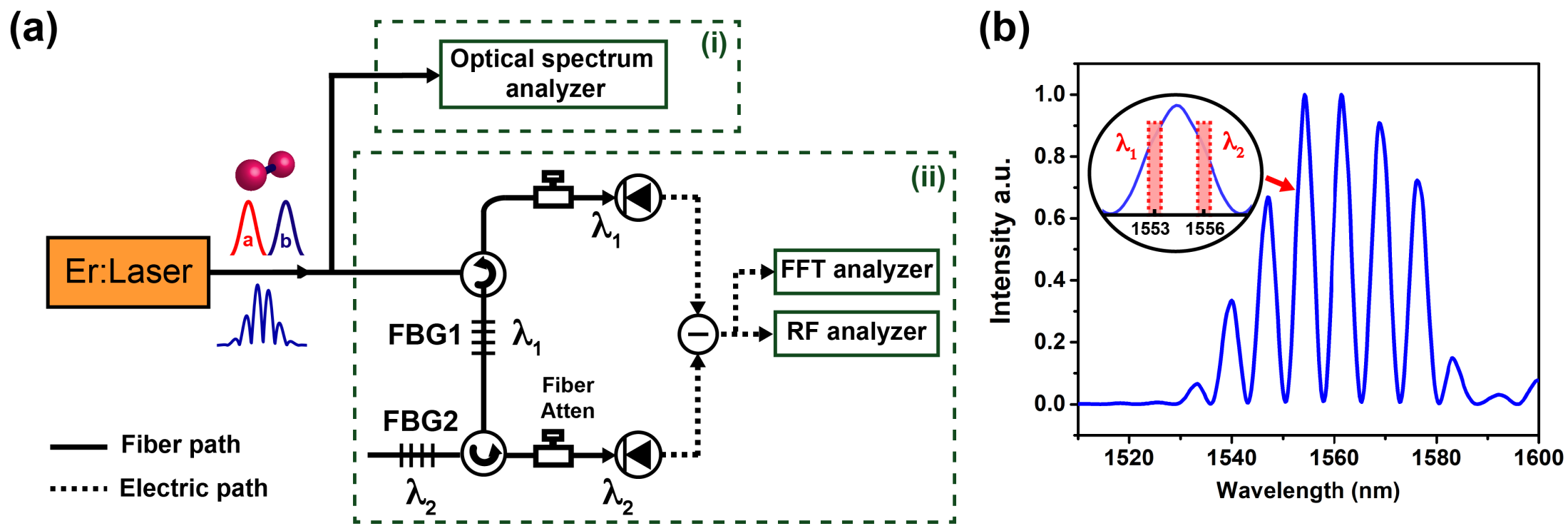

图 4-9　（a）孤子分子对相位噪声测量装置示意图。（b）孤子分子对光谱

孤子分子对相对相位噪声测量的装置如图 4-9。待测的孤子分子对来源于一个基于非线性放大环形镜锁模的全保偏掺铒光纤激光器。在单脉冲运转状态下，

激光器输出功率为 20 mW，脉冲宽度为 120 fs。在加高泵浦功率后，通过调整腔内波片的角度，可以得到稳定的孤子分子对输出，孤子分子对的光谱见图 4-9（b）。相位噪声测量的装置分为两部分，第一部分为使用光谱仪对孤子分子对的光谱进行记录。在实验装置的第二部分中，孤子分子对经过一个光纤环形器后，中心滤波波长在 $\lambda_1$=1553 nm 的反射式光纤布拉格光栅将孤子分子对中 1553 nm 部分的光谱滤出，入射进入平衡探测器（Thorlabs，PDB420C）的一个输入端。从 FBG1 透射过的光谱经过另一个光纤环形器和中心波长在 $\lambda_2$=1556 nm 的反射式光纤布拉格光栅，波长为 1556 nm 的光谱成分被滤出，入射进入平衡探测器的另一个输入端。在平衡探测器的两个输入端，有两个可调光纤衰减器，以保证进入平衡探测器两个输入端口的功率相同。两个波长的信号在平衡探测器中相减后，测量输出的电压信号的功率谱密度图，经过校准，便可以得到孤子分子对相位噪声的功率谱密度图。

### 4.4.3 束缚态相位噪声测量结果

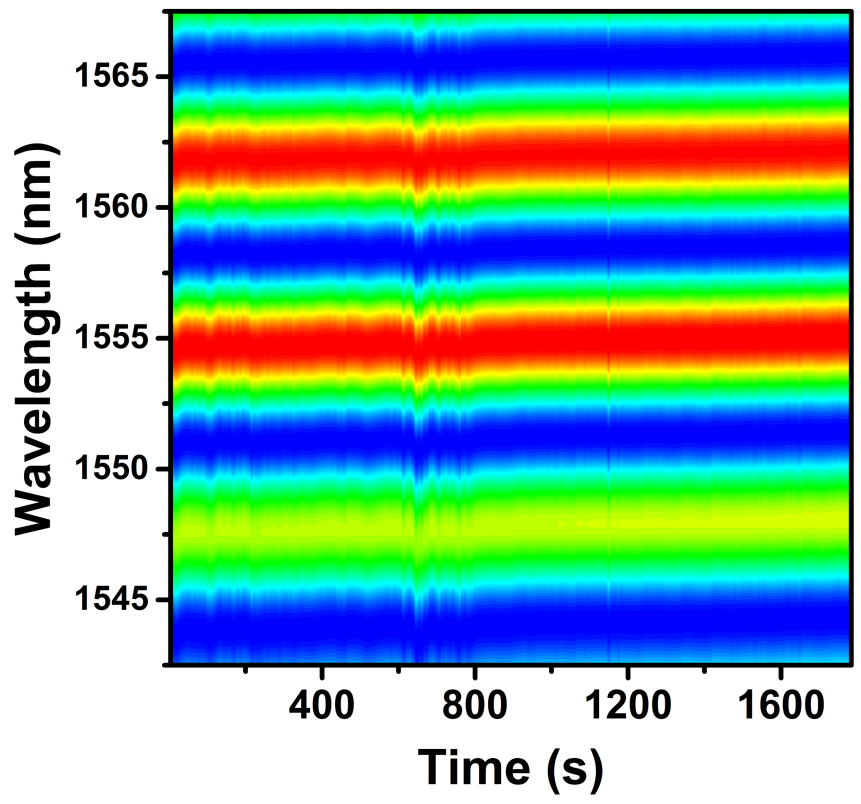

图 4-10　30 分钟时间内孤子分子对光谱变化图

首先，使用光谱分析仪对孤子分子对的光谱进行 30 分钟的光谱数据采集。每个光谱的测量和数据存储时间为 1 s。30 分钟内，孤子分子对的光谱变化见图 4-10。经过计算，可得到孤子分子对的相位噪声在 1 mHz 至 100 mHz 频率范围的功率谱密度，见图 4-11（a）中曲线（i）。

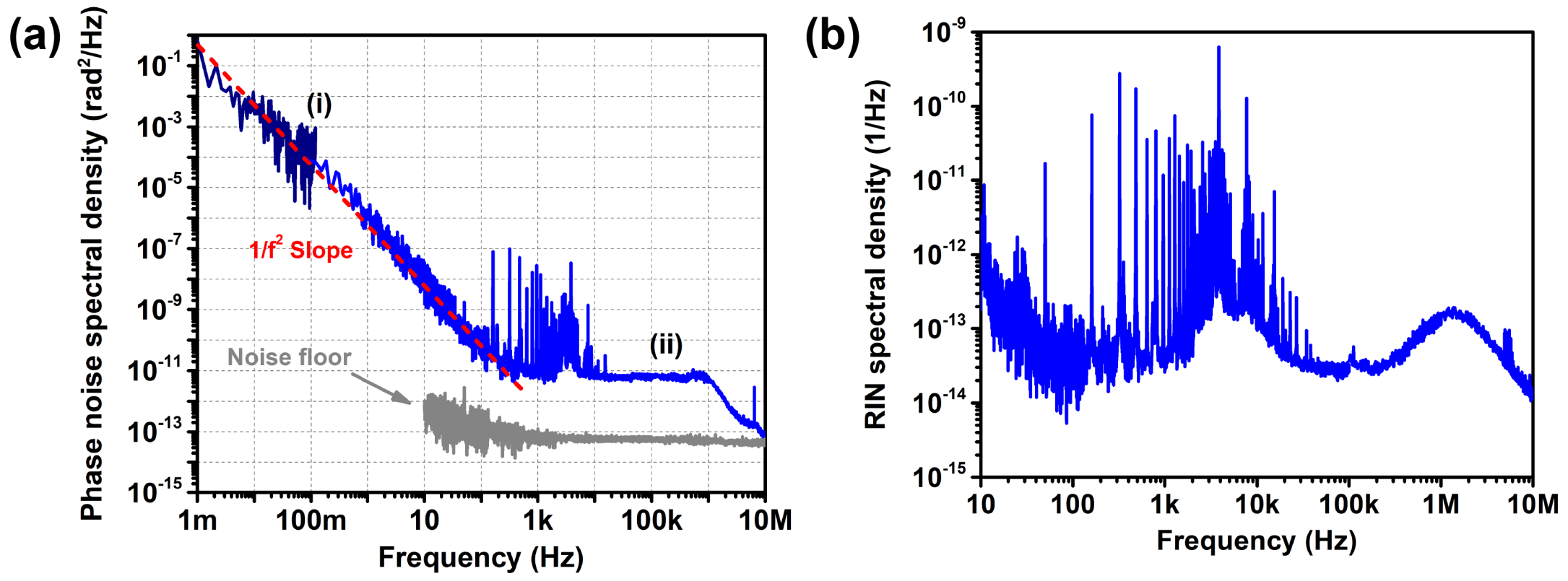

图 4-11　孤子分子对相位噪声功率谱图测量结果。（a）相位噪声功率谱（b）相对强度噪声功率谱

平衡探测器输出的电压反映了孤子分子对相位噪声的变化。分别使用频谱分析仪（Agilent，8560EC）和快速傅里叶变换分析仪（Standford Research Systems，SR770）对平衡探测器输出的误差信号的 100 kHz 至 2 MHz 频率范围和 100 Hz 至 100 kHz 频率范围的功率谱进行了测量，得到误差电压的功率谱密度图。使用图 4-11（a）中的曲线（i）对该电压功率谱图进行校准，得到了曲线（ii）。图中灰色曲线为平衡探测器的热噪声基底。该相位噪声测量的精度为 $10^{-13}$ rad²/Hz。这个测量精度是已报道的文献中相位测量精度最高的。需要指出的是，通过改变激光器的泵浦功率，两光学孤子的相对相位会发生改变。因此，理论上讲，可以在激光器泵浦上加以一个正弦的电流调制，使得光学孤子的相对相位在±π/2 范围内发生周期性改变，此时，记录平衡探测器输出的电压信号的线性部分作为相位鉴别曲线，可以实现电压功率谱至相位噪声功率谱的转换。但是，在实验过程中，给激光器的泵浦电流加以调制后，孤子分子对的锁模状态极易消失，很难记录到有效的相位鉴别曲线。所以，本文才使用曲线（i）校对曲线（ii）的方法进行电压功率谱至相位噪声功率谱的转换。另外，对孤子分子对的强度噪声功率谱进行了测量，结果见图 4-11（b）。

从图 4-11（a）中可以看出，对于傅里叶频率低于 1 kHz 的部分，相位噪声功率谱以 $1/f^2$ 的斜率下降。这表明了，在低频部分，两个光孤子的相位特性为随机游走的相位噪声。两个光学孤子在 1 kHz 的带宽以上，被紧密束缚在一起。在 1 kHz 至 1 MHz 傅里叶频率范围内，两个光孤子的相位噪声特性主要为白噪声。kHz 频率处的尖峰是由声学噪声引入的。在高于 1 MHz 的频率范围内，相位噪声的功率谱逐渐降低。通过将该功率谱图与强度噪声的功率谱图进行对比，发现在 1 MHz 处有相同的拐点。强度波动在激光腔内的非线性作用下，耦合成为光孤子的相位波动。因此，在 1 kHz 以上的高频部分，相位噪声的主要来源为由光

孤子强度波动引入的振幅-相位调制噪声。

将相位噪声的功率谱转化为频率噪声的功率谱，使用 $\beta$-line 分析法，估算两个光学孤子的相对线宽，见图 4-12。可以得到两光学孤子的相对线宽应远小于 1 mHz，如果将红色虚线与蓝色曲线继续延伸，二者应在 μHz 处相交，因此，孤子分子对中两光学孤子的相对线宽非常窄，应当仅为 μHz 量级。

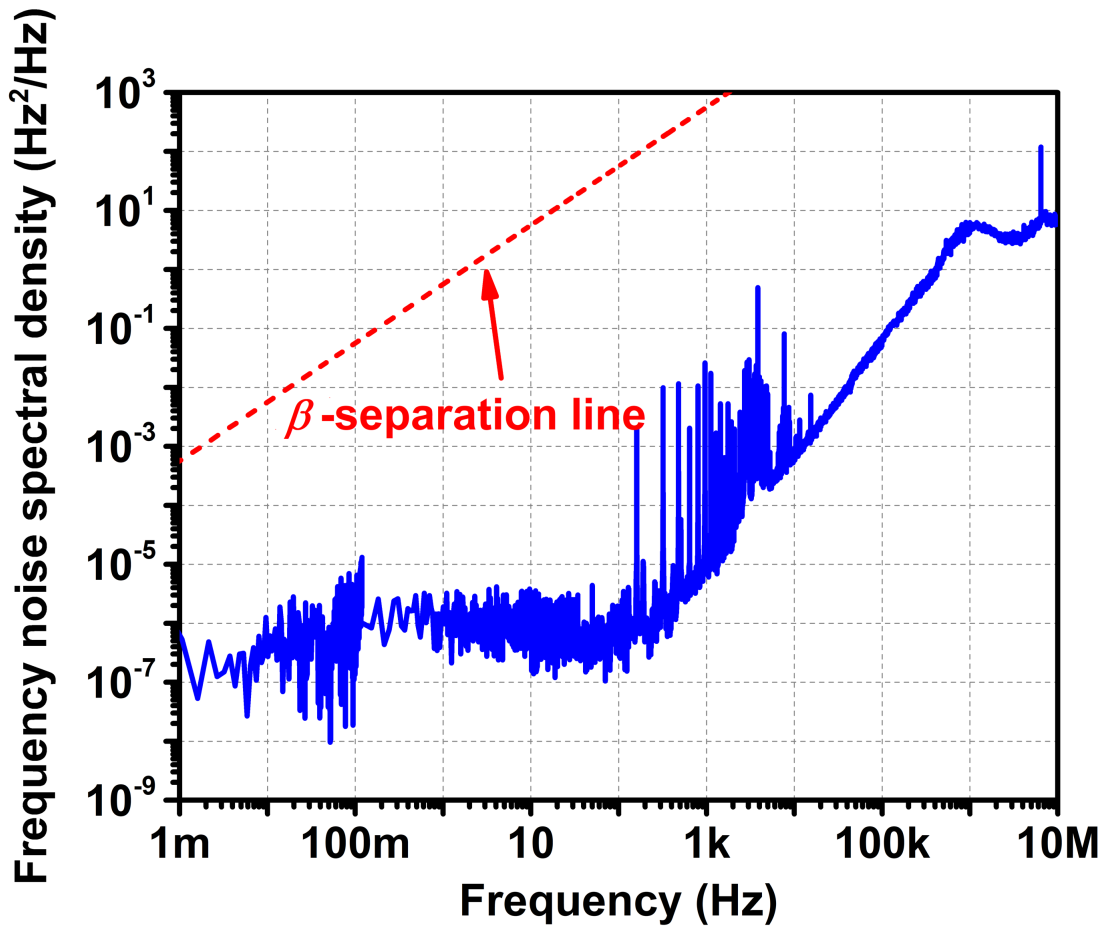

图 4-12　孤子分子对频率噪声功率谱图

## 4.5 本章小结

本章介绍了飞秒激光光学频率梳中的载波-包络相位噪声的高精度测量及分析。该测量方法在两台重复频率锁定的掺镱光纤飞秒激光器的基础上，首次使用光学外差探测法实现了载波-包络相位噪声功率谱的超大范围、超高精度测量。该方法无需使用 $f$-$2f$ 干涉仪，摆脱了 $f$-$2f$ 干涉仪中的非线性扩谱步骤，很大程度地提高了噪声测量的信噪比。该方法可以实现傅立叶频率范围在 5 mHz 至 8 MHz，动态范围高于 270 dB 的载波-包络相位噪声功率谱测量。相位噪声测量基底低于 1μrad/√Hz。这是在报道过的文献中最高灵敏度的载波-包络相位噪声测量。从噪声功率谱中曲线的不同斜率可以看出，傅里叶频率低于 20 Hz 的部分，该噪声主要由环境扰动引入；20 Hz 至 35 kHz 的部分主要为泵浦激光器的波动引入的相位噪声；高于 35 kHz 的部分为放大自发辐射引入的量子噪声。随后的 Hadamard 方差分析也同样发现了泵浦波动引入的噪声和量子噪声的分界点。然后，本工作使用 Kendall 互相关分析对不同飞秒激光器的载波-包络相位噪声进行了分析。通过对钛宝石飞秒激光器、基于非线性放大环形镜锁模的掺铒光纤飞秒激光器和基于可饱和吸收镜锁模的掺铒光纤飞秒激光器的自由运转的载波-包络偏移频率信

号进行 Kendall 互相关分析，发现在不同激光器中，载波-包络相位对振幅-相位调制噪声影响是不同的。其中，基于非线性放大环形镜锁模的掺铒光纤飞秒激光器受到该因素的影响最小。

最后，本工作首次对孤子分子对中的两个光学孤子的相对相位噪声进行了测量。通过两个窄带滤波器滤出孤子分子的输出光谱中的一个干涉峰的两翼的两个波长，将两个光学脉冲的相对相位改变转化为这两个波长相对强度的改变，使用平衡探测器探测，就可以得到光孤子的相对相位噪声的功率谱密度。从 10 Hz 至 10 MHz 积分，相对相位噪声的积分值为 3.5 mrad。相位噪声测量的精度为 $10^{-13}$ $rad^2/Hz$。使用 $\beta$-line 分析法，估算两个光学孤子的相对线宽仅为 μHz 量级。

# 第5章 基于光纤延迟线的光纤频率梳噪声测量及分析

本章工作对掺铒光纤飞秒激光光学频率梳的$f_{rep}$、$f_{ceo}$和$\nu_n$的频率噪声功率谱进行了测量及分析。采用全保偏非线性放大环形镜锁模的掺铒光纤激光器作为待测光源。使用掺铒光纤放大器对激光振荡器输出的脉冲进行放大后，使用非对称光纤延迟线干涉仪，将光纤延迟线与光学外差探测法相结合作为相位鉴别器，实现了无需外部参考的$f_{rep}$噪声功率谱和$\nu_n$噪声功率谱测量。使用$f$-$2f$干涉仪和相位噪声分析仪直接测量$f_{ceo}$的噪声功率谱。另一方面，本工作使用文献[31-32, 100]中的理论模型，对$f_{rep}$、$f_{ceo}$、$\nu_n$和光学频率梳中所有梳齿的噪声来源进行了理论分析。在实验中，本工作还验证了$f_{rep}$与$f_{ceo}$噪声功率谱之间的反相关特性。最后，本工作对锁相环之间的串扰进行了实验验证。基于光纤延迟线的光学频率梳噪声分析是获得低噪声光学频率基准的重要理论基础。该系统不仅可以用于飞秒激光光学频率梳的噪声测量，还可以用于研究其他不同类型光学频率梳如微环谐振腔、电光调制光学频率梳、半导体激光器等的内在噪声动力学分析。

## 5.1 光纤延迟线测量噪声原理

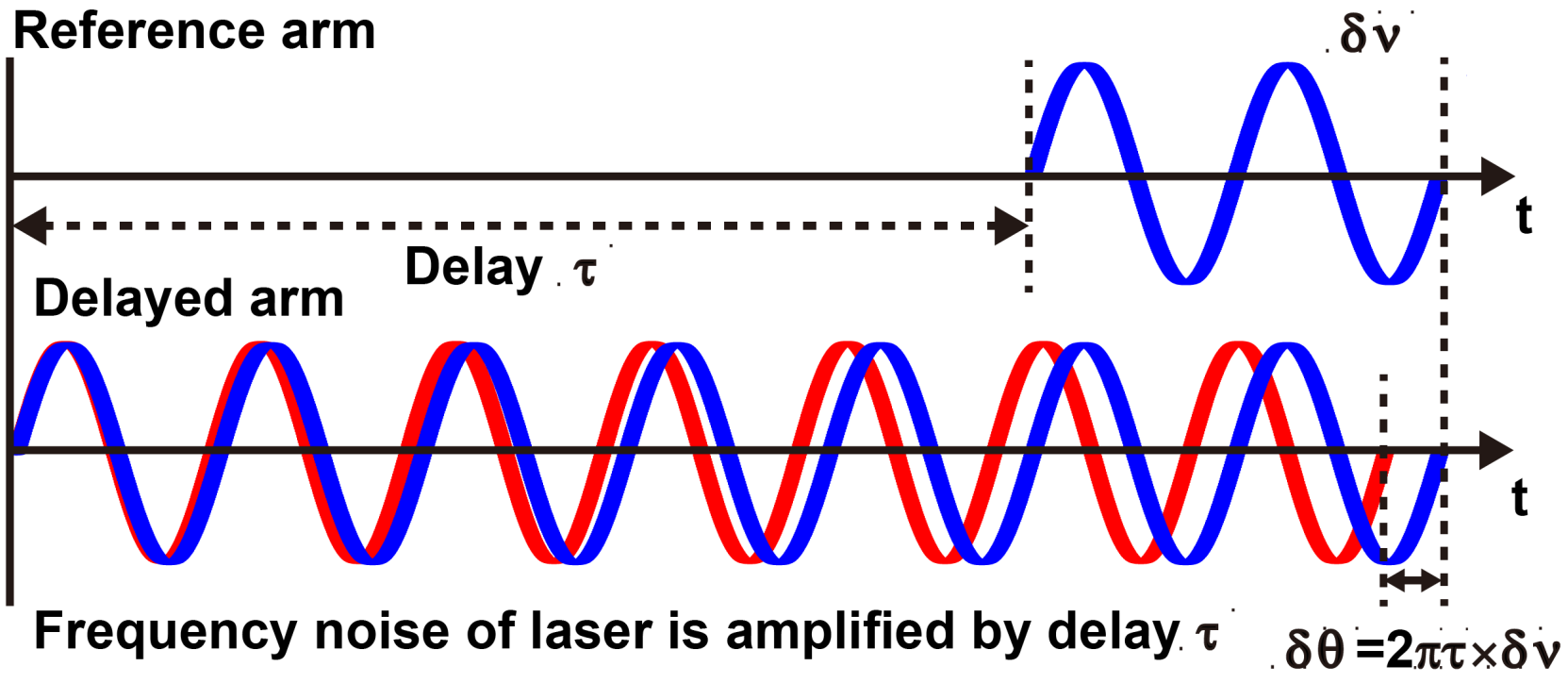

图 5-1　光纤延迟线测量频率噪声原理示意图

在连续光激光器中，首先实现了使用光纤延迟线技术对激光器的频率噪声的测量。其原理如图5-1所示。将连续光激光器的输出分为两部分，一部分作为参考光，另一部分经过公里级长度的光纤延迟线，连续光激光器的相位噪声被放大$\tau$倍，将延迟光与参考光的相位进行比较，就可以得到待测激光的相位噪声。显然的，相位噪声测量的灵敏度正比于延迟$\tau$。在飞秒激光光学频率梳中，基于相同的原理，使用窄带光学滤波器选取频率梳中的两个波长$\lambda_1$和$\lambda_2$附近的光学梳齿，

由于在窄滤波器带宽内光学模式的频率噪声差异可以忽略，参考光与延迟光的拍频信号所对应的频率噪声为$p\cdot\delta[\tau(mf_{rep}+f_{ceo})]$，$p\cdot\delta[\tau(nf_{rep}+f_{ceo})]$。其中$p$代表滤波带宽内的光学模式的数量，$m$与$n$分别代表$\lambda_1$和$\lambda_2$波段的梳齿模式数。使用如下传递函数将测量到的电压噪声功率谱转化为频率噪声功率谱：

$$T_F(f)=V_{pk}\frac{1-\exp(-\mathrm{i}2\pi f\tau)}{if}[\mathrm{V/Hz}] \tag{5-1}$$

其中，$V_{pk}$为鉴相信号的振幅，$\tau$为光纤延迟线的长度。也可以使用简化的表达式：

$$T_F(f)=2\pi\tau V_{pk}[\mathrm{V/Hz}] \tag{5-2}$$

由于$V_{pk}$包含滤波带宽内 $p$ 个梳齿的贡献，使用传递函数进行转化后，便可得到 $p$ 个梳齿的平均频率噪声 $\delta[\tau(mf_{rep}+f_{ceo})]$，$\delta[\tau(nf_{rep}+f_{ceo})]$。由于滤波带宽内光学模式的频率噪声差异很小，该噪声可视为第 $m$ 根和第 $n$ 根梳齿的频率噪声。将两个拍频信号用混频器混频，使用传递函数进行转化，即可消除$f_{ceo}$噪声影响，得到 $\delta(\mathrm{m-n})f_{rep}$ 的功率谱，该功率谱代表飞秒激光器的重复频率噪声，即飞秒激光器的定时抖动。

## 5.2 实验装置

### 5.2.1 待测光纤飞秒激光器

待测量的飞秒激光器为全保偏的非线性放大环形镜锁模的掺铒光纤飞秒激光器。激光器结构如图 5-2。光纤环路部分、法拉第旋光器、QWP1 和 PBS2 构成了非线性放大环形镜。考虑脉冲从 PBS2 处入射，经过相位偏置器后耦合至非线性放大环形镜。在环形镜中，脉冲被分解为沿快、慢轴传输的两个垂直偏振分量。经 PBS1 分束后，这两个偏振分量分别沿光纤环路的顺时针和逆时针方向传输。其中，顺时针传输的脉冲先被放大，后经过保偏光纤产生一定程度的非线性相移。逆时针传输的脉冲先经过保偏光纤产生非线性相移，后被增益光纤放大。因此，两方向的脉冲会积累不同的非线性相移。环外光纤中，快、慢轴上传输的光脉冲再次经相位偏置器作用后，两偏振分量在 PBS2 透射的水平偏振方向和反射的竖直偏振方向上的投影分别发生干涉。由于脉冲通过 PBS2 的透射率与脉冲在非线性放大环形镜中积累的非线性相移差相关，因此，调节激光器腔内法拉第旋光器和波片角度，使能量高的脉冲中心部分透射率高而两翼透射率低时，非线

性放大环形镜起到等效可饱和吸收体的作用，从而实现锁模。

激光器输出脉冲的重复频率为 78 MHz，平均功率 20 mW。光谱中心波长为 1580 nm，光谱半高宽为 30 nm。激光器直接输出脉冲的时域脉宽为 100 fs。激光器输出的光谱和脉冲自相关图见图 5-3。在激光器的端镜 M 粘在一个长度为 2 mm 的压电陶瓷上，以此来实现高带宽的重复频率调节。

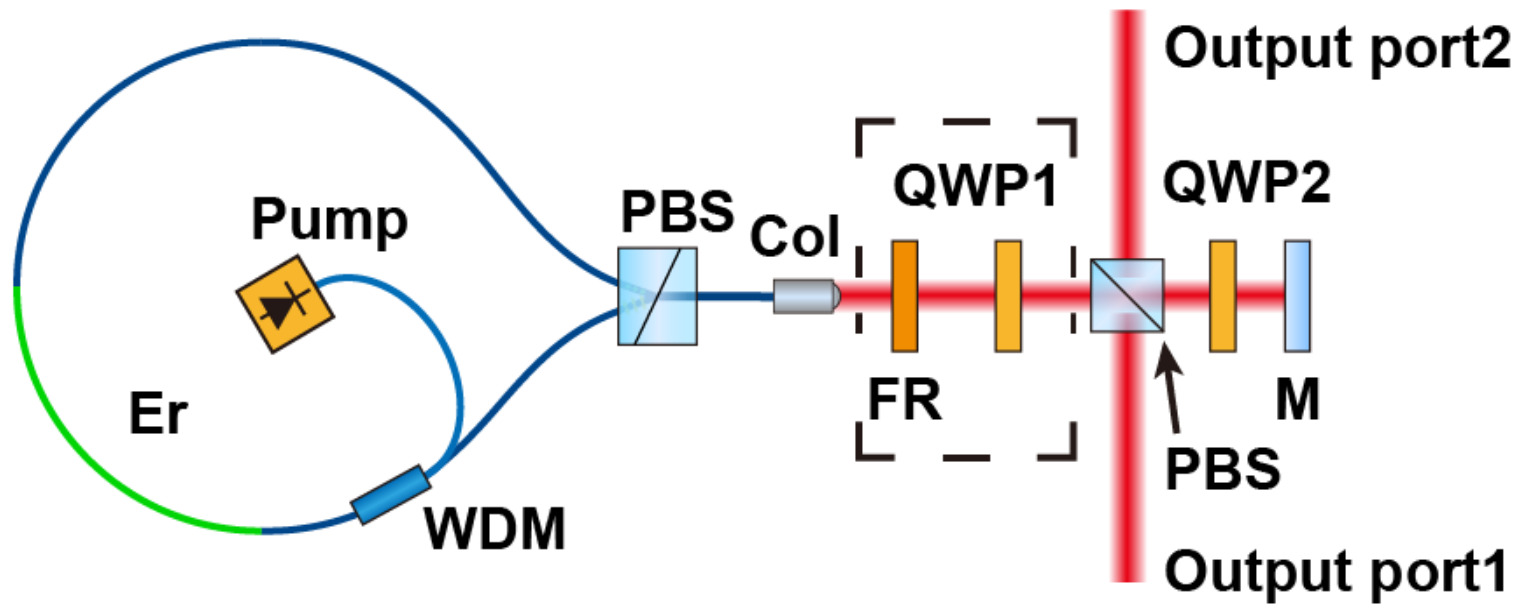

图 5-2　基于非线性放大环形镜锁模的全保偏掺铒光纤激光器结构示意图。WDM，波分复用器；Er，掺铒光纤；Pump，泵浦激光二极管；PBS，偏振分束器；COL，准直器；FR，法拉第旋光器；QWP，四分之一波片；M，反射镜

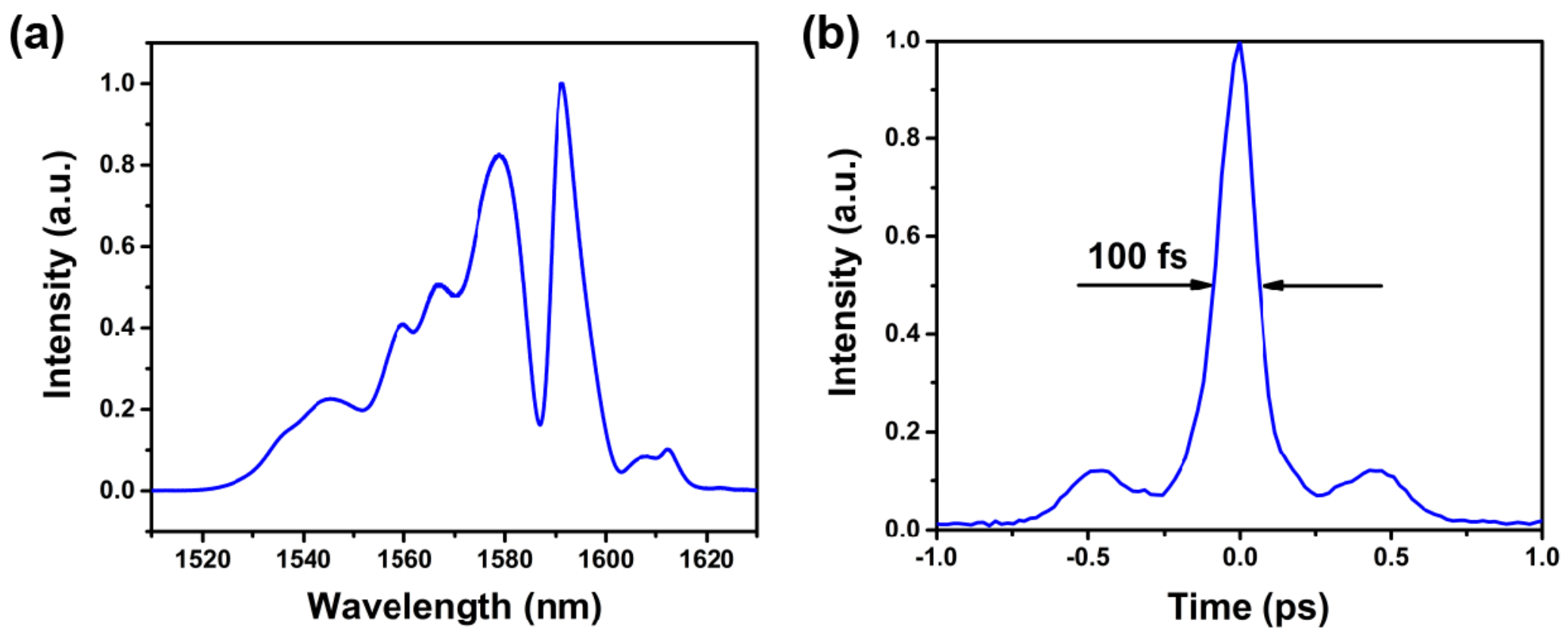

图 5-3　激光器输出脉冲的光谱（a）及自相关曲线（b）

激光振荡器输出的脉冲经过一个自行搭建的掺铒光纤放大器进行放大。放大器的结构如图 5-4（a）。脉冲从激光振荡器输出后，经过一个准直器耦合进入光纤。随后，一个光纤隔离器置于振荡级与放大级之间，用于防止放大器中 1550 nm 波段的激光经光纤的端面反射后，反馈进入振荡级，影响振荡级锁模的正常运转。由于光纤准直器和光纤隔离器会提供长度约为 50 cm 的光纤尾纤。脉冲经过这两个器件后会产生一定程度的负啁啾。在此，使用色散补偿光纤（Thorlabs，DCF-38）对这段光纤尾纤引入的二阶色散进行补偿。如果不进行色散补偿，当负啁啾的脉

冲经过正色散的增益光纤时，脉冲在增益光纤中被放大的同时，脉冲宽度也在被压缩。这时，过高的峰值功率会产生非常强的非线性效应，导致脉冲分裂。放大器的泵浦采用双向泵浦结构。前向泵浦通过波分复用器耦合进入增益光纤，后向两泵浦经过泵浦合束后，通过波分复用器耦合进入增益光纤。放大器使用的增益光纤为 Nufern 公司生产的 Er-80-4/125，长度为 80 cm。进入放大器的种子光功率为 13 mW，在前向泵浦功率为 500 mW，后向泵浦总功率为 800 mW 的条件下，放大器可输出大约 150 mW 功率的脉冲。放大器输出脉冲的光谱见图 5-4（b）。可以看出，由于放大器的增益光纤的二阶色散为正，脉冲在传输过程中在自相位调制效应的作用下，光谱覆盖将近 100 nm 范围。放大级的宽光谱输出可以支持更窄的傅里叶变换极限脉冲，这对之后的 $f_{ceo}$ 的频率噪声功率谱探测是非常有益的。

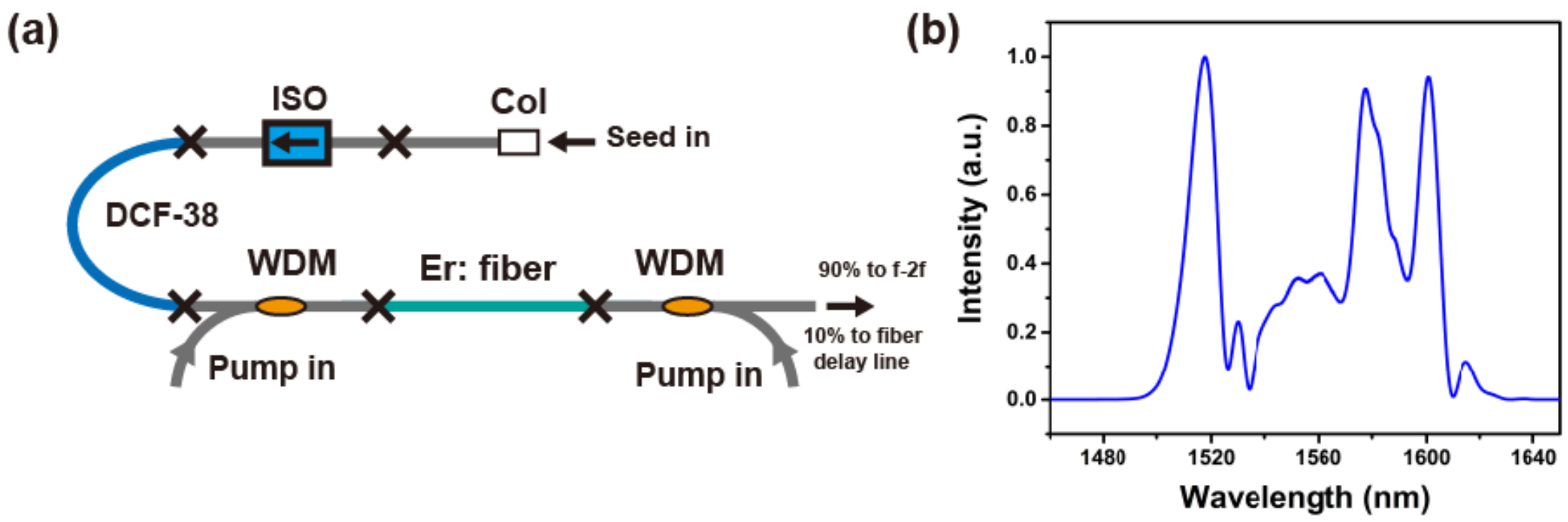

图 5-4 （a）光纤放大器结构示意图。（b）放大器输出脉冲光谱

激光振荡器和放大系统的搭建过程中，有如下注意事项：

一、在激光器的 QWP2 和端镜之间，可以加入一个双折射滤波片。在激光器内适当地引入耗散机制，可以一定程度地减少激光器输出脉冲光谱的波动，降低激光器输出脉冲的定时抖动。

二、激光器输出脉冲的中心波长最好在 1540 nm 或者 1560 nm 附近，放大级的种子光波长与铒光纤的增益峰所在的波长相匹配。这样，种子光才能在放大器中最大效率地被放大。如果种子光的中心波长偏离增益峰，例如，中心波长波长为 1580 nm。那么在放大过程中，当泵浦功率增加到一定程度时，在 1540 nm 和 1560 nm 波段会非常容易产生放大自发辐射。继续增加泵浦功率，种子光不会再被放大，放大器中的上能级粒子均以自发辐射的形式输出至放大器外。

三、种子光进入放大器前，过长的单模光纤对放大器的放大效果十分有害。一方面，具有很强负啁啾的脉冲在正色散的增益光纤中被放大时，脉冲在增益光纤中被放大的同时，脉冲也在被压缩。过高的峰值功率会导致脉冲在增益光纤中

发生分裂。另一方面，负啁啾的脉冲在正色散的增益光纤中传输，会受到光谱窄化效应的影响，导致放大器输出光谱变窄。对此，更理想的解决方法是，使用空间的隔离器和分束器代替光纤器件，以减少脉冲的负啁啾。

四、如果能使用保偏光纤来搭建该放大器，其对环境的稳定性会更好。

### 5.2.2 噪声测量及分析系统

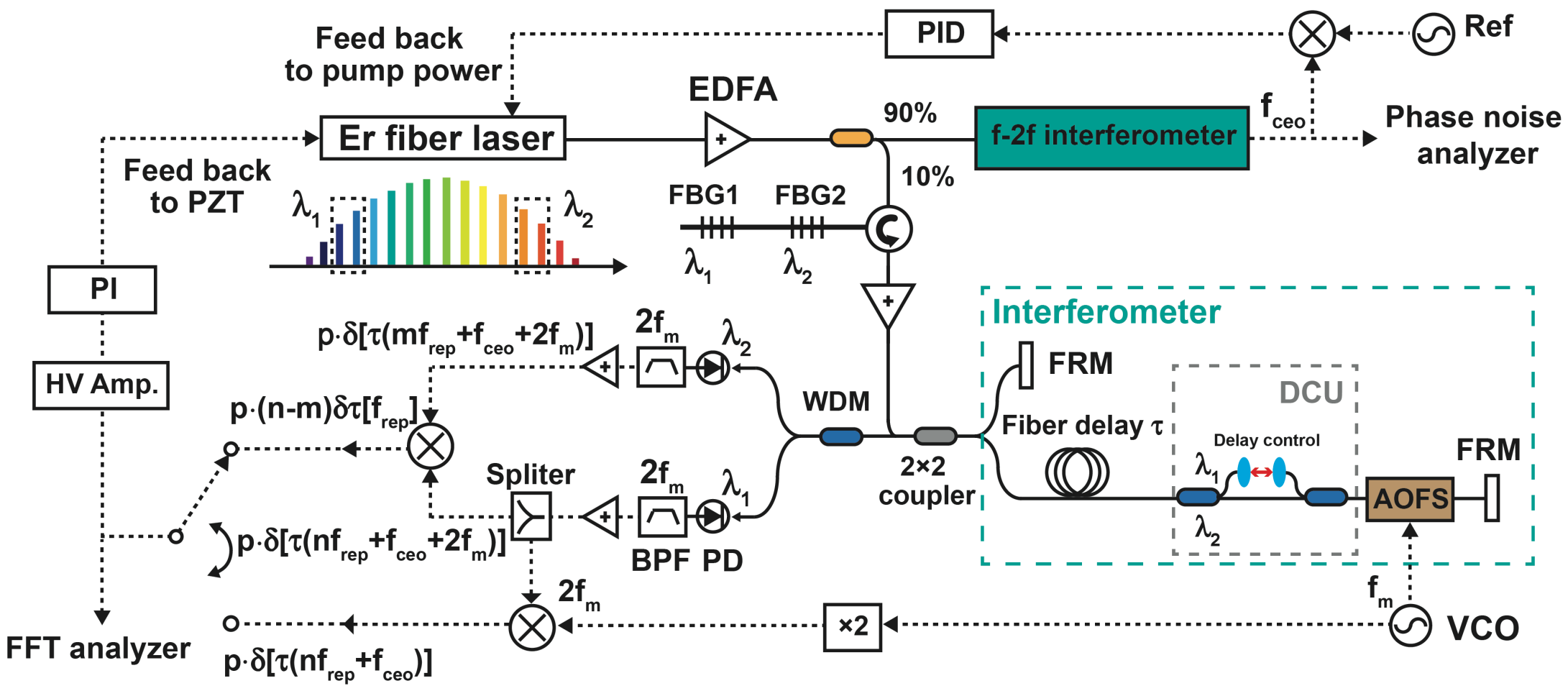

图 5-5　基于光纤延迟线的光学频率梳噪声测量及分析系统示意图。PZT，压电陶瓷；EDFA，掺铒光纤放大器；FBG，光纤布拉格光栅；FRM，法拉第旋光镜；WDM，波分复用器；VCO，压控振荡器；AOFS，声光频移器；BPF，带通滤波器；PD，光电探测器；PI，比例积分伺服控制器；HV amp，高压放大器；PID，比例积分微分伺服控制器；DCU，色散补偿单元

基于光纤延迟线的光学频率梳噪声测量及分析的系统结构见图 5-5。经过放大后的脉冲被一个光纤 9:1 分束器分为两部分。脉冲能量的 10%依次被两个反射式光纤布拉格光栅滤波。两个布拉格光栅的滤波中心波长分别在 $\lambda_1$=1540 nm 和 $\lambda_2$=1560 nm，滤波带宽为 2 nm。两个波长的总功率小于 1 mW。使用一个自行搭建的掺铒光纤放大器对两个波长进行放大。该放大器的增益光纤为低掺杂的掺铒光纤，长度为 3 米。在输出功率约为 200 mW 的单个 980 nm 激光二极管泵浦下，可将两个波长的总功率放大至 20 mW。放大后的两个波长进入一个非对称的 Michelson 干涉仪，进行激光器重复频率噪声与梳齿噪声的探测。构成 Michelson 干涉仪的主要器件为一个 2×2 的光纤耦合器。两个波长的脉冲进入耦合器的输入端，50%的脉冲能量进入干涉仪的参考臂后，被法拉第旋光镜原路反射。另 50%的脉冲能量进入干涉仪的延迟臂，依次经过 140 m 长的光纤延迟链路、独立的色

散补偿单元、压控振荡器（Brimrose，FFF-50-B1-F1）控制的声光调制器（Brimrose，AMF-50-1560-2FP）后，被法拉第旋光镜原路反射。该实验中使用的 140 m 长光纤链路是由北京拓普光电公司制作的光纤延迟模块。在该模块中，特定长度的单模光纤和色散补偿光纤被熔接在一起后，使得该 140 m 的光纤链路的总色散在零色散附近。使用该模块，可以很大程度地保证两个波长的脉冲在延迟臂中的光程相同。尽管如此，该系统仍旧需要一个可调的色散补偿单元来补偿延迟臂中其他光纤器件（如声光调制器的尾纤）的色散引入的两脉冲的光程差。色散补偿单元由两个 1540/1560 的波分复用器和一个可调光纤延迟线组成。通过调节光纤延迟线的延迟量，可以使得两个波长的脉冲在延迟臂中的光程相同。从参考臂和延迟臂返回的光学脉冲在 50/50 耦合器的另一个输入端口处出射。通过微量调节激光器的腔长与延迟臂光纤的长度，可以使得参考臂与延迟臂的光程差为激光器腔长的整数倍，在该实验中，从激光器输出的脉冲与其之后的第 120 个脉冲相遇，并发生干涉。由于色散补偿单元的作用，1540 nm 和 1560 nm 的脉冲在延迟臂中的传输不产生额外的由光纤色散引入的时间延迟。因此，1540 nm 和 1560 nm 的脉冲可以同时与其之后的第 120 个脉冲在 50/50 耦合器处相遇。声光频移器的作用是将两个波长的脉冲的光频进行 50 MHz 的频移。由于在该实验中，脉冲两次通过声光频移器，参考脉冲与其之后的第 120 个脉冲的干涉结果是产生频率为 100 MHz 的射频信号。该干涉信号包含了光学频率梳中单个波长的（如 1540 nm）相位噪声。对于 $\lambda_1$=1540 nm，干涉信号包含的噪声可以表示为 $p\cdot\delta[\tau(nf_{rep}+f_{ceo}+2f_m)]$，对于 $\lambda_2$=1560 nm，干涉信号包含的噪声可以表示为 $p\cdot\delta[\tau(mf_{rep}+f_{ceo}+2f_m)]$，其中 $p$ 为滤波带宽内的梳齿数量。将两个波长的脉冲经过一个 1540/1560 波分复用器后，分别用两个光电探测器（MenloSystems，FPD510）对两个波长的 100 MHz 干涉信号进行探测。随后，两个干涉信号经过中心频率为 100 MHz 的带通滤波器（K&L Microwave，6LB30-100/T24-0/0）后，被低噪声的电放大器放大至至少 0 dBm 的电功率。将 $\lambda_1$ 的干涉信号与压控振荡器输出信号的倍频进行混频后，就可以得到与 $\lambda_1$ 的梳齿噪声相关的误差信号 $p\cdot\delta[\tau(nf_{rep}+f_{ceo})]$。同理，将 $\lambda_2$ 的干涉信号与压控振荡器输出信号的倍频进行混频后，就可以得到与 $\lambda_2$ 的梳齿噪声相关的误差信号 $p\cdot\delta[\tau(mf_{rep}+f_{ceo})]$。将此误差信号经过比例积分伺服控制器和高压放大器后，加载到激光器腔内的压电陶瓷上，便可实现激光器梳齿噪声的消除。对误差信号的功率谱进行测量，使用传递函数进行转换后，便可得到激光器梳齿噪声的功率谱。两一方面，将两个波长的干涉信号混频后，就可以消除载波-包络相位噪声和压控振荡器输出信号的影响，得到只与激光器重复频率噪声相关的噪声 $p\cdot(m\text{-}n)\delta[\tau f_{rep}]$。同样的，将此误差信号经过比例积分伺服控制器和高压放大器后，加载到激光器腔内的压电陶瓷上，便可实现激光器重复频率的稳定。对误差信号

的功率谱进行测量，使用传递函数进行转换后，便可得到激光器重复频率的功率谱。

非对称的 Michelson 干涉仪的搭建过程中，有如下注意事项：

一、光纤链路长度的选择：在该实验中，光纤链路的长度为 140 m。较短长度的光纤长度适合于噪声的测量。如果选择更长的光纤链路，则更适合激光器噪声的消除。在文献[222]中，最长选用了 10 km 长度的单模光纤用于激光器重复频率的稳定。稳定后激光器的重复频率稳定度相比于自由运转情况下，最大可提高 4 个数量级。

二、在进入 Michelson 干涉仪前的放大器中，由于待放大的两个波长的功率之和只有不到 1 mW，属于小信号放大。因此，放大器泵浦源的功率不宜过高，否则，1540 nm 和 1560 nm 处易产生放大自发辐射，恶化噪声测量的信噪比。

三、对于整个系统中光纤器件的连接：应尽量减少使用光纤连接头，如 FC-APC 和 FC-UPC 光纤接头。由于光纤接头表面存在一定程度的反射，应采用直接熔接的方法进行光纤之间的连接。

四、将 100 MHz 的干涉信号放大至 0 dBm 以上再进入混频器，这一点非常重要，因为理论上讲，混频器的 LO 端口需要至少 7 dBm 的电功率，才可以保证混频器的低损耗、高效率运转。

五、在调节系统过程中，通常先调节激光器重复频率，使得一个波长的干涉信号的信噪比达到最大后，再调节色散补偿单元中另一个波长的延迟，使得另一个波长的干涉信号信噪比达到最大。

六、如果想要使用频谱分析仪对误差信号的噪声谱进行测量，则需要一个噪声基底较低的频谱分析仪。RIGOL 公司生产的 DSA815 无法实现该噪声谱的测量。

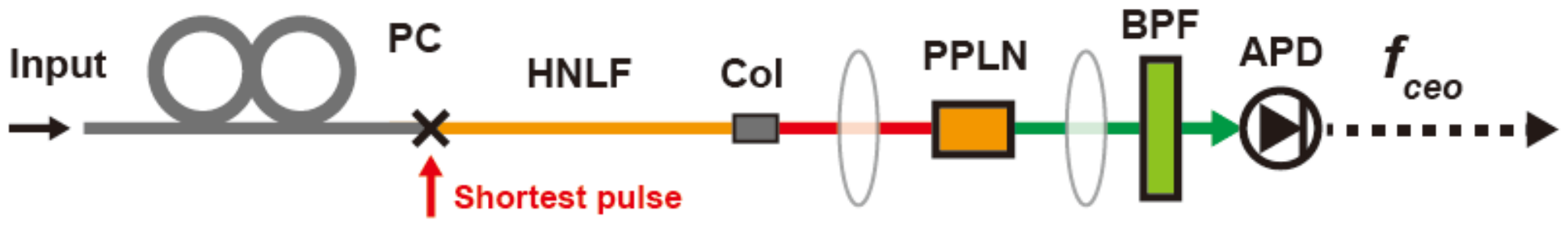

图 5-6　*f-2f* 干涉仪系统结构示意图。PC，偏振控制器；HNLF，高非线性光纤；PPLN，周期极化铌酸锂晶体；BPF，带通滤波器；APD，雪崩光电探测器

脉冲能量的 90%进入一个自行搭建的 *f-2f* 干涉仪中，实现 $f_{ceo}$ 信号的探测。*f-2f* 干涉仪的主要结构如图 5-6 所示。首先，适当长度的单模光纤对放大后的脉冲进行脉冲宽度的压缩。压缩后，脉冲的时域宽度在 70 fs 以下。随后脉冲进入一段长度为 35 cm 的高非线性光纤中，将脉冲的光谱拓展至超过一个倍频程。高

非线性光纤之后，一个准直器用于将光束准直。准直后的光束被一个焦距为 100 mm 的聚焦透镜聚焦在 PPLN 晶体上。该 PPLN 晶体为 CTL Phtonics 公司生产的倍频晶体，周期为 30~34 μm，厚度为 5 mm。在 PPLN 晶体中，超连续光谱中的 1.1 μm 部分透过晶体，2.2 μm 的部分被晶体倍频，倍频后的 1.1 μm 部分与未被倍频的 1.1 μm 部分做拍，经过一个中心波长为 1.1 μm 的光学滤波片（Thorlabs，FB1100-10）后，入射进入雪崩探测器（Thorlabs，APD110C/M）。需要使用两个光纤偏振控制器对进入高非线性光纤之前的脉冲的偏振态进行调节。在仔细调节偏振控制器的角度、放大器泵浦功率、PPLN 的周期等一系列因素后，可以得到 100 kHz 分辨率带宽下，信噪比为 35 dB 的 $f_{ceo}$ 信号。对该信号使用电学带通滤波器滤波后，用电学放大器将其放大至 0 dBm 以上。使用数字鉴相器和比例积分微分伺服控制系统（Vescent Photonics，D2-135），通过激光器的泵浦电流反馈控制，可以将 $f_{ceo}$ 信号锁定至一个射频参考源（Stanford Research Systems，SG382）上。使用一个相位噪声分析仪（Rohde & Schwarz，R&S FSWP26）可以实现自由运转的和锁定后的 $f_{ceo}$ 信号的相位噪声谱的测量。

*f-2f* 干涉仪的搭建过程中，要注意以下几点：

一、不同偏振态的脉冲进入高非线性光纤，所产生的超连续光谱不同。因此对于非保偏的 *f-2f* 干涉仪系统，在进入高非线性光纤前，脉冲的偏振态调节非常重要。可以选择将脉冲用准直器耦合出光纤后，用一个半波片和一个四分之一波片来调节，再通过准直器进入光纤。当然，这样会不可避免地引入 10%左右的损耗。

二、可以加入一个可调的光学延迟，对超连续中的 1.1 μm 和 2.2 μm 光谱的时间延迟进行更精密的调节。具体结构可以参考文献[62]。1.1 μm 和 2.2 μm 部分的光谱在时域的重合程度对 $f_{ceo}$ 信号信噪比的影响十分明显。

三、如果使用全保偏光纤的放大器和高非线性光纤来搭建 *f-2f* 干涉仪，其对环境的稳定性会更好。

### 5.2.3 频率噪声功率谱

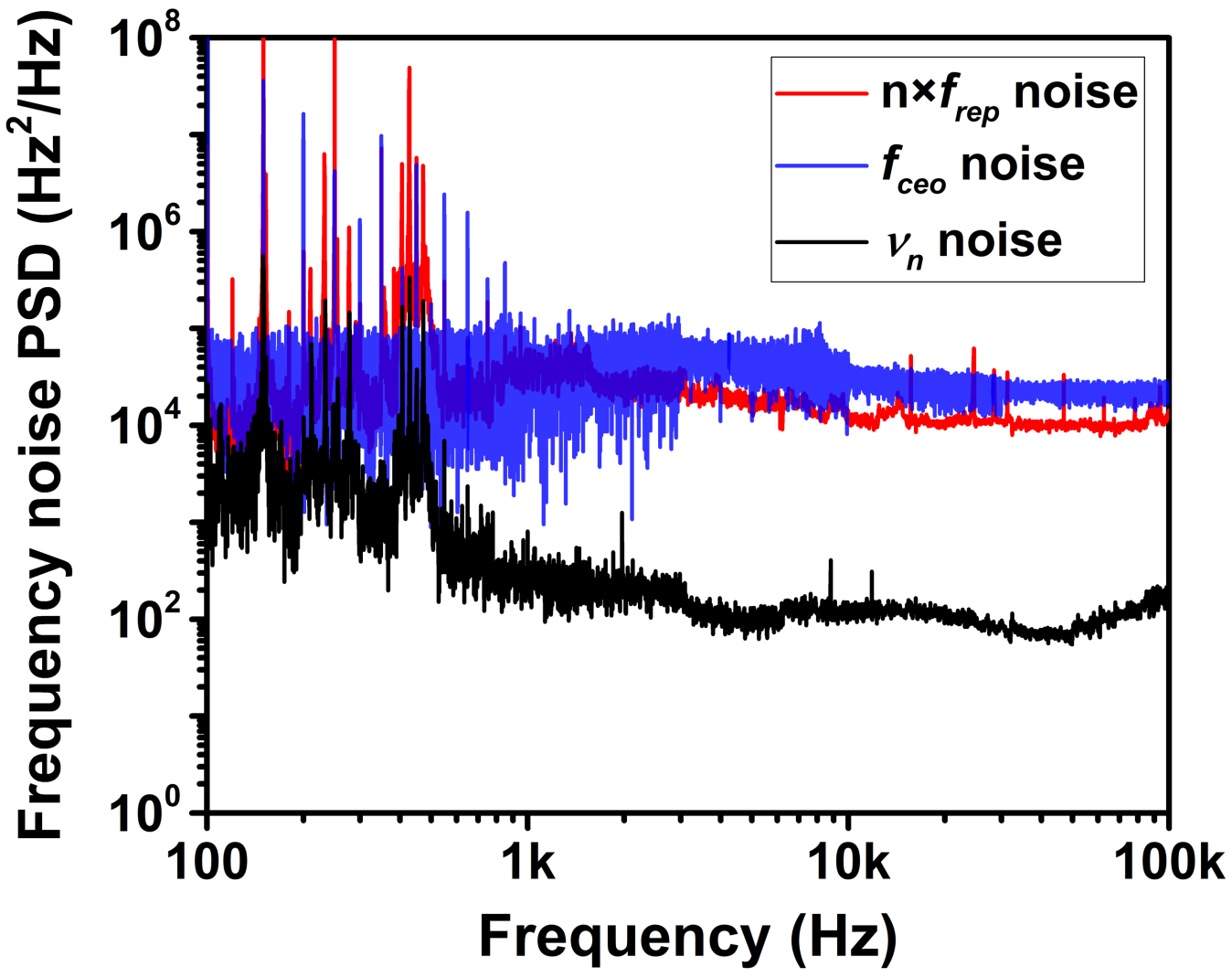

图 5-7　n×$f_{rep}$（红色曲线）、$f_{ceo}$（蓝色曲线）和 $\nu_n$（黑色曲线）的频率噪声功率谱

首先，将激光器的重复频率锁定至光纤延迟线，锁定带宽为~1 kHz。使用快速傅里叶变换分析仪（Stanford Reasearch Systems，SR770）测量$(m\text{-}n)\times f_{rep}$的频率噪声功率谱，然后使用鉴相斜率将其转换至 $n\times f_{rep}$ 的频率噪声。在该实验中 $n = 2.50\times10^5$，为 1540 nm 光学梳齿的模式数。$m = 2.47\times10^5$，为 1560 nm 光学梳齿的模式数。$n\times f_{rep}$ 的频率噪声的功率谱密度图见图 5-7 中红色曲线。将其转换至相位噪声后，对噪声功率谱的 1 kHz 至 100 kHz 范围内进行积分，得到 $n\times f_{rep}$ 的相位噪声的均方根值为 2 μrad，对应的定时抖动为 4.35 fs，其中 $f_{rep}$ = 78 MHz。随后，关闭重复频率锁定的锁相环，将光学频率梳的第 $n$ 根梳齿（$n\times f_{rep}+f_{ceo}$）锁定至光纤延迟线，锁定带宽为~ 200 Hz。采用同样的方法，可以得到光学频率梳的第 $n$ 根梳齿的频率噪声，见图 5-7 中黑色曲线。梳齿的光学频率为$\nu_n$=194.8 THz。将频率噪声功率谱图转换为相位噪声功率谱图，在 1 kHz 至 100 kHz 范围内进行积分，得到的梳齿的相位噪声的均方根值为 0.4 rad。最后，本工作使用相位噪声分析仪对自由运转的 $f_{ceo}$ 的频率噪声进行测量，结果见图 5-7 中蓝色曲线，其中 $f_{ceo}$ = 8 MHz。在 1 kHz 至 100 kHz 范围内进行积分，对应的 $f_{ceo}$ 的相位噪声的均方根值为 5.5 rad。从图中可以看出，$n\times f_{rep}$、$f_{ceo}$ 和 $\nu_n$ 的频率噪声功率谱高频部分的噪声特性均为白噪声。

### 5.2.4 频率噪声的反相关特性

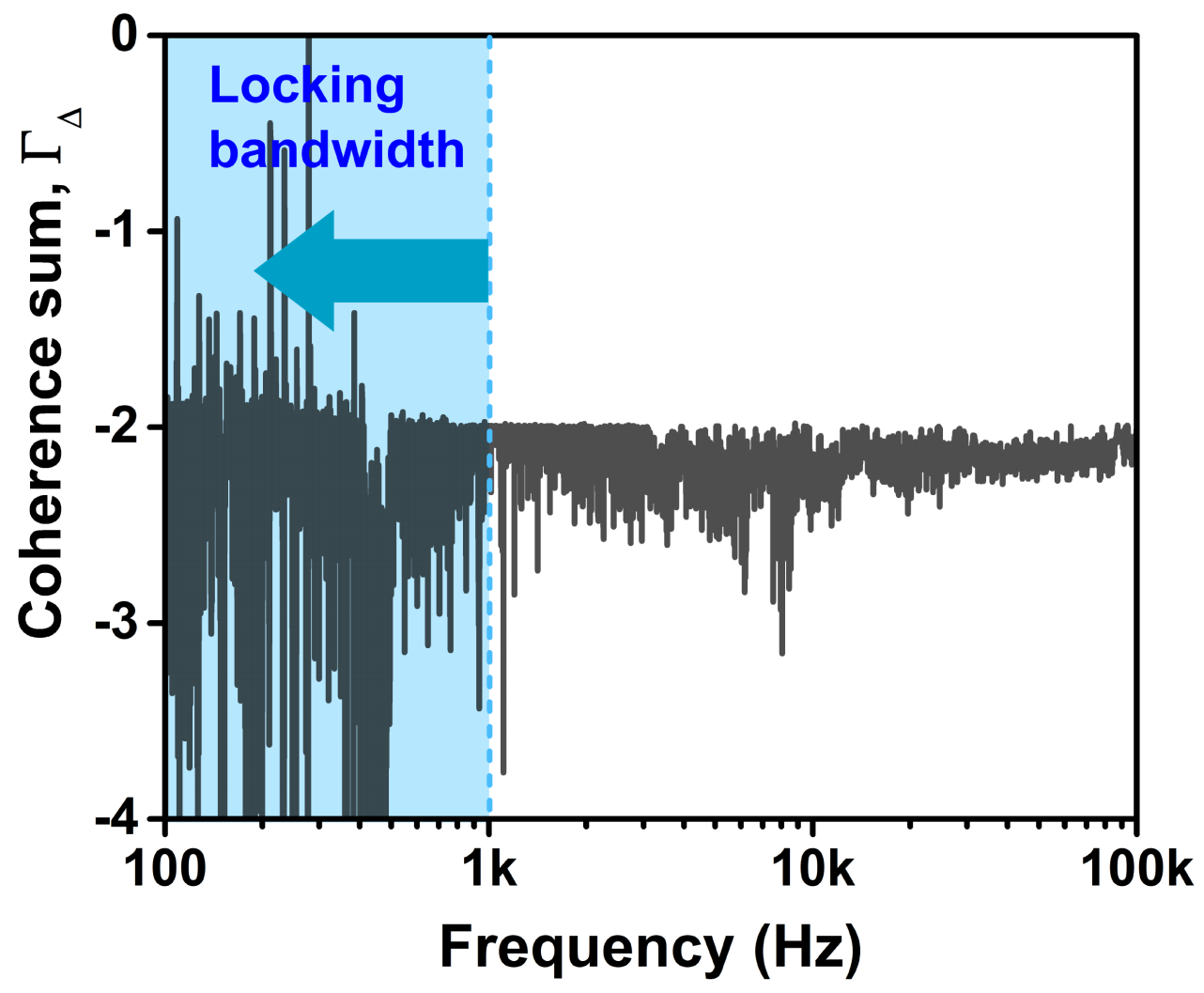

图 5-8　n×$f_{rep}$、$f_{ceo}$ 和 $\nu_n$ 的频率噪声功率谱的复相干系数随频率变化曲线

本工作提出的噪声测量与分析系统可以用于光学频率梳中，$n\times f_{rep}$、$f_{ceo}$ 和 $\nu_n$ 的频率噪声功率谱的复相干系数分析。根据光学频率梳方程，$f_{rep}$、$f_{ceo}$ 和 $\nu_n$ 的关系为 $\nu_n = n\times f_{rep} + f_{ceo}$。但是，三者的频率噪声谱之间的关系并不是简单的重复频率噪声谱与载波-包络相位噪声谱之和等于梳齿噪声。在噪声谱相关分析中，如果变量 $A$、$B$ 和 $C$ 的噪声功率谱满足关系：

$$S_C^2 = \left(S_A + S_B\right)^2 = S_A^2 + S_B^2 + 2S_A \cdot S_B \tag{5-3}$$

则 $A$ 与 $B$ 之间为 100%正相关。如果满足

$$S_C^2 = \left(S_A - S_B\right)^2 = S_A^2 + S_B^2 - 2S_A \cdot S_B \tag{5-4}$$

则 $A$ 与 $B$ 之间为 100%反相关。相对应的，根据文献[219]中所述，$f_{rep}$ 的噪声与 $f_{ceo}$ 的噪声之间存在着反相关关系，三者频率噪声功率谱的关系如下公式表示：

$$S_{\Delta\nu_n\Delta\nu_n} = S_{\Delta f_{ceo}\Delta f_{ceo}} + S_{N\cdot\Delta f_{rep}N\cdot\Delta f_{rep}} + \Gamma_\Delta(\omega)\times\sqrt{S_{\Delta f_{ceo}\Delta f_{ceo}} \cdot S_{N\cdot\Delta f_{rep}N\cdot\Delta f_{rep}}} \tag{5-5}$$

其中，$S_{N\cdot\Delta f_{rep}N\cdot\Delta f_{rep}}$、$S_{\Delta f_{ceo}\Delta f_{ceo}}$ 和 $S_{\Delta\nu_n\Delta\nu_n}$ 分别为 n×$f_{rep}$、$f_{ceo}$ 和 $\nu_n$ 的频率噪声功率谱。$\Gamma_\Delta(\omega)$ 为 n×$f_{rep}$ 的频率噪声功率谱与 $f_{ceo}$ 的频率噪声功率谱之间的复相干系数。该复相干系数的值在-2 至 2 之间，-2 对应 100%反相关，2 对应 100%正相关。图 5-8 曲线为由图 5-7 中 $n\times f_{rep}$、$f_{ceo}$ 和 $\nu_n$ 的频率噪声功率谱计算得到的复相干系数随傅里叶频率变化曲线。可以看出，在 1 kHz 至 100 kHz 频率范围内，$n\times f_{rep}$ 的

频率噪声功率谱与 $f_{ceo}$ 的频率噪声功率谱的复相干系数保持在-2 左右。这表明二者噪声功率谱存在着 100%的反相关的关系。这与文献[219]中的光学频率梳噪声理论相符合。并且在文献[219]中，Dolgovskiy 等人在商用光学频率梳（MenloSystems，FC1500）中也观察到这一现象。

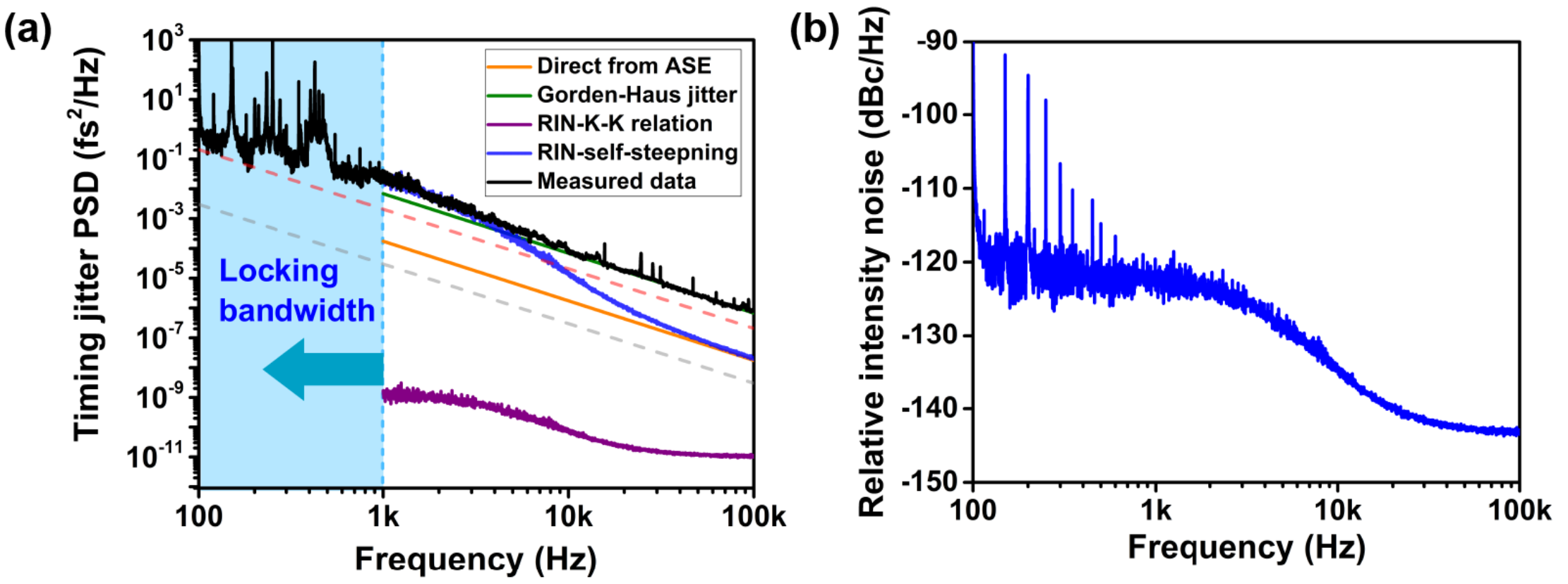

图 5-9　（a）激光器定时抖动功率谱密度图（黑色曲线）、放大自发辐射引入的定时抖动（橙色曲线）、Gorden-Haus 定时抖动（绿色曲线）、激光器强度噪声通过 Kramer-Krönig 关系引入的定时抖动（紫色曲线）和通过自陡峭效应引入的定时抖动（蓝色曲线）。红色虚线代表 100 MHz 干涉信号的强度噪声引入的测量基底。灰色虚线代表测量系统中光电探测器、电放大器、带通滤波器和混频器引入的测量噪声基底。（b）激光器的相对强度噪声曲线。

## 5.2.5 噪声来源分析

本小节将使用理论模型对测量到的 $n\times f_{rep}$、$f_{ceo}$ 和 $\nu_n$ 的噪声谱进行噪声来源分析。对噪声来源的理论计算可以对光学频率梳内的动力学过程进行更为深入的了解。

$f_{rep}$ 的噪声在时域表现为脉冲时域包络的抖动。与 $f_{ceo}$ 噪声和 $\nu_n$ 噪声不同的是，$f_{rep}$ 的噪声不对应光学频率梳内任何梳齿的噪声（$f_{ceo}$ 噪声为光学频率梳第 0 根梳齿的噪声，$\nu_n$ 噪声为第 $n$ 根梳齿的噪声）。将激光器 $n\times f_{rep}$ 的频率噪声谱转化为定时抖动功率谱，见图 5-9（a）。在 1 kHz 至 100 kHz 傅里叶频率范围内，定时抖动的功率谱的斜率近似为-20 dB/decade。为了对定时抖动的噪声谱进行噪声来源的分析，本文使用文献[32]中提出的理论模型对激光器中各种因素引入的定时抖动进行计算。在计算中，使用的激光器的参数为：重复频率为 78 MHz、腔内净色散为-0.0086 ps$^2$、腔内平均脉冲宽度 200 fs、饱和增益 0.693（假设激光器输出率为 50%）、腔内单脉冲能量 0.26 nJ、激光器输出光谱中心波长 1580 nm、增益介质上能级粒子寿命为 0.25 ms、增益介质的增益带宽为 4.8 THz。基于这些

参数，使用非线性薛定谔方程对脉冲在激光器腔内的动力学过程进行了模拟。采用分步傅里叶算法求解非线性薛定谔方程，得到飞秒脉冲在激光器腔内传输一周所积累的非线性相移为 2.2 π。本工作还对激光器输出脉冲的相对强度噪声进行了测量，结果见图 5-9（b）。从图中可以看出，激光器的相对强度噪声的功率谱在低频部分（100 Hz 至 1 kHz）保持平坦，在-120 dBc/Hz 量级。自 1 kHz 开始，随着傅里叶频率的增加，相对强度噪声的功率谱开始逐渐下降，至 100 kHz 处，变得再次平坦。100 kHz 处的强度噪声功率密度为-145 dBc/Hz。产生这一现象的原因主要是掺铒光纤中，上能级粒子的寿命为 0.25 ms，对应频率为 4 kHz。增益介质中的粒子相当于一个等效的低通滤波器，将其高于 4 kHz 的强度波动滤除。使用以上数据和第二章中的公式：（2-1）—（2-8）。可以计算出激光器中由放大自发辐射引入的定时抖动、激光器腔内色散耦合的 Gorden-Haus 定时抖动、激光器强度噪声通过 Kramers-Krönig 关系引入的定时抖动和激光器强度噪声通过自陡峭效应引入的定时抖动，见图 5-9。可以看出，在该光纤激光器中，高频部分（> 10 kHz）的定时抖动主要由激光器输出脉冲光谱的波动通过腔内色散引入的 Gorden-Haus 定时抖动，而低频部分（1~10 kHz）的定时抖动主要为激光器强度噪声通过自陡峭效应引入的定时抖动。整个噪声测量系统的测量基底主要体现在两方面：100 MHz 干涉信号的强度噪声与电学器件（光电探测器、电放大器、带通滤波器和混频器）的噪声基底。将测量到的这两项噪声用相同的鉴相斜率转化至定时抖动，绘制在图 5-9（a）中，即可得到的该测量的噪声基底。可以看出，两项噪声基底均小于激光器的定时抖动功率谱，这表明了本工作对于定时抖动功率谱的测量真实有效。

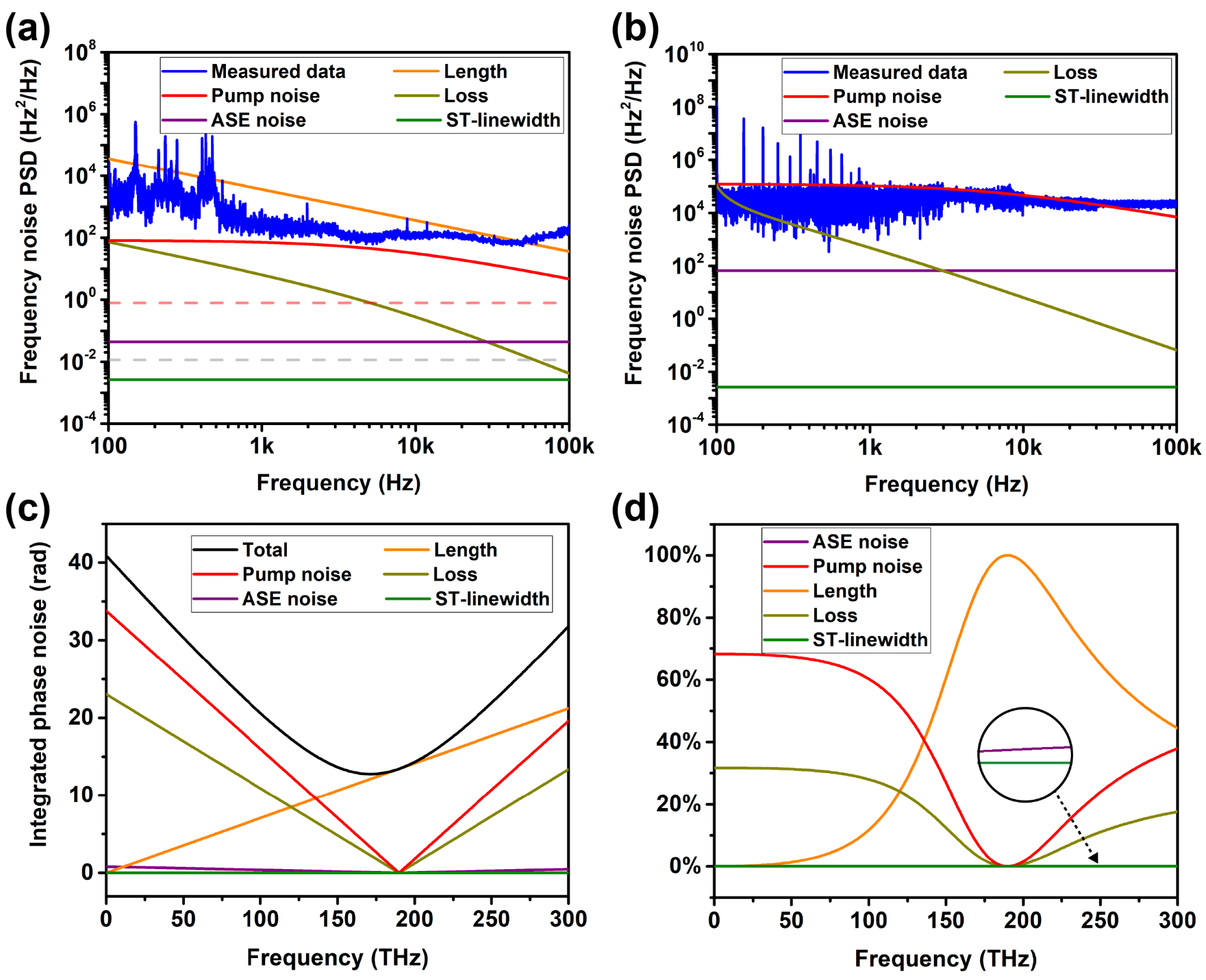

图 5-10　（a）$f_{ceo}$ 和（b）$\nu_n$ 的频率噪声功率谱。图中曲线分别代表：泵浦功率波动引入的频率噪声（红色曲线）、放大自发辐射引入的频率噪声（紫色曲线）、激光器腔长波动引入的频率噪声（橙色曲线）、激光器腔损引入的频率噪声（深黄色曲线）和 Shawlow-Townes 线宽极限（墨绿色曲线）。红色虚线代表 100 MHz 干涉信号的强度噪声引入的测量基底。灰色虚线代表测量系统中光电探测器、电放大器、带通滤波器和混频器引入的测量噪声基底。（c）不同梳齿中，每个噪声源引入梳齿噪声的均方根值。（d）不同噪声源引入的频率噪声占每个梳齿总噪声之比

对于 $f_{ceo}$ 噪声和 $\nu_n$ 噪声，文献[100]中提出的光学频率梳齿的“橡皮筋”模型很好地描绘了梳齿间的“呼吸”过程。某种特定的扰动，例如激光器腔长的波动、泵浦功率的波动等，会对整个光学频率梳中一个特定的频率梳齿影响最小，该频率被定义为“固定点”，随着其他梳齿的频率偏离“固定点”，受到该扰动对梳齿产生的噪声逐渐增大。不同的扰动源，在光学频率梳内的“固定点”不同，因此，多种扰动相结合，导致光学频率梳中存在着非常丰富的动力学过程。

使用第二章中的公式（2-10）—（2-14），可以对 $f_{ceo}$ 和 $\nu_n$ 的频率噪声谱的来源进行分析，见图 5-10（a）和（b）。除了前文中提到的参数，本实验还测量了泵浦二极管的相对强度噪声和泵浦功率变化导致脉冲重复频率的改变比率。泵浦二极管的相对强度噪声在 100 Hz 至 100 kHz 傅里叶频率内基本保持平坦，为-128

dBc/Hz。泵浦功率变化导致脉冲重复频率的改变比率 $B=(P_0\mathrm{d}f_r/f_r\mathrm{d}P)^2=1.21\times10^{-11}$。对于激光器腔长改变所产生的扰动，一般地，该扰动的“固定点”频率约为 0 THz。对于泵浦扰动、放大自发辐射引入的量子噪声和腔损引入的噪声，假设这些扰动的“固定点”频率约为光谱中心的光学频率。从图 5-10（a）和（b）中可以看出，对于 $f_{ceo}$ 和 $\nu_n$ 的频率噪声谱，腔损扰动、放大自发辐射引入的量子噪声和 Schawlow-Townes 线宽极限均对噪声功率谱的贡献非常小。对于 $\nu_n$ 的噪声，其噪声谱的主要构成为腔长的波动和泵浦二极管功率的波动。对于 $f_{ceo}$ 的噪声，其噪声谱则基本上完全由泵浦二极管功率的波动产生的。激光器腔长的波动对其的贡献十分低（$<10^{-11}$ Hz$^2$/Hz）。

进一步地，本工作对每一种噪声源（放大自发辐射、泵浦功率等）对每个光学频率梳齿（0 THz 至 300 THz）的影响计算出来。将得到的每个频率噪声谱转换为相位噪声谱后，在 100 Hz 至 100 kHz 范围进行积分，得到每个噪声源对每个光学频率梳齿引入的相位噪声的均方根值，将结果绘制在图 5-10（c）中。图 5-10（c）中黑色曲线为将各种噪声源相加后，每个光学频率梳齿的相位噪声均方根值。图 5-10（d）为不同噪声源在每个光学频率梳齿中所占比例。从图 5-10（c）中可以看出，整个光学频率梳范围内，噪声最低的，最“安静”的梳齿位于 174 THz 光频附近，与光谱中心大约相差 20 THz。对于 0 THz 附近的光学频率梳齿，梳齿的噪声主要由泵浦噪声和腔损引起。然而，腔长的波动的影响对这些梳齿的影响几乎消失。相反的，对于光频在 200 THz 附近的频率梳齿，梳齿的噪声主要由腔长的波动引入。任何腔长的波动都会在光频波段被“放大”。因此，在光学频率梳的锁定和光频分频系统中，通常用腔长控制来锁定 $f_{rep}$ 和 $\nu_n$ 噪声，用泵浦功率调制来锁定 $f_{ceo}$ 噪声。

## 5.3 锁相环串扰的研究

### 5.3.1 重复频率锁定与梳齿线宽锁定的串扰

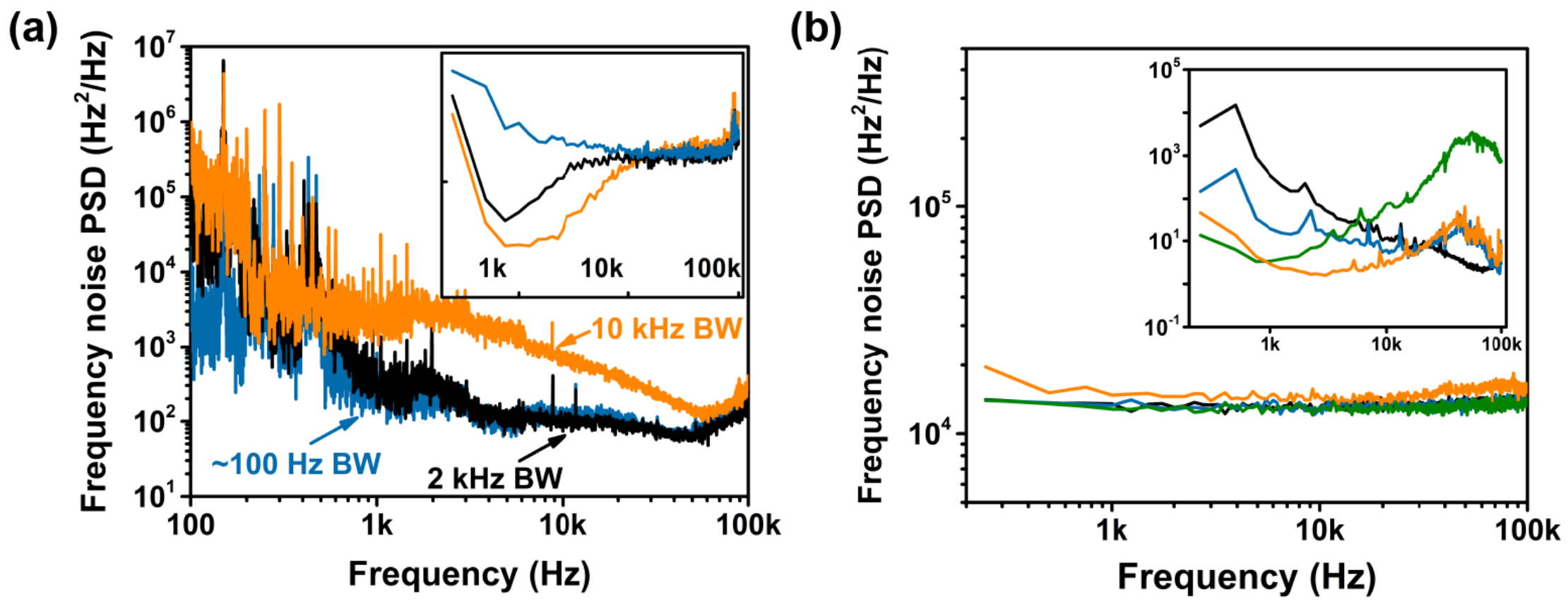

图 5-11　（a）在三种不同的 $f_{rep}$ 的锁定条件下，1540 nm 的梳齿噪声功率谱。插图为对应的锁定后的 $f_{rep}$ 噪声功率谱。（b）四种不同的梳齿锁定条件下，$n\times f_{rep}$ 的噪声功率谱。插图为对应的锁定后的梳齿噪声功率谱

本小节中，对 $f_{rep}$ 的稳定与 $\nu_n$ 的稳定之间的串扰进行了实验验证。首先，在 $f_{rep}$ 噪声被锁相环消除的条件下，测量 1540 nm 的梳齿噪声。不同 $f_{rep}$ 的锁定带宽下，1540 nm 的梳齿噪声见图 5-11（a）。可以看出，当 $f_{rep}$ 的锁定带宽小于 2 kHz 时，$f_{rep}$ 的锁定对梳齿线宽没有明显的影响。然而，当 $f_{rep}$ 的锁定带宽逐渐变高，接近 kHz 量级，达到 10 kHz 量级时，梳齿噪声谱的高频部分会引入额外的噪声，见图 5-11（a）中橙色曲线。这一梳齿噪声谱的恶化现象可以用上一节中的噪声来源分析进行解释。当腔内压电陶瓷仅仅补偿由环境中低频振动和声学振动引入的腔长波动时，光学频率梳中每个梳齿的噪声谱不会受到影响。然而，如果压电陶瓷的锁定带宽过大，压电陶瓷开始补偿由放大自发辐射引入的量子噪声时，压电陶瓷会对激光器腔长产生“过量的”调制。由于梳齿的噪声极易受到激光器腔长扰动的影响，因此，该调制会对光学频率梳的梳齿引入额外的频率噪声。

接下来，本工作在锁定了 $\nu_n$ 的条件下，对 $f_{rep}$ 的噪声谱进行了测量分析，结果见图 5-11（b）。可以看出，在四种不同的 $\nu_n$ 锁定条件下（低带宽锁定、高带宽锁定和有过量调制的锁定），$f_{rep}$ 的噪声谱并没有明显的区别。并且，$f_{rep}$ 的噪声谱在快速傅里叶变换分析仪上呈现出上下波动的现象。图 5-11（b）中的功率谱均为将待测量功率谱平均 500 次后的结果。从该图中，可以看出直接对 $\nu_n$ 的噪声锁定并不会对 $f_{rep}$ 噪声的消除产生任何作用。$\nu_n$ 的噪声主要都是由激光器腔

长波动引入的，因此直接锁定 $\nu_n$ 时，锁相环消除的便都是激光器腔长波动引入的噪声。至于其他噪声源例如泵浦源的噪声，放大自发辐射引入的噪声等，在 $\nu_n$ 的噪声谱中没有明显体现。而激光器 $f_{rep}$ 的噪声在高频部分的主要来源是放大自发辐射引入的噪声。所以，直接锁定 $\nu_n$ 时，$f_{rep}$ 的噪声基本上处于自由运转状态，$f_{rep}$ 的噪声不会得到消除。另外，本工作对比了 $\nu_n$ 被锁定和未被锁定时，$f_{rep}$ 的艾伦方差。在两种情况下，$f_{rep}$ 的艾伦方差均保持在相同的量级不变（1 s 门时间下的稳定度为 $10^{-9}$ 量级）。

## 5.3.2 重复频率锁定与载波-包络相位锁定的串扰

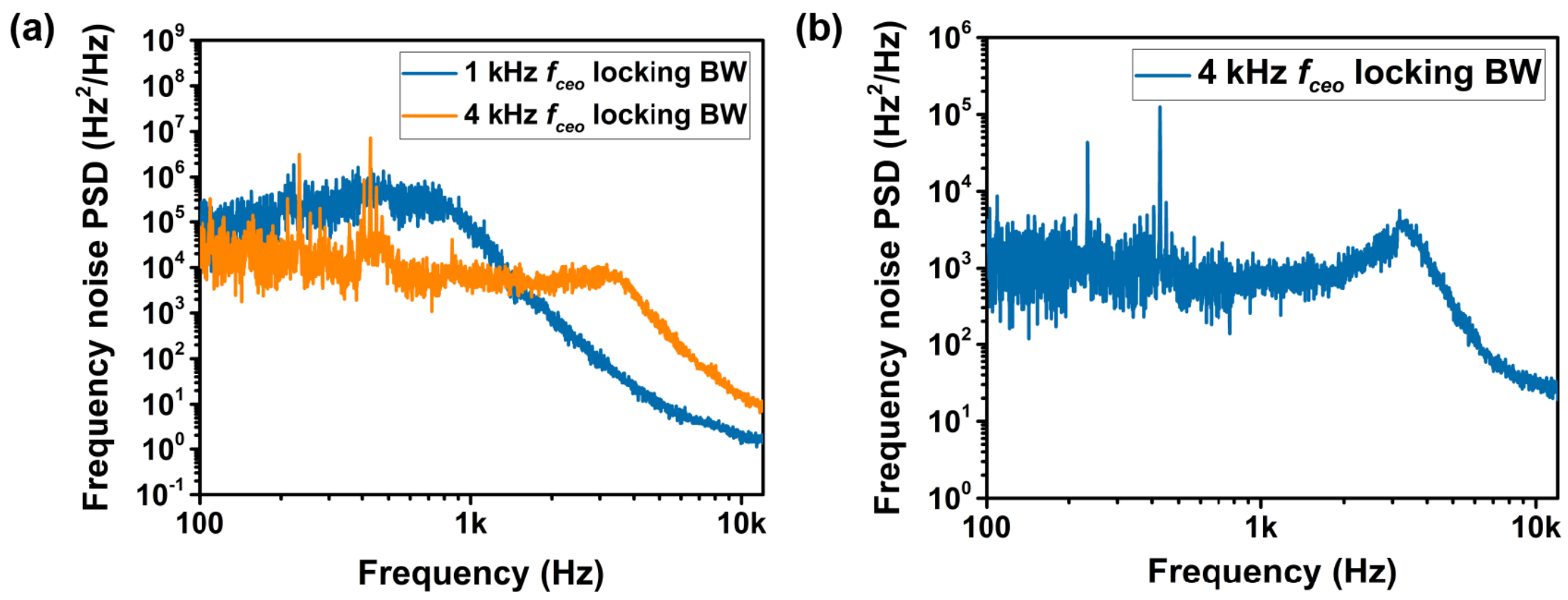

图 5-12　在 $f_{ceo}$ 被锁定的条件下，（a）1540 nm 的梳齿噪声功率谱（b）n×$f_{rep}$ 的噪声功率谱

最后，本文实验验证了 $f_{ceo}$ 的锁定对 1540 nm 的梳齿噪声功率谱和 $n\times f_{rep}$ 的噪声功率谱的影响。使用数字鉴相器对 $f_{ceo}$ 信号和参考射频信号进行鉴相，将得到的误差信号用高速比例积分微分伺服控制器进行锁定，可以实现 kHz 带宽的 $f_{ceo}$ 信号的锁定。

在 $f_{ceo}$ 被锁定后，$\nu_n$ 噪声和 $n\times f_{rep}$ 噪声的功率谱分别见图 5-12（a）和（b）。可以看到，在 $\nu_n$ 噪声和 $n\times f_{rep}$ 噪声的功率谱中，均有明显的隆起。$\nu_n$ 噪声和 $n\times f_{rep}$ 噪声的功率谱在高于 10 kHz 频率的部分仍然有很明显的波动，但是功率谱隆起的部分却十分稳定。泵浦源的噪声对 $\nu_n$ 噪声和 $n\times f_{rep}$ 噪声的影响均是不可忽略的，因此，在 $f_{ceo}$ 锁定过程中引入的泵浦调制会在 $\nu_n$ 噪声和 $n\times f_{rep}$ 噪声的功率谱中显现出来。相反地，当 $\nu_n$ 或 $n\times f_{rep}$ 被锁定时，$f_{ceo}$ 的噪声功率谱并没有明显的变化。同样的现象也被 Dolgovskiy 等人报道。需要指出的是，这个结论仅仅在激光器的重复频率是通过腔内压电陶瓷调节的情况下成立。当腔内使用电光调制器调节重复频率时，$f_{ceo}$ 和 $\nu_n$ 的锁定会发生串扰，继而 $f_{ceo}$ 的噪声特性会遭到恶化。这

是因为如果电光调制器中的晶体失谐会引起脉冲偏振态和激光器腔长的耦合，进一步的，会引入 $f_{ceo}$ 锁定的锁相环和 $\nu_n$ 锁定的锁相环之间的串扰。

## 5.4 本章小节

光学频率梳中的噪声测量是深入了解腔内噪声动力学过程和实现低噪声光学频率梳运转的前提。在本章中，基于非对称光纤延迟线干涉仪和自行搭建的 $f$-$2f$ 干涉仪，$f_{rep}$、$f_{ceo}$ 和 $\nu_n$ 的频率噪声谱得到了高精度的测量。并且噪声测量结果与 Paschotta 的理论模型和“固定点”模型相吻合。使用理论模型对光学频率中所有梳齿的噪声来源进行了分析，发现噪声最低的梳齿位于 174 THz 频率附近。噪声的功率谱分析揭示了 $n\times f_{rep}$ 噪声与 $f_{ceo}$ 噪声之间的反相关特性。反相关的强度与光学频率梳中每个梳齿的固有噪声直接相关。反相关特性的研究为光学频率梳的噪声性能提升提供了另一个独特的角度。

该系统验证了 $f_{rep}$、$f_{ceo}$ 和 $\nu_n$ 的锁相环环路之间的串扰。实验发现，如果 $f_{rep}$ 的锁定带宽过高，压电陶瓷会对 $\nu_n$ 引入额外的噪声。反之，$\nu_n$ 的锁定不会对 $f_{rep}$ 的噪声特性产生任何影响。另一方面，在 $f_{ceo}$ 的锁定过程中，泵浦电流调制会在 $n\times f_{rep}$ 和 $\nu_n$ 的噪声功率谱上产生明显的调制。特别的是，$f_{rep}$ 与 $\nu_n$ 的两锁相环之间的串扰从一个独特的角度揭示了压电陶瓷调节激光器重复频率的过程。该实验得到的结论为使用 Pound-Drever-Hall 技术、电子锁定技术以及光学平衡互相关技术锁定激光器的重复频率提供了理论指导。最后，该噪声测量与分析方法不仅能被用于飞秒激光器中，还可以用于其他新型光学频率梳，例如电光调制频率梳、微环谐振腔等光源的噪声动力学研究中。

# 第6章 两台独立运转掺镱光纤飞秒激光器的相干合成

在本论文的第三章与第四章中，实现了对飞秒激光器输出脉冲的定时抖动和载波-包络相位的高精度的功率谱测量及精密控制。在此基础上，本章将对两台掺镱光纤激光器进行持续时间为 1 小时的稳定的相干脉冲合成。首先，通过激光器腔内净色散优化，可以将激光腔内的量子噪声降至最低。随后，使用单非线性晶体的光学平衡互相关装置将两台激光器输出的飞秒脉冲进行同步，得到的两脉冲的相对剩余定时抖动为 380 as（从 100 Hz 到 10 MHz 积分）。最后，通过平衡探测法探测两脉冲的相对载波-包络相位，并通过腔外声光频移器将其锁定为零。剩余的相对载波-包络噪声为 375 mrad（从 100 Hz 到 2 MHz 积分）。相干合成后的脉冲有着良好的光谱干涉和光斑干涉特性。

## 6.1 相干合成的实验装置图

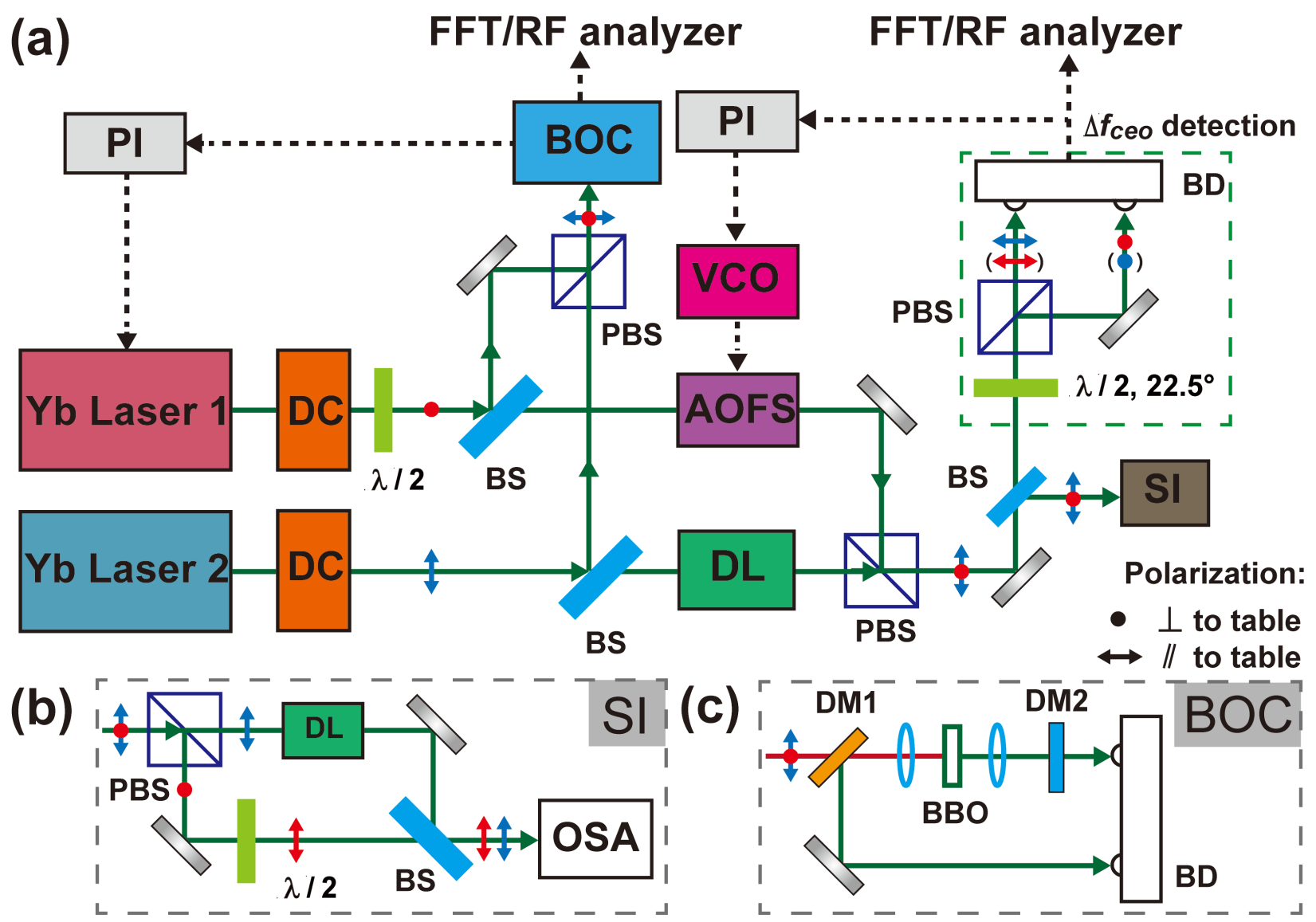

图 6-1　相干合成系统实验装置图。(a) 实验装置　(b) 光谱干涉系统　(c) 光学平衡互相关系统。AOFS，声光频移器；BOC，光学平衡互相关系统；BD，平衡探测器；BS，50/50 分束器；DC，腔外色散补偿器；DL，可调时间延迟；DM，双色镜；OSA，光谱分析仪；PBS，偏振分束器；PI，比例积分伺服控制器；SI，光谱干涉系统；VCO，压控振荡器

相干合成系统的实验装置图见图 6-1。本工作使用两台结构相似的基于 NPR

旋转锁模的掺镱光纤激光器实现相干合成。激光器工作在近零色散域，重复频率为 157 MHz。激光器的结构在 3.2 节中已详细介绍。由于两台激光器工作在近零色散域，因此由激光器光谱中心波动通过腔内非零色散产生的 Gordon-Haus 定时抖动被降至最低。一对啁啾镜分别在两台激光器腔外提供负色散，将待合成的两脉冲压缩至傅里叶变换极限。激光器 1 输出的脉冲的偏振态经过一个半波片后，偏振态被旋转 90 度。被压缩至变换极限的两脉冲分别被两个 50/50 分束器等功率地分为两部分。分别被两分束器反射的两脉冲在一个偏振合束器上合束之后，进入平衡光学互相关系统。平衡光学互相关系统的结构示意图见图 6-1（c），其结构与 3.2.1 节中的环内平衡光学互相关结构相同。平衡光学互相关系统输出与两台激光器重复频率差相关的鉴相曲线。将该误差信号经过比例积分伺服控制器后，对伺服控制器输出的纠正信号进行放大，加载到激光器 1 腔内的压电陶瓷上，以实现两台激光器重复频率的高速锁定。激光器的重复频率锁定与时间同步在第三章中已经详细介绍。另一方面，来自激光器 1 的，从 50/50 分束器透射的脉冲经过一个声光频移器（Gooch & Housego，33080-16-.7-I-TB）。激光器 2 输出的，从另一 50/50 分束器透射的脉冲经过一个可调时间延迟后，与经过声光频移器后的脉冲在一个偏振分束器上合束。合束后的两脉冲再次被一个 50/50 分束器分束。其中 50%能量的脉冲进入相对载波-包络相位探测系统，另外 50%能量的脉冲进入光谱干涉系统。相对载波-包络相位探测系统的结构与 4.2.2 小节中的结构相同。相对载波-包络相位探测系统输出的误差信号经过比例积分伺服控制器后，将伺服控制器输出的纠正信号输入至压控振荡器，压控振荡器的输出驱动声光频移器，完成锁相环的闭环，实现两台激光器相对载波-包络相位的锁定。

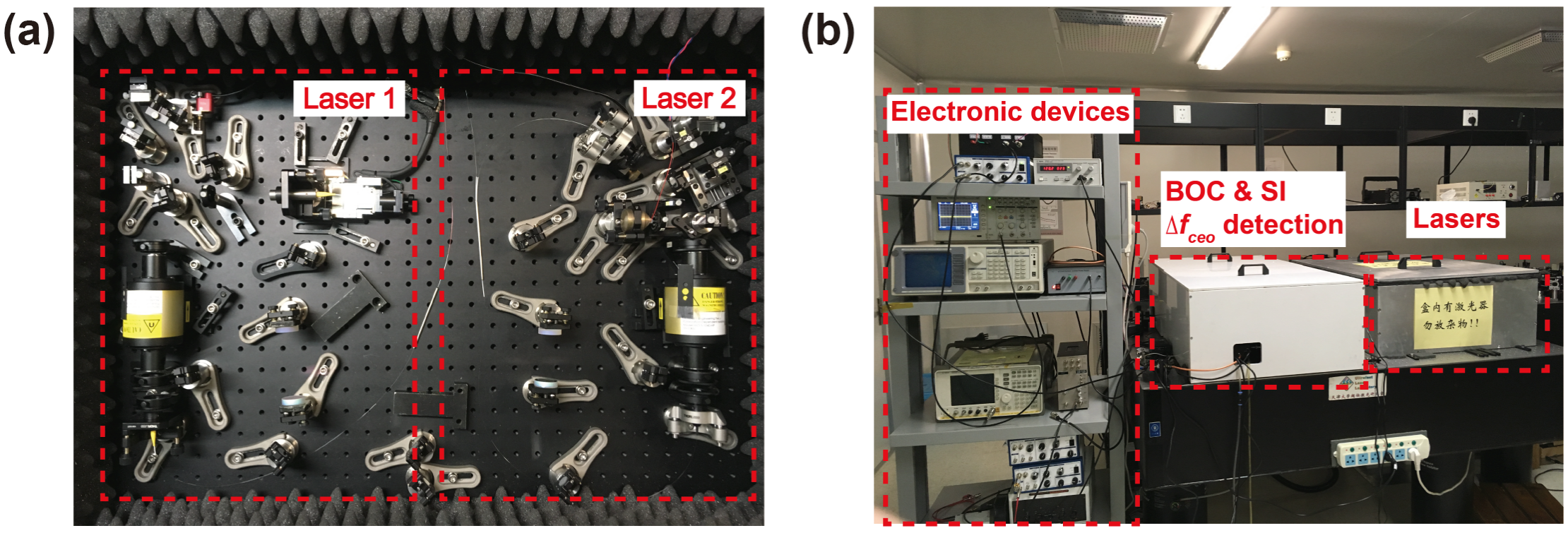

图 6-2　相干合成系统的环境噪声隔离。（a）对于两台激光器的隔离　（b）整个合成系统的示意图

特别指出的是，对整个光学系统的环境噪声隔离是实现长时间稳定的相干合成的重要前提。本工作对待合成的两台飞秒激光器进行了环境噪声隔离，如 3.3

节中介绍的，搭建两台激光器的面包板下方有多层橡胶、铅板和隔音海绵，两台激光器的四周以及顶部由一个内部粘了隔音海绵的铝制大盒子包裹。平衡光学互相关系统、相对载波-包络相位探测系统和光谱干涉系统均搭建在同一块面包板上，被一个聚合物材料制成的盒子罩住。其他的电子设备如比例积分伺服控制器、压电陶瓷的放大器、平衡探测器的电源、压控振荡器、稳压电源和测量设备如快速傅里叶变换分析仪和频谱分析仪均放在同一个仪器架上。声光频移器的水冷系统置于隔震光学平台底部。整个系统的装置图见图 6-2（b）。

在整个相干合成系统的搭建过程中，有如下注意事项：

一、关于声光调制器的晶体选择：晶体的材料应当选择石英，而不是 $TeO_2$。原因是后者的二阶色散较大，不宜应用于飞秒脉冲的频率调制。

二、声光调制器的调制带宽与声波在光斑中的渡越时间有关，为了实现高带宽的载波-包络相位锁定，激光需要聚焦入射进入声光调制晶体。使用将两个焦距为 90 mm 的聚焦透镜将激光先聚焦后再准直。将声光调制器置于透镜聚焦焦点附近，实现高带宽的频率调制。应当注意，搭建过程中需保证激光光斑全部通过晶体，即晶体不切光。

三、声光调制器运转后，将其置于聚焦焦点处，旋转晶体角度，在适当角度范围内，可以看到零级衍射光与 1 级衍射光。这时，固定声光调制器的支架。使用感光卡遮挡住零级衍射光，测量 1 级衍射光的功率。调节声光调制器支架水平方向的角度，优化角度至 1 级衍射光的效率最大。需要指出的是，经过声光调制器后，零级衍射光与 1 级衍射光均无法保持完美的高斯光斑。并且，由于零级衍射光与 1 级衍射光均通过同一个透镜进行准直，无法保证每个光斑都严格经过透镜中心，因此，两个衍射光的波前非常容易发生畸变。

四、声光调制器在运转过程中，晶体发热严重，需要冷却。不建议使用风扇冷却，因为风扇转动会产生额外的机械噪声。因此本工作中，在声光调制器的晶体下方设计了一个铜制的循环水冷却系统，可以非常有效地对晶体进行降温冷却。使用一个水泵将去离子水抽运至铜制冷却系统中，由于水泵会产生一定的声学噪声，需要将其置于隔震台底部，尽量保证与系统隔离。

五、在相对载波-包络相位探测系统搭建过程中，两脉冲的时域重合和光斑重合是非常重要的。对于光斑重合，建议使用两个光学小孔，将一个小孔置于偏振合束器之后的近端，另一个小孔置于偏振合束器之后的远端，两小孔之间的距离越大越好。同时使得两台激光器输出的脉冲通过两个小孔，即可实现较为理想的光斑重合。

六、关于两脉冲时域重合的调节：为了保证能够探测到两脉冲的相对载波-包络相位，需要调节延迟线的长度，以保证两脉冲在空间相干长度之内。理论上

讲，使用如下方法均可以寻找两光学脉冲的零延迟：

1、时域光电探测法。锁定两激光器重复频率。同时使用带宽在 GHz 量级的高速光电二极管和高速采样示波器对合束后的脉冲进行探测，可以粗略地确定两脉冲时域是否重合，其精度在厘米量级的空间长度。

2、频域光电探测法。锁定两激光器重复频率。使用带有放大的光电二极管和频谱分析仪对合束后信号的相对载波-包络频率信号进行探测。在一定范围内扫描光学延迟的位置，直到在频谱分析仪上可以观察到相对载波-包络频率信号，该信号的频率范围在 1/2 重复频率范围内。由于在刚开始寻找零延迟的时候，光斑重合的调节可能不是很理想，相对载波-包络频率信号的强度会很弱，因此建议将频谱分析仪的显示调整为线性坐标，更易于弱信号的寻找。

3、环外光学互相关法。不锁定两激光器重复频率。将两脉冲聚焦至一个 BBO 晶体上，通过和频过程产生两脉冲的光学互相关。同时，用示波器探测环外与环内光学互相关的曲线，调节光学延迟线，使得两互相关曲线在时域上没有延迟，这时的位置便是脉冲零延迟的位置。使用双光子吸收的光电探测器代替 BBO 晶体，同样可以产生光学互相关信号，用于脉冲零延迟的调节。

七、相对载波-包络频率偏移信号的锁定。不同于第四章中的载波-包络相位噪声测量，在相干合成的实验中，需要将两台激光器的相对载波-包络相位锁定至 0 Hz，即两台激光器的相位是完全同步的。该实验中使用的比例积分伺服系统为 Newfocus 公司的 LB1005。首先在 LB1005 中，将锁相环的比例增益调节至零，调节伺服控制器的输出偏置，使得 $\Delta f_{ceo}$ 信号的频率在 0 Hz 附近。将伺服控制器的积分拐点设置在 30 kHz 或者 100 kHz 档位，伺服控制器的开关拨至 LFGL 档位，随后逐渐增加锁相环的比例增益。在此过程中，不断地调节伺服控制器的输出偏置电压，使得 $\Delta f_{ceo}$ 信号一直在 0 Hz 附近。当增益增加到一定程度时，可以发现 $\Delta f_{ceo}$ 信号在主动地向 0 Hz 靠拢，并且 100 kHz 频率附近会产生鼓包。这时将伺服控制器的开关推至 Lock on 档位，即可实现锁相环的锁定。如果无法实现锁相，可以尝试将输入信号从 A 通道换至-B 通道，将输入信号反相，再尝试锁定。

八、更为理想的设计为所有光学部分均在一整块大面包板上。但是由于相干合成系统的光路十分复杂。本工作中并未采用这一理想化设计。

## 6.2 相干合成的实验结果

### 6.2.1 锁定后的功率谱

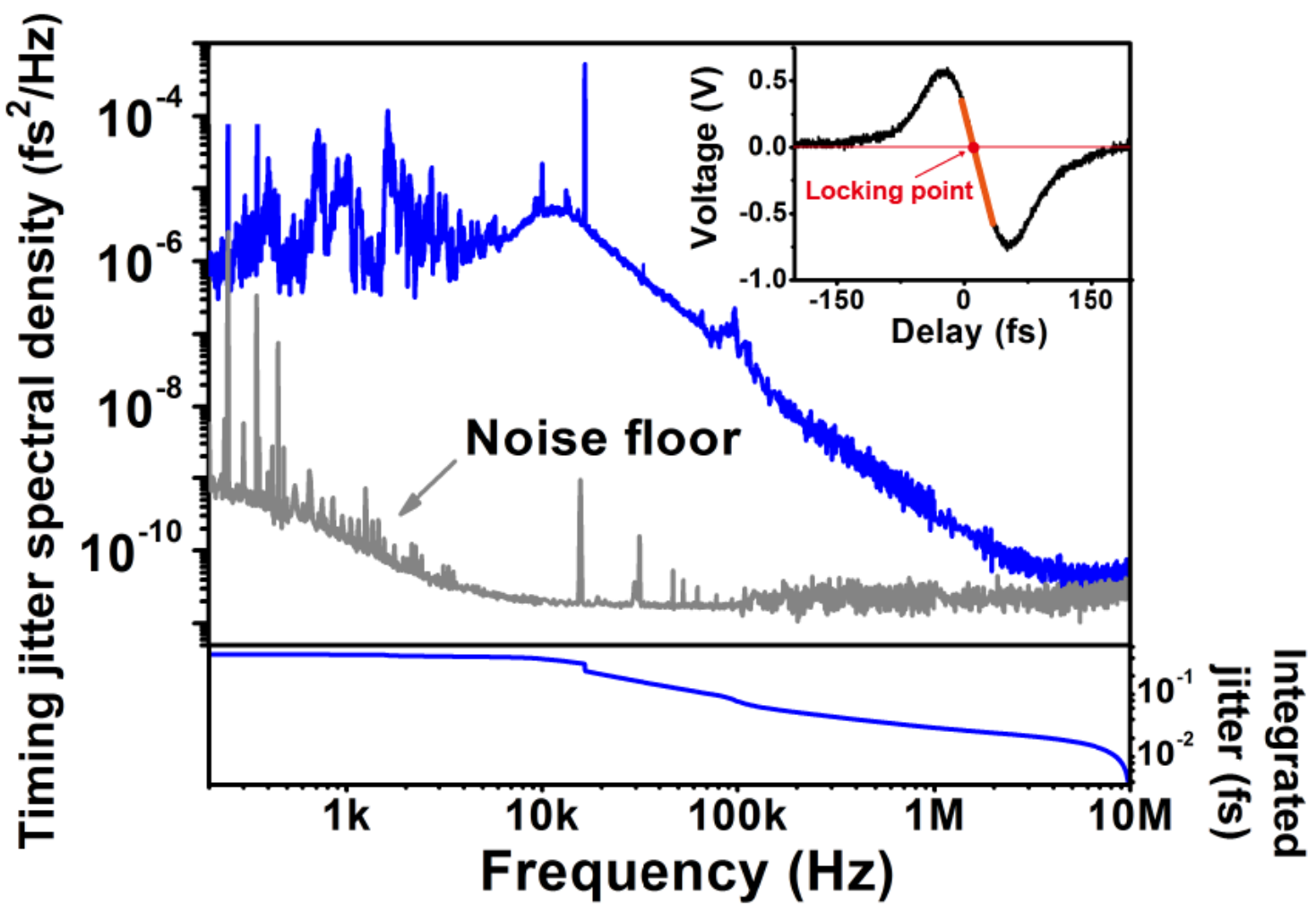

图 6-3　重复频率锁定后的剩余定时抖动功率谱密度图

使用图 6-3 插图中的 S 型曲线的线性部分作为两台激光器重复频率差的误差信号，鉴相斜率为 25.4 mV/fs。将误差信号锁定在曲线中间的红色点处，使两台激光器具有相同的重复频率。重复频率锁定后的剩余定时抖动功率谱密度见图 6-3。从该曲线中可以看出，重复频率的锁定带宽为 10.5 kHz。锁定带宽主要受限于粘有反射镜的压电陶瓷的响应带宽。对于高于锁定带宽的功率谱，功率谱曲线以 20 dB/decade 的斜率下降。这表明，该部分定时抖动的主要来源为激光器中增益介质的放大自发辐射引入的量子噪声。将功率谱密度曲线在 100 Hz 至 10 MHz 范围内积分后，得到的剩余定时抖动的均方根值为 380 as。该定时抖动值远小于 1.04 μm 波段的 1/10 个光学周期，是实现有效相干合成的必要条件。需要指出的是，在文献[214]中，Cox 等人在使用腔内压电陶瓷锁定重复频率的同时，在激光器腔外使用了一个电光调制器来进一步锁定两激光器输出的脉冲的相位延迟。使用电光调制器的锁定带宽可以达到 1 MHz 以上，因此，放大自发辐射引入的量子噪声可以被进一步得到抑制。相比之下，本工作使用腔内压电陶瓷进行主动锁定，通过优化激光器腔的净色散被动地降低高频部分的量子噪声。在简化了实验装置的同时，达到了低于[214]中结果的剩余定时抖动。

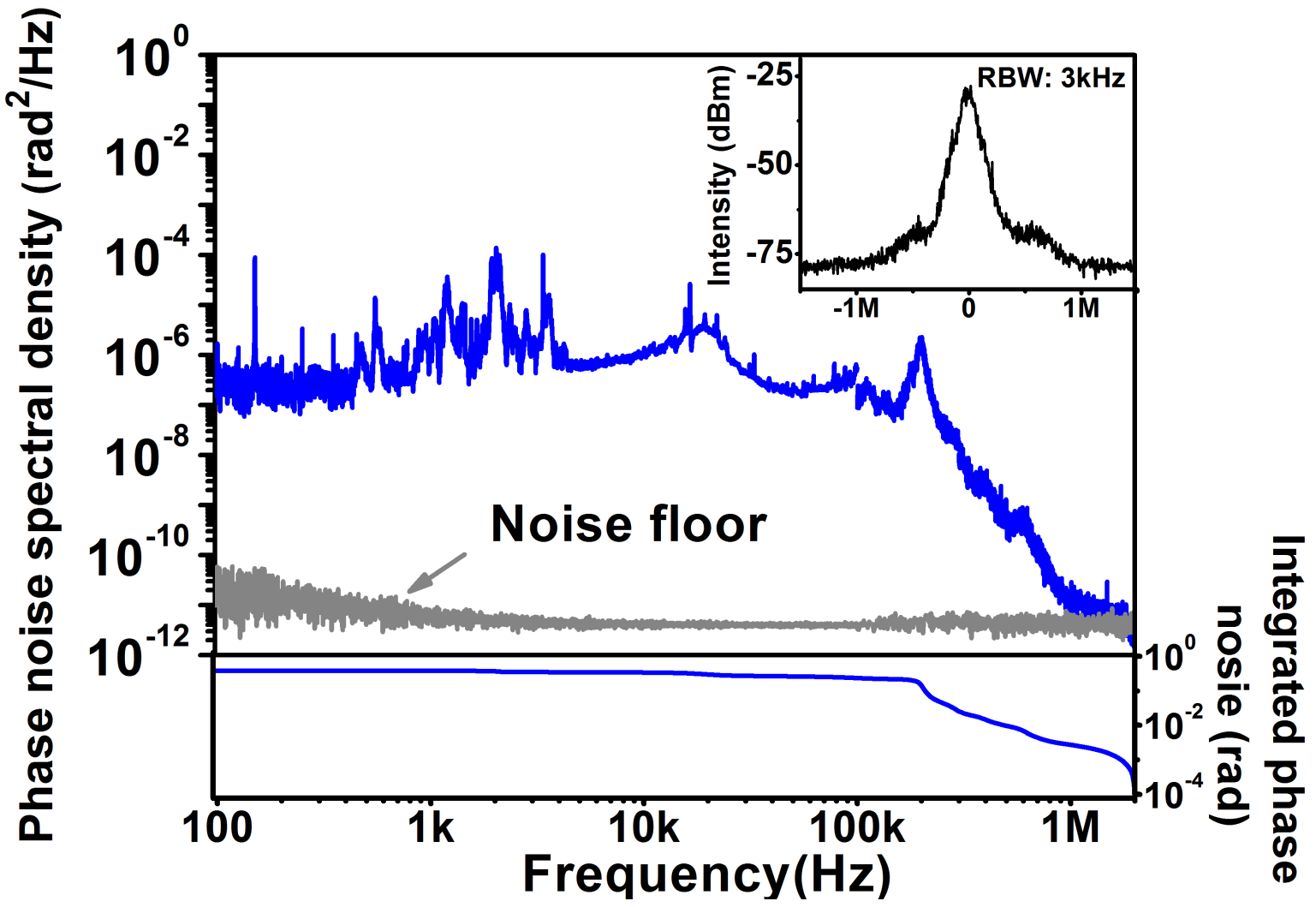

图 6-4　相对载波-包络相位锁定后的相位噪声功率谱密度图

在实现了两台激光器的重复频率锁定后，本工作使用相对载波-包络相位探测系统对两台激光器的 $\Delta f_{ceo}$ 信号进行探测。调节光学延迟，使得 $\Delta f_{ceo}$ 信号的信噪比达到最大。$\Delta f_{ceo}$ 信号的信噪比在 3 kHz 分辨率带宽下为 50 dB。对应的 3 dB 带宽为 68 kHz，见图 6-4 中插图。使用声光频移器对一台激光器输出的脉冲进行频移。声光频移器的通光孔径为 0.7 mm。激光经聚焦后入射进入声光调制器，聚焦后的光斑直径约为 113 μm，对应声波在光斑内的渡越时间小于 13 ns，确保了高带宽的 $\Delta f_{ceo}$ 信号的相位锁定。比例积分伺服控制器接收到相对载波-包络相位探测系统输出的误差信号后，其输出的纠正信号加载到一个压控振荡器上，压控振荡器产生可调的射频信号，该射频信号驱动声光频移器，实现 $\Delta f_{ceo}$ 信号的相位锁定。锁定后的 $\Delta f_{ceo}$ 信号的剩余相位噪声功率谱见图 6-4。锁定带宽为 110 kHz。在 100 Hz 至 2 MHz 频率范围内积分，剩余相位噪声的均方根值为 375 mrad。需要指出的是，如果使用比例积分微分伺服系统进行 $\Delta f_{ceo}$ 信号的相位锁定，高频部分的相位噪声会得到更好地抑制。剩余相位噪声的均方根值也会更低。

### 6.2.2 相干合成后的光谱干涉图

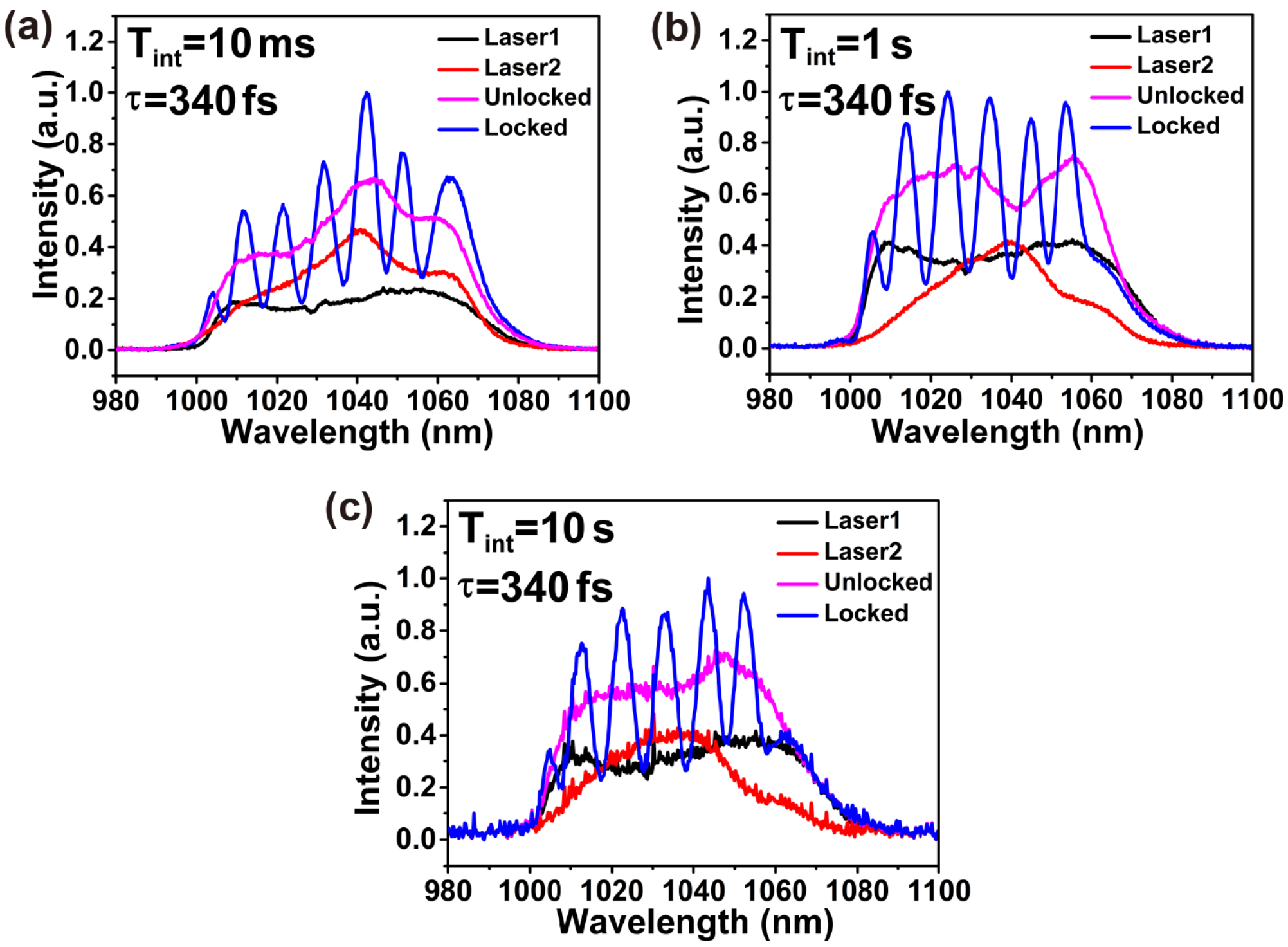

图 6-5　相干合成后的光谱干涉图。积分时间分别为（a）10 ms（b）1 s（c）10 s 的光谱图

相干合成后的两台激光器的频域相干性由光谱干涉系统进行验证。光谱干涉系统的主要结构为图 6-1（b）中的 Mach-Zehnder 干涉仪。待测量的两脉冲的偏振态需要完全相同才可以产生光谱干涉。偏振态互相垂直的两脉冲经过一个偏振分束器分束后，一路脉冲经过一个半波片将偏振态旋转 90 度。最后由一个偏振无关的分束器进行合束。合束后的脉冲用光谱仪（OceanOptics，HR4000CG-UV-NIR）进行光谱测量。干涉仪的一臂中加入了可调光学延迟来对两脉冲的时间延迟进行调节。将两脉冲的时域延迟调节至 340 fs 后，分别对激光器 1、激光器 2、相位未锁定的合成光谱和相位锁定后的合成光谱进行了测量。图 6-5 给出了光谱仪积分时间分别为 10 ms、1 s 和 10 s 时的干涉光谱测量结果。从图中可以看出，当两台激光器的重复频率和相对载波-包络相位未锁定时，由于两脉冲的相对相位有很大的波动，用光谱仪测量得到的光谱没有干涉条纹。当重复频率和相对载波-包络相位都锁定后，两脉冲的相对相位波动被很大程度地降低了。两脉冲的包络和相位均保持了很好地同步，因此两脉冲的光谱有很明显的干涉条纹。通过改变 Mach-Zehnder 干涉仪中的延迟，可以调节两脉冲的时域

延迟。两脉冲的时域延迟越大，光谱干涉的条纹越密集。反之，两脉冲的时域延迟越小，光谱干涉的条纹越稀疏。另一方面，干涉条纹的对比度能够反映相位锁定后的两脉冲的相干性。干涉条纹对比度由如下公式定义：

$$I(\omega) = I_1(\omega) + I_2(\omega) + 2C[I_1(\omega)I_2(\omega)]^{1/2} \bullet \cos(\omega\tau + \alpha) \qquad (6\text{-}1)$$

其中，$C$ 为对比度，$I(\omega)$、$I_1(\omega)$和 $I_2(\omega)$分别为合成后的光谱、激光器 1 的光谱和激光器 2 的光谱，$\omega$ 为角频率，$\tau$ 为两脉冲的时间间隔，$\alpha$ 为相位常数。对比度越高，则相位短期相干性越好。如果增加光谱仪的积分时间，干涉条纹的对比度有所下降，则说明在长时间尺度范围内，两脉冲的相对相位有漂移。本实验中，10 ms、1 s 和 10 s 积分时间内的光谱干涉条纹的对比度均在 58%左右。干涉条纹的对比度不随光谱仪积分时间的增加而下降，这说明了该实验的相位锁定结果有着很好的长时间稳定性。

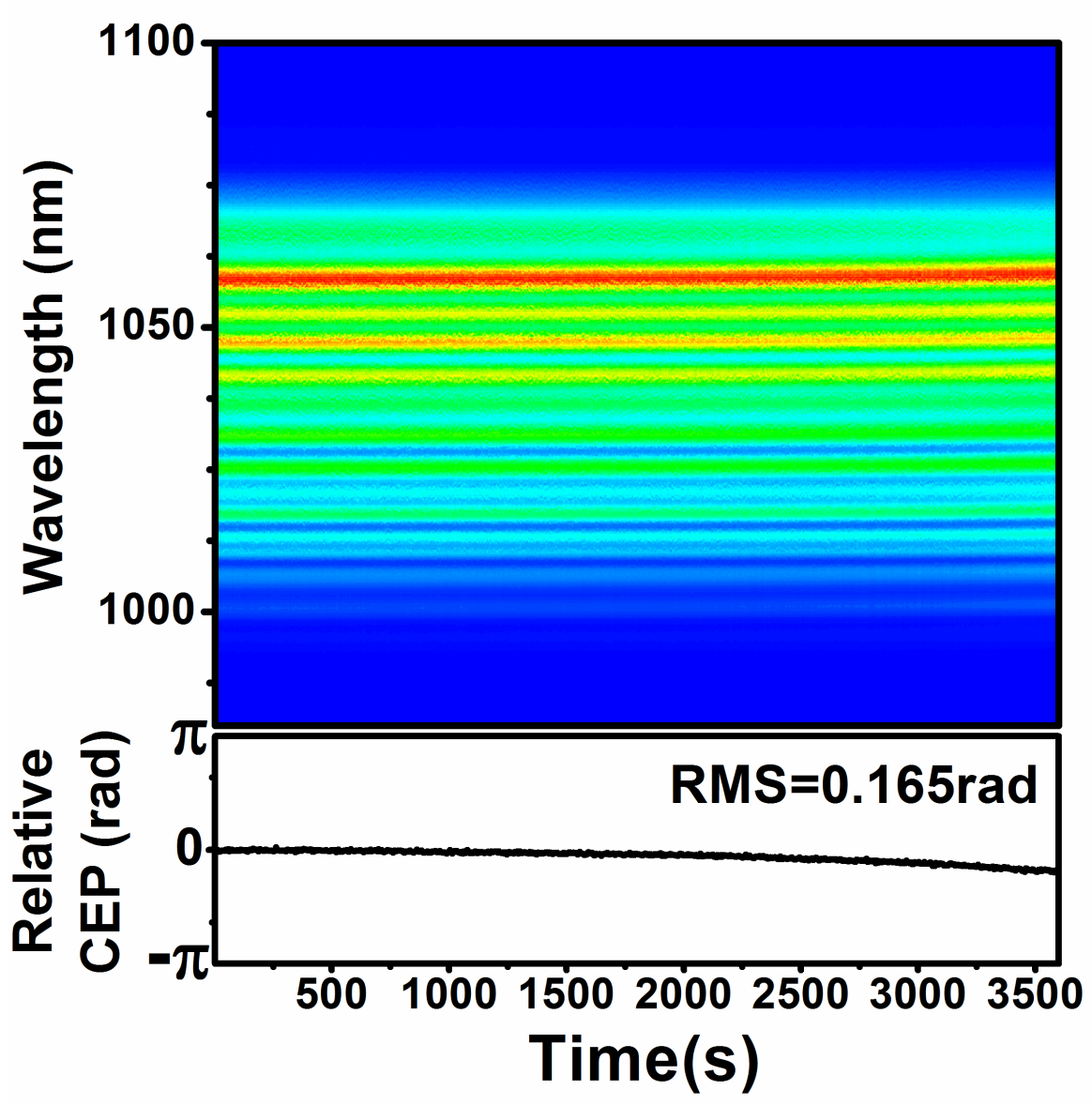

图 6-6　光谱干涉条纹在 60 分钟时间内的漂移

图 6-6 给出了 60 分钟内，两脉冲在相位锁定条件下的光谱干涉条纹变化。使用光谱仪（Yokogawa，AQ6370B）对合成后的光谱每 5 s 进行一次测量，得到了 720 个光谱干涉图。两脉冲的时域延迟被调至 599 fs。60 分钟内，两脉冲的相对载波-包络相位漂移的均方根值为 165 mrad。由于光谱干涉系统并没有进行特殊的温度控制，因此，两脉冲相对相位的长期漂移可能源自由于环境引入的干涉仪的漂移。

### 6.2.3 相干合成后的光斑干涉图

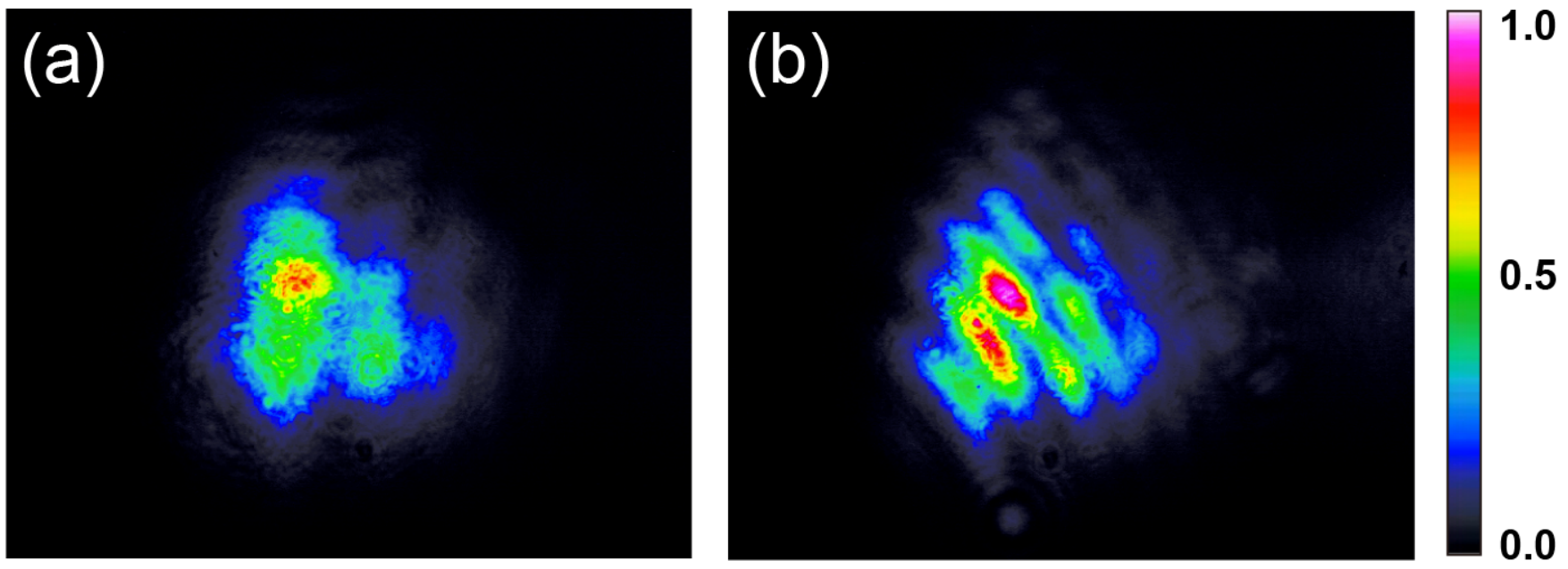

图 6-7　合成后两脉冲的远场光斑图。(a) 相位锁定前。(b) 相位锁定后

为了进一步验证合成后两脉冲的相干性，使用光束质量分析仪（LaserCam Hr System，Coherent）对两脉冲在合成后的远场光斑进行了测量，见图 6-7。可以看出，两脉冲的相对相位锁定后，相比于未锁定时，光斑有着明显的干涉条纹。当两脉冲的时域间隔被改变时，可以观测到光斑干涉条纹的移动。另外，可以看出，经过声光调制器调制后的激光光斑与高斯型相差甚远。声光调制器对零级和 1 级衍射光波前的恶化是无法避免的。若使用基于光纤输入和输出的声光调制器，该问题会得到解决。但伴随而来的是声光调制器输出效率的下降。

## 6.3 本章小节

本章主要介绍了两台独立的掺镱光纤飞秒激光器的长期稳定相干脉冲合成。通过拥有阿秒分辨率的平衡光学互相关系统锁相，在实现了脉冲的高精度同步之后，通过零差检测提取两脉冲的相对的载波-包络相位，并通过腔外声光调制器的反馈控制将其锁定至零。合成后，两脉冲的相对剩余定时抖动的积分值为 380 as，剩余的相对载波-包络噪声的积分值为 375 mrad。使用 Mach-Zehnder 干涉仪对相干合成后的脉冲的相干性进行评估，合成后的光谱干涉的对比度为 58%。60 分钟内，两脉冲的相对载波-包络相位漂移的均方根值为 165 mrad。并且两激光器的光斑有明显的干涉条纹。与其他光纤飞秒激光器之间的相干脉冲合成的最新研究相比，通过优化光纤激光器腔内色散实现的量子噪声降低使锁相系统的结构更加简单。独立光纤激光器之间的长期稳定相干合成是亚单周期脉冲合成、光学频率梳远距离传输等应用的基础。

# 第7章 总结和展望

## 7.1 总结

本文围绕飞秒激光光学频率梳的重复频率与载波-包络相位的精密控制进行了研究：系统研究了飞秒激光光学频率梳的定时抖动特性，实现了两台飞秒激光光学频率梳的时间同步；系统研究了飞秒激光光学频率梳的载波-包络相位噪声特性，实现了飞秒激光光学频率梳的载波-包络相位噪声的高精度测量；系统研究了两台独立的掺镱光纤激光器的长期稳定相干脉冲合成；最后，系统研究了光学频率梳中 $f_{rep}$、$f_{ceo}$ 和 $\nu_n$ 的频率噪声功率谱，实验验证了 $f_{rep}$、$f_{ceo}$ 和 $\nu_n$ 的锁相环环路之间的串扰。这些成果从理论和实验方面为低噪声的光学频率梳运转提供了基础。

本文的工作具体总结如下：

1、对飞秒激光光学频率梳输出脉冲的定时抖动、载波-包络相位噪声和梳齿噪声的理论进行了系统研究，分析了放大自发辐射、泵浦功率波动、环境引起的腔长波动和腔损耗等因素对光学频率梳中不同种类噪声的影响。这是实现稳定光学频率梳运转的重要理论基础。详细研究了功率谱分析法和方差分析法，两种方法分别从频域和时域的角度对相位锁定后的光学频率梳进行稳定性评价。特别地，分析了不同种类的方差分析法，例如标准方差、艾伦方差、交叠艾伦方差、改进的艾伦方差、时间方差和 Hadamard 方差适用于不同的实验系统。功率谱分析和方差分析可以从不同角度反映出不同类型的噪声，例如白相位调制噪声、闪烁相位调制噪声、白频率调制噪声等对待测量频率源的影响。研究了在光电探测过程中，光电探测器产生的热噪声和散粒噪声对噪声探测的信噪比的影响。飞秒激光器的理论建模与稳定性评价方法均为实现低噪声光学频率梳运转的重要基础。

2、系统研究了两台独立飞秒激光光学频率梳的定时抖动测量与时间同步。在优化了激光器量子噪声引入的定时抖动特性后，使用光学平衡互相关技术首次实现了掺镱光纤激光器的连续 5 天（120 小时）的阿秒量级的精密时间同步。实现时间同步后，环内光学互相关系统输出的剩余时间误差信号的均方根值为 103 as，环外光学互相关系统输出的剩余时间误差信号的均方根值为 733 as。经分析，环外定时抖动结果的恶化主要是由温度和湿度变化导致的光路失谐造成的。首次以平均约 $10^5$ s 的门时间评估了飞秒激光器之间的时间同步的长期稳定性。环内交叠的艾伦方差的 $1.31\times10^5$ s 稳定度为 $8.76\times10^{-22}$。环外交叠的艾伦方差的 $1.31\times10^5$ s 稳定度为 $1.36\times10^{-20}$。从方差分析可得知，环内的剩余定时抖动主要为白相位噪声，环外的剩余定时抖动受到环境因素的影响，是无规律的长期漂移。环内噪声功率谱在 10 kHz 至 10 MHz 频率范围的积分值为 430 as。研究了飞秒脉冲在光纤传播过程中，光纤引入的额外定时抖动的测量。使用低噪声的重复频率为 500 MHz 的 1550 nm 波段的固体激光器作为光源，放大后的脉冲经过两段 5 米长的单模光纤，用光学外差探测技术对单模光纤引入的定时抖动进行了测量。得到了积分值为 64.7 as 的定时抖动结果。分别使用电学降频法和互相关算法对光学外差探测法进行改进，改进后定时抖动的测量仍受限于系统中仪器的噪声。飞秒激光器的高精度长时间稳定时间同步是迈向新兴的亚周期光脉冲合成、光学频率基准的传输和基于 X 射线自由电子激光器的高时间分辨技术的关键一步。光纤引入的定时抖动的高精度测量是实现阿秒量级时间、光学频率基准远距离传输的重要理论和实验基础。

3、研究了飞秒激光光学频率梳中的载波-包络相位噪声的高精度测量及噪声分析。使用光学外差探测法首次实现了超大范围、超高精度的载波-包络相位噪声的功率谱测量。该测量方法无需使用 *f-2f* 干涉仪，摆脱了 *f-2f* 干涉仪中的非线性扩谱步骤，很大程度地提高了噪声测量的信噪比。该方法可以实现傅立叶频率范围在 5 mHz 至 8 MHz 内的，动态范围高于 270 dB 的载波-包络相位噪声功率谱测量。相位噪声测量基底低于 1 μrad/√Hz。对自由运转的载波-包络相位噪声特性进行了研究，傅里叶频率低于 20 Hz 的部分，该噪声主要由环境扰动引入；20 Hz 至 35 kHz 的部分主要为泵浦激光器的波动引入的相位噪声；高于 35 kHz 的部分为放大自发辐射引入的量子噪声。Hadamard 方差分析也同样发现了泵浦波动引入的噪声和量子噪声的分界点。使用 Kendall 互相关分析对不同飞秒激光器的载波-包络相位噪声进行了分析。发现在不同激光器中，载波-包络相位对振幅-相位调制噪声影响是不同的。得到了基于非线性放大环形镜锁模的掺铒光纤飞秒激光器受到该因素的影响最小的结论。首次实现了孤子分子对中的两个光学孤子的相对相位噪声的测量。将两个光学脉冲的相对相位改变转化为这两个波长相对强度的改变，使用平衡探测器探测，就可以得到光孤子的相对相位噪声的功率谱密度。从 10 Hz 至 10 MHz 积分，相对相位噪声的积分值为 3.5 mrad。相位噪声测量的精度为 $10^{-13}$ rad²/Hz。使用 $\beta$-line 分析法，估算两个光学孤子的相对线宽仅为 μHz 量级。

4、研究了基于非对称光纤延迟线干涉仪的飞秒激光光学频率梳噪声测量系统。对 $f_{rep}$ 和 $f_{ceo}$ 的频率噪声功率谱进行了高精度的测量。首次使用光纤延迟线系统对飞秒激光光纤频率梳梳齿的频率噪声功率谱进行了测量。噪声测量结果与 Paschotta 的理论模型和“固定点”模型相吻合。研究了光学频率中所有梳齿的噪

声来源，噪声最低的梳齿位于 174 THz 频率附近。研究了 $n \times f_{rep}$ 噪声与 $f_{ceo}$ 噪声之间的反相关特性。反相关特性的研究为光学频率梳的噪声性能提升提供了另一个独特的角度。研究了 $f_{rep}$、$f_{ceo}$ 和 $\nu_n$ 的锁相环环路之间的串扰。实验发现，如果 $f_{rep}$ 的锁定带宽过高，压电陶瓷会对 $\nu_n$ 引入额外的噪声。反之，$\nu_n$ 的锁定不会对 $f_{rep}$ 的噪声特性产生任何影响。在 $f_{ceo}$ 的锁定过程中，泵浦电流调制会对 $n \times f_{rep}$ 和 $\nu_n$ 的噪声功率谱上产生明显的调制。$f_{rep}$ 与 $\nu_n$ 的两锁相环之间的串扰从一个独特的角度揭示了压电陶瓷调节激光器重复频率的过程。该实验得到的结论为使用 Pound-Drever-Hall 技术、电子锁定技术以及光学平衡互相关技术锁定激光器的重复频率提供了理论指导。该噪声的测量与分析方法不仅能被用于飞秒激光器中，还可以用于其他新型光学频率梳，例如电光调制频率梳、微环谐振腔频率梳等光源的噪声动力学研究中。

5、首次实现了两台独立的掺镱光纤飞秒激光器的长期稳定相干脉冲合成。通过平衡光学互相关系统锁定重复频率，在实现了脉冲的高精度同步之后，通过零差检测提取两脉冲的相对的载波包络相位，并通过腔外声光调制器的反馈控制将其锁定至零。合成后，两脉冲的相对剩余定时抖动的积分值为 380 as，剩余的相对载波-包络噪声的积分值为 375 mrad。使用 Mach-Zehnder 干涉仪对相干合成后的脉冲的相干性进行评估，合成后的光谱干涉的对比度为 58%。60 分钟内，两脉冲的相对载波-包络相位漂移的均方根值为 165 mrad。并且两激光器的光斑有明显的干涉条纹。独立光纤激光器之间的长期稳定相干合成是亚单周期脉冲合成、光学频率梳远距离传输等应用的基础。

## 7.2 展望

对本工作的展望主要包括以下几点：

1、在光纤链路引入的定时抖动测量实验中，基于互相关算法的光学外差法测量定时抖动的噪声基底主要受限于声光调制器的驱动器引入的相位噪声。因此，可以将该实验改进为测量两台单片固体飞秒激光器的定时抖动，以消除声光调制器对测量结果的影响。搭建两套相同的光学外差探测系统，将测量得到的两个定时抖动功率谱使用互相关算法消除多个独立探测器所产生的噪声基底，实现优于幺秒量级分辨率的定时抖动测量。

2、非对称光纤干涉仪作为一种噪声测量技术的同时，也可以对飞秒激光器的重复频率或梳齿线宽进行稳定。因此，可以使用光纤延迟线稳频技术，替代昂贵的光学超稳腔，实现窄线宽的光学频率梳运转。具体方案有如下几种：（i）可以将飞秒激光器的重复频率用公里量级的光纤延迟线稳定，再将载波-包络偏移

频率锁定至重复频率的分频上，以实现无需外部射频频率基准参考的光学频率梳系统。（ii）可以在参考至射频基准的光学频率梳系统中（重复频率和载波-包络偏移频率均锁定至射频基准），使用光纤延迟线稳频技术进一步压缩梳齿线宽，替代昂贵的光学超稳腔。（iii）可以将光学频率梳中的两个波长，如 1540 nm 和 1560 nm 用同一个光纤延迟线进行稳频。将 1540 nm 梳齿的误差信号反馈至腔内压电陶瓷和电光调制器，实现锁定激光器腔长的快速和慢速锁定、将 1560 nm 梳齿的误差信号反馈至泵浦电流和腔外声光频移器，实现偏移频率的快速和慢速锁定。

3、可以将两台飞秒激光器使用同一光纤延迟线稳频，利用光纤的色散，产生具有重复频率差的双飞秒激光光学频率梳，具体实验方案有如下两种：（i）将两台激光器的载波-包络偏移频率分别锁定至射频基准。将一台激光器的 1540 nm 波段的梳齿用光纤延迟线稳频，将另一台激光器的 1560 nm 波段的梳齿用同一套光纤延迟线稳频，利用公里量级光纤的色散，可以实现具有重复频率差的两台光学频率梳。（ii）使用一台激光器的 1530 nm 波段和 1550 nm 波段的误差信号将重复频率锁定至光纤延迟线，使用另一台激光器的 1540 nm 波段和 1560 nm 波段的误差信号将重复频率锁定至同一光纤延迟线。同样地，利用公里量级光纤的色散，可以实现具有重复频率差的两飞秒激光器的运转。具有重复频率差的双飞秒激光光学频率梳系统可以作为距离测量和光谱学测量的光源，具有广阔的应用前景。

4、可以使用非对称光纤干涉仪对孤子分子对中的两个光学孤子的相对相位噪声进行测量。具体原理为：孤子分子对中的两个光学脉冲分别为脉冲 a 与脉冲 b，将两个光学脉冲用分束器分为两路，一路经过参考臂，另一路经过光纤延迟臂后与参考臂合束。调节延迟臂的光纤长度，使得参考臂中的脉冲 a 与延迟臂中的脉冲 b 做拍。这样得到的拍频信号包含脉冲 a 与脉冲 b 之间的相对相位噪声，实现相对相位噪声的测量。

# 参考文献

# 发表论文和参加科研情况说明

已经发表和录用论文：

已公开专利：